\documentclass[preprint,usenatbib,twocolumn]{aa}  
\usepackage{graphicx}
\usepackage{verbatim}
\usepackage{threeparttable}
\usepackage{subcaption}
\usepackage{enumitem}
\setlist[itemize]{noitemsep, topsep=0pt}
\usepackage{tabu}
\usepackage{gensymb}
\usepackage{txfonts}
\usepackage{caption}
\usepackage{makecell}
\usepackage{afterpage}
\usepackage{multirow}
\usepackage[]{hyperref}
\usepackage[normalem]{ulem}
\usepackage{subcaption}
\newcommand{\edits}[1]{#1}             
\usepackage{color}
\newcommand{\editstwo}[1]{#1} 	
\newcommand{\editsjmg}[1]{#1} 	

\usepackage{afterpage}
\defcitealias{Turner2019}{T19}

\begin{document} 
\title{The search for radio emission from the exoplanetary systems 55 Cancri, $\upsilon$~Andromedae, and $\tau$~Bo\"{o}tis using LOFAR beam-formed observations
}
   \author{Jake D. Turner\inst{1,2},
           Philippe Zarka\inst{3,4},
           Jean-Mathias Grie{\ss}meier\inst{3,5},
           Joseph Lazio\inst{6},
           Baptiste Cecconi\inst{3,4},
           J. Emilio Enriquez\inst{7,8},
           Julien N. Girard\inst{9,10},
           Ray Jayawardhana\inst{1},
           Laurent Lamy\inst{4},
           Jonathan D. Nichols\inst{11},
           Imke de Pater\inst{12}
          }
   \institute{Department of Astronomy and Carl Sagan Institute, Cornell University, Ithaca, NY, USA\\
              \email{jaketurner@cornell.edu}
     \and Department of Astronomy, University of Virginia, Charlottesville, VA, USA  
    \and  Station de Radioastronomie de Nan\c{c}ay, Observatoire de Paris, PSL Research University, CNRS, Univ. Orl\'{e}ans, OSUC, 18330 Nan\c{c}ay, France
     \and LESIA, Observatoire de Paris, CNRS, PSL, Meudon, France
     \and Laboratoire de Physique et Chimie de l'Environnement et de l’Espace (LPC2E) Universit\'{e} d’Orl\'{e}ans/CNRS, Orl\'{e}ans, France
    \and Jet Propulsion Laboratory, California Institute of Technology, Pasadena, CA, USA  
     \and Department of Astronomy, University of California, Berkeley, 501 Campbell Hall $\#$3411, Berkeley, CA, 94720, USA 
    \and Department of Astrophysics/IMAPP, Radboud University, P.O. Box 9010, NL-6500 GL Nijmegen, The Netherlands 
    \and Department of Physics and Electronics, Rhodes University, PO Box 94, Grahamstown 6140, South Africa 
    \and AIM, CEA, CNRS, Universit\'{e} Paris-Saclay, Universit\'{e} Paris Diderot, Sorbonne Paris Cit\'{e}, F-91191 Gif-sur-Yvette, France 
    \and Department of Physics and Astronomy, University of Leicester, Leicester, UK
    \and Department of Astronomy, University of California at Berkeley, Berkeley, CA, USA  
        }
   \date{}

\abstract
   {
The detection of radio emissions from exoplanets will open up a vibrant new research field. Observing planetary auroral radio emission is the most promising method to detect exoplanetary magnetic fields, the knowledge of which will provide valuable insights into the planet's interior structure, atmospheric escape, and habitability.
   }
   {
  We present LOFAR (LOw Frequency ARray) Low Band Antenna (LBA: 10-90 MHz) circularly polarized beamformed observations of the exoplanetary systems 55 Cancri, $\upsilon$~Andromedae, and $\tau$~Bo\"{o}tis. All three systems are predicted to be good candidates to search for exoplanetary radio emission. 
   }
   {
   We applied the \texttt{BOREALIS} pipeline that we have developed to mitigate radio frequency interference and searched for both slowly varying and bursty radio emission. Our pipeline has previously been quantitatively benchmarked on attenuated Jupiter radio emission.
   }
   {
   We tentatively detect circularly polarized bursty emission from the $\tau$~Bo\"{o}tis system in the range 14-21 MHz with a flux density of $\sim$890 mJy and with a statistical significance of $\sim$3$\sigma$. For this detection,
   \edits{we do not see any signal in the OFF-beams, \editsjmg{and we do not find any potential causes which might cause false positives}.}
   \edits{We also tentatively detect} slowly variable \edits{circularly polarized} emission from $\tau$~Bo\"{o}tis in the range 21-30 MHz with a flux density of $\sim$400 mJy \editsjmg{and} with a statistical significance of \editsjmg{$>$8$\sigma$}. The slow emission is structured in the time-frequency plane and \edits{shows an excess} in the ON-beam with respect to the two simultaneous OFF-beams. 
   \editsjmg{While the bursty emission seems rather robust, close examination} casts some doubts on the reality of the 
   \editsjmg{slowly varying} signal. We discuss in detail all the arguments for and against an actual detection, and derive methodological tests that will \editsjmg{also} apply to future searches. Furthermore, a $\sim$2$\sigma$ marginal signal is found from the $\upsilon$~Andromedae system \edits{in one} observation of bursty emission in the range 14-38 MHz and no signal is detected from the 55 Cancri system, on which we placed a 3$\sigma$ upper limit of 73 mJy for the flux density at the time of the observation.
   }
  {
  Assuming the detected signals are real, we discuss their potential origin. 
  Their source probably is the $\tau$ Bootis planetary system, and a
  possible explanation is radio emission from the exoplanet $\tau$ Bootis b via the cyclotron maser mechanism. 
  Assuming a planetary origin, we derived limits \editsjmg{for} the \editsjmg{planetary} polar surface magnetic field strength, finding values compatible with theoretical predictions.
 Further observations with LOFAR-LBA and other low-frequency telescopes, such as NenuFAR or UTR-2, are required to confirm this possible first detection of an exoplanetary radio signal. 
}


   \titlerunning{Search for radio emission from 55 Cancri, $\upsilon$~Andromedae, and $\tau$~Bo\"{o}tis}
   \authorrunning{J.D. Turner, P. Zarka, J.-M. Grie{\ss}meier, et al} 
   \maketitle



\section{Introduction} 
The direct detection of exoplanetary magnetic fields has been elusive despite decades of searching.
All the planets in our Solar System, except Venus, have or used to have a magnetic field \citep{Stevenson2003}, and theoretical scaling laws predict that many exoplanets might have one as well (e.g., \citealt{SL2004,Gr2007,Christensen2009}). Measuring the magnetic field of an exoplanet will give valuable information to constrain its interior structure (composition and thermal state), its atmospheric escape, and the nature of any star-planet interaction \citep{Hess2011,Zarka2015SKA,G2015,Lazio2016,Griessmeier17PREVIII,Lazio2018,Griessmeier2018haex,Zarka2018haex,Lazio2019}. Historically, some of the first constraints on the interior structure of the Solar System gas giants came from the knowledge that they had magnetic fields (\citealt{Hubbard1973}). Magnetic drag caused by an exoplanet's magnetic field on its atmosphere (e.g., \citealt{Perna2010a,Rauscher2013,Rogers2014b}) could be an important factor for the atmosphere's dynamics and evolution, and it could contribute to the anomalously large radii of hot Jupiters (\citealt{Laughlin2018haex}). Additionally, the magnetic field of Earth-like exoplanets might contribute to their sustained habitability by deflecting energetic stellar wind particles and cosmic rays (e.g., \citealt{Gr2005,Gr2009,Gr2015_cosmic,Gr2016,Lammer2009,Kasting2010,Owen2014,Lazio2009,Lazio2016,McIntyre2019}).\\
\indent Many methods have been proposed to study the magnetic fields of exoplanets. A full description of all available methods (excluding the recent method described by \citealt{Oklopcic2019} using spectropolarimetric transits of the helium line) can be found in \citet{G2015}. The methods most extensively discussed in the literature are observations of radio emission (\citealt{Farrell1999,Zarka2001,Zarka2007}), optical signatures of star-planet interactions (SPI; \citealt{Cuntz2000,Shkolnik2003,Shkolnik2005,Shkolnik2008,Cauley2019}), and near-ultraviolet light curve asymmetries (\citealt{Vidotto2010,Vidotto2011a,Llama2011}). The latter two methods have many astrophysical scenarios that can create false-positives (e.g., \citealt{Preusse2006,Lai2010,Kopp2011,Miller2012,Miller2015,Bisikalo2013A,Saur2013,Alexander2016,Kislyakova2016,Turner2016a,Gurumath2018IAUS,Route2019}). By contrast, radio observation can constrain the magnetic field amplitude directly, without invoking complex model assumptions, and is less susceptible to false positives (\citealt{G2015}). In this paper, we focus on the detection of exoplanetary magnetic fields \edits{via} radio emission.  

\begin{figure}[!htb]
    \centering
    \includegraphics[width=0.5\textwidth]{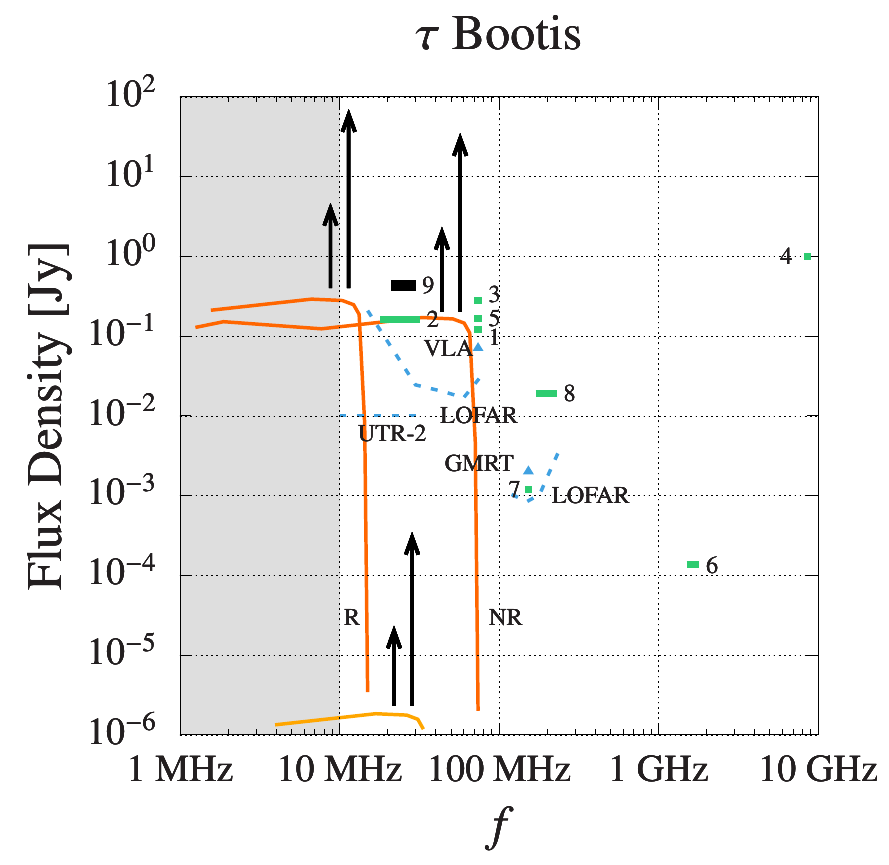}
    \caption{Predictions and observations for the exoplanet $\tau$~Bootis~b. 
    Gray area: Emission below 10 MHz is not detectable for ground-based observations (ionospheric cutoff).
    Lower solid line (light orange): 
    \edits{Typical spectrum of Jupiter's radio emissions} at 15.6 pc distance. 
    Two upper solid line (orange): Jupiter's radio emission scaled for values expected for $\tau$~Bootis~b according to models \edits{R} (left) and  \edits{NR} (right) from \citet{Griessmeier17PREVIII}. 
    \editsjmg{For Jupiter, the radio flux increases 
    during periods of high activity
    \citep[frequently by one order of magnitude, and exceptionally by two orders of magnitude, see e.g.][]{Zarka2004,Zarka2004b}; the same variability is assumed for exoplanetary emission, as indicated by the vertical black arrows.}
    Dashed lines and triangles (light blue) show the theoretical sensitivity limit of the radio-telescopes UTR-2, LOFAR (low band), VLA, LOFAR (high band), and GMRT
    \edits{(for 1 h of integration time and a bandwidth of 4 MHz, or an equivalent combination)}. 
    Numbered lines and points (\edits{green}): sensitivity achieved in previous observations of $\tau$~Bootis. Care must be taken when comparing the theoretical sensitivity limit to previous observations. The sensitivity limits of the radio telescopes as well as most upper limits were calculated for continuous emission, whereas the values for observations 2 and 9 take into account the expected ``bursty'' nature of the emission.
    }
    \tablebib{(1) \citet{Farrell2003}; (2) \citet{Ryabov2004}; (3) \citet{Lazio2004}; (4) \citet{Shiratori2006}; (5) \citet{Lazio2007}; (6) \citet{Stroe2012}; (7) \citet{Hallinan2013}; (8) \citet{Lynch2018}; (9) this article, Section \ref{sec:tauboo_detection}.}
    \label{fig:radioprediction:tauBootis}
\end{figure}

 All the magnetized planets and moons in our Solar System emit or induce radio emissions via the Cyclotron Maser Instability (CMI) mechanism (\citealt{Wu1979,Zarka1998,Treumann2006}). The first proof and measurement of Jupiter's magnetic field, the first measured magnetic field of a planet other than Earth, came from observing its decametric radio emission (\citealt{Burke1955}). Planetary CMI radio emission is caused by electrons accelerated to energies of several keV by the interaction of the stellar wind or coronal mass ejections with the magnetosphere or by acceleration processes inside the magnetosphere, resulting from magnetosphere-ionosphere or magnetosphere-satellite coupling (\citealt{Cowley2003,Griessmeier07PSS,Zarka2007,Griessmeier17PREVIII,Zarka2018haex}).
 CMI emission is highly circularly polarized, beamed, and time-variable (e.g., \citealt{Zarka1998,Zarka2004}). 
 It is produced at the local electron cyclotron frequency (or gyrofrequency) in the source region;
 \editsjmg{its spectrum sharply drops off at a maximum gyrofrequency $\nu_{g}$, which is determined by the maximum 
 magnetic field $B_{p}$ near the planetary surface, as}
  $\nu_{g} [MHz] =  2.8 \times B_{p} [G]$
 (\citealt{Farrell1999}).
 \editsjmg{This sharp drop-off can also be seen, such as, in Fig.~\ref{fig:radioprediction:tauBootis}}. 

Recent reviews by \citet{Zarka2015SKA} and \citet{G2015,Griessmeier17PREVIII} summarize the observational campaigns and theoretical work concerning radio emission from exoplanets. Following \edits{several seminal} works \editsjmg{(\citealt{Winglee1986,Zarka1997pre4,Farrell1999,Zarka2001,Zarka2007})}, 
an extensive body of theoretical work has been published (e.g., \citealt{Pater2000pras,Farrell2004,Lazio2004,Stevens2005,Griessmeier05AA,Griessmeier07PSS,Gr2007,Jardine2008,Vidotto2010r,Hess2011,Nichols2011,Nichols2012,Vidotto2012,See2015,Vidotto2015,Nichols2016,Fujii2016,Weber2017,Weber2017pre8,Weber2018,Lynch2018,Zarka2018haex,Wang2019,Kavanagh2019,Shiohira2020}).

One of the goals of these \editsjmg{theoretical} studies is to predict the radio flux and frequency of emission as observed from Earth. However, these predictions are hardly more than educated guesses, with associated uncertainties estimated as one order of magnitude for the flux density and a factor of 2-3 for the maximum emission frequency (\citealt{Gr2007}). For example, different
rotational-independent \editsjmg{and} rotational-dependent scaling laws have been employed to predict exoplanetary magnetic-field strengths (e.g., \citealt{SL2004,Gr2007,Christensen2009,Reiners2010,Griessmeier17PREVIII} -- see \citealt{Christensen2010} for an overview of all scaling laws in the literature). \editsjmg{Similarly, the} Radiometric Bode's law, successfully used to predict 
the radio fluxes from Uranus and Neptune (\citealt{Desch1984,Warwick1986,Desch1988,Million1988,Warwick1989}), is often used to estimate the intensity of emission from exoplanets (e.g., \citealt{Farrell1999,Lazio2004}), especially in its Radio--Magnetic version (\citealt{Zarka2007,Gr2007,Zarka2018,Zarka2018haex}). In particular, using \editsjmg{a rotational-dependent scaling law} \cite{Griessmeier17PREVIII} finds 15 exoplanets with flux densities \edits{potentially} above the theoretical detection limit \editsjmg{\citet{Turner2017pre8} derived for the beamformed mode of the Low-Frequency Array (LOFAR; \citealt{vanHaarlem2013}).}

\begin{table*}[!tbh]
    \centering
        \caption{\editsjmg{
        Stellar and planetary parameters for the exoplanets observed in this study.
         Row 2: spectral type of host star. Row 3: stellar distance. Row 4: stellar age. 
           Row 5: orbital distance. Row 6: planetary mass. Row 7: planetary radius.
           Row 8: expected maximum emission frequency (under the hypothesis planetary rotation does not have any influence on the planetary magnetic field). 
           Row 9: maximum expected radio flux density at Earth (under the hypothesis planetary rotation does not have any influence on the planetary magnetic field). 
           Row 10: expected maximum emission frequency (under the hypothesis planetary rotation has a strong influence on the planetary magnetic field). 
           Row 11: maximum expected radio flux density at Earth (under the hypothesis planetary rotation has a strong influence on the planetary magnetic field).
        }}
    \begin{tabular}{lc|cccc}
    \hline
    \hline
    & units & 55 Cnc b & 55 Cnc e & $\upsilon$~And b & $\tau$~Boo b \\
    \hline 
    * type                   &                           & G8V  \tablefootmark{a}            & G8V  \tablefootmark{a}  &  F9V     & F7V    \\
    $d$                         & [pc]                      & 12.5 \tablefootmark{b}           & 12.5 \tablefootmark{b}   &  13.5 \tablefootmark{b}     &  15.6\tablefootmark{b}   \\ 
    $t_\star$                   & [Gyr]                     & $10.2\pm2.2$ \tablefootmark{a}        & $10.2\pm2.2$ \tablefootmark{a}  & 3.8$\pm1$ \tablefootmark{b}     &  1.0$\pm$0.6\tablefootmark{b}   \\
    \hline
    $a$                         & [AU]                      & $0.114$\tablefootmark{a}  & $0.0156$ \tablefootmark{a} &      0.057\tablefootmark{c} &  0.0462\tablefootmark{c}   \\    
    $M_\text{p}$                & [$M_\text{J}$]            & 0.81\tablefootmark{a}   & 0.024\tablefootmark{a}     & $\ge$0.68\tablefootmark{c} & $\ge$3.87\tablefootmark{c} \\
    $R_\text{p}$                & [$R_\text{J}$]            & unknown                 & 0.194\tablefootmark{a}    &    unknown    & unknown \\
    \hline    
    $\nu_{max}^{NR}$            & [MHz]                     & 20 \tablefootmark{d}   & 30\tablefootmark{d}        & 14\tablefootmark{d}       & 74\tablefootmark{d} \\  
    $\Phi_{max}^{NR}$           & [mJy]                     & 2.9\tablefootmark{d}    & 150\tablefootmark{d}       & 75\tablefootmark{d}       & 170\tablefootmark{d} \\   
    \hline
    $\nu_{max}^{R}$             & [MHz]                     & 3.3\tablefootmark{d}    & 19\tablefootmark{d}      & 2.2\tablefootmark{d}       & 15\tablefootmark{d} \\  
    $\Phi_{max}^{R}$            & [mJy]                     & 5.3\tablefootmark{d}   & 170\tablefootmark{d}       & 140\tablefootmark{d}       & 290\tablefootmark{d} \\  
    \hline
    \end{tabular}
    \tablefoot{
            \tablefoottext{a}{\citet{Fischer2018haex}}
            \tablefoottext{b}{\citet{Fuhrmann1998}}
            \tablefoottext{c}{\citet{Butler1997}}
            \tablefoottext{d}{\citet{Griessmeier17PREVIII}}
            
        }
    \label{tb:predictions}
\end{table*}

In parallel to these theoretical studies, a number of ground-based observations have been conducted to find radio emission from exoplanets, most of which have resulted in clear non-detections
(\citealt{Yantis1977,Winglee1986,Zarka1997pre4,Bastian2000,Farrell2003,Ryabov2004,Shiratori2006,George2007,Lazio2007,Smith2009,Lecavelier2009,Lecavelier2011,Lazio2010a,Lazio2010b,Stroe2012,Hallinan2013,Murphy2015,Lynch2017,Turner2017pre8,Gorman2018,Lenc2018,Lynch2018,Green2020,Narang2020,Gasperin2020}). Most of these studies have involved imaging observations and only span a small fraction of the planetary orbit. 
\editsjmg{A number of possible reasons could account for the non-detections}
(see, e.g., \citealt{Bastian2000,Hallinan2013,Zarka2015SKA,G2015,Griessmeier17PREVIII}): (1) the observations were not sensitive enough, (2) the planetary magnetic field is not strong enough for emission at the observed frequencies, (3) 
\editsjmg{The} Earth was outside the beaming pattern of the radio emission at the time of the observations (\citealt{Hess2011}), 
and (4) CMI quenching due to the plasma frequency of the planet's ionosphere being greater than the cyclotron frequency \citep{Gr2007,Weber2017,Weber2017pre8,Lamy2018,Weber2018}.

There have been a few tentative detections \citep{Lecavelier2013,Sirothia2014,Vasylieva2015,Bastian2018} but none of these have been confirmed by follow-up observations. The recent detection of 8--12 GHz emission from SIMP0136 is the first radio signal of an object near the brown dwarf and planetary boundary ($12.7 \pm 1.0 M_{Jup}$; \citealt{Saumon2008}) and opens up the possibility of detecting free-floating planets (\citealt{Kao2016,Kao2018}). Additionally, V830$\tau$ is the first non-degenerate exoplanet host-star detected to emit radio emission (\citealt{Bower2016}) and continued radio monitoring of the system may allow for the detection of star-planet interactions (see \citealt{Loh2017}, Loh et al. in prep). \edits{Recent LOFAR observations from the LoTSS survey (\citealt{Shimwell2017,Shimwell2019}) \edits{revealed} low-frequency radio emission from the M-dwarf GJ 1151 in the range 120-167 MHz, suggested to be caused by the interaction with a close-in Earth-size planet (\citealt{Vedantham2020,Pope2020,Foster2020}).}

In this study, we analyze LOFAR Low Band Antenna (LBA) beamformed observations for three exoplanetary systems (55 Cancri, Upsilon Andromedae and $\tau$~Bo\"{o}tis). Our data are the first beamformed observations of exoplanets performed with LOFAR. This work extends the preliminary analysis of the LOFAR observation of 55 Cnc published in \cite{Turner2017pre8}. 

\indent In this paper, 
\editsjmg{Section \ref{sec:targets} gives relevant information on the three planetary systems we observed.}
Sections \ref{sec:obs} and \ref{sec:pipeline} describe the observations and data processing, respectively. The analysis of individual observations can be found in Section \ref{sec:dataAnalysis}. The implications of our tentative detections \editsjmg{are discussed in Sections \ref{sec:discussion}. Section \ref{sec:conclusion} contains conclusions and suggestions for future steps}. \edits{Extensive supporting material is presented in the appendices.} 

\section{\editsjmg{Target selection}} \label{sec:targets}

In this study, we analyze LOFAR-LBA beamformed observations of the planetary systems
55 Cancri, $\upsilon$ Andromedae, and $\tau$~Bo\"{o}tis. Based on their proximity to the Solar System, the stellar age, the estimated planetary mass, and the small orbital distance of the planet, these systems were predicted to be good candidates to search for radio emission (\citealt{Gr2007,Griessmeier17PREVIII}). All three systems are also considered prime candidates for the search for star-planet interactions (\citealt{Shkolnik2005,Shkolnik2008,Folsom2020}).

\editsjmg{The relevant parameters of the planets and their host stars are summarised in Table \ref{tb:predictions}.
The values for the expected emission frequency 
and the expected radio flux are taken from \citet{Griessmeier17PREVIII}, the data of which are also available electronically at CDS\footnote{ftp://cdsarc.u-strasbg.fr/pub/cats/VI/151/)}. 
We estimate the uncertainties of $\nu_{max}$ to be factor 2-3 and the uncertainties on $\Phi_{max}$ to be approximately one order of magnitude (\citealt{Gr2007}). 
For the magnetic field estimate (required for the calculation of both $\nu_{max}$ and $\Phi_{max}$), we have explored two different models.
In model NR, the planetary rotation does not have any influence on the planetary magnetic field. In model R, the planetary rotation has an influence on the planetary magnetic field (via tidal locking of the planet with its parent star).
Details are given in \citet{Griessmeier17PREVIII}.}.

\editsjmg{An observation campaign makes sense if the expected emission frequency is above the terrestrial ionospheric cutoff ($\sim$10 MHz) and if the expected flux density is higher than the telecope sensitivity. 
\citet[hereafter T19]{Turner2019} 
estimate that the sensitivity of LOFAR is $\sim$50 mJy for circularly polarized flux (Stokes V) in integrations of 2 minutes and 10 MHz. In the following, we compare these numbers to the expected values for the three observed systems.}

\paragraph{55 Cancri (55 Cnc)} 
The G8V star 55 Cancri A hosts one of the first known exoplanets; 
today, five planets are known in this system, and more are expected to exist. 
The star has a visual binary companion of spectral type M4.5V, $\rho^1$ Cnc B, with a projected distance of 1100 AU. The two stars are expected to be gravitationally bound \citep{Fischer2018haex}. For this system, two planets are interesting with respect to radio emission, namely 55 Cnc b and 55 Cnc e.
If the planetary magnetic field does strongly depend on planetary rotation (model R), the emission of 55 Cnc b is below the ionospheric cutoff, but not the emission from 55 Cnc e. If the planetary magnetic field does not strongly depend on planetary rotation (model NR), both planets have emission above the ionospheric cutoff. The expected flux density for 55 Cnc b is low, but the estimate for 55 Cnc e is above the LOFAR detection threshold. As an added benefit, the orbital period of 55 Cnc e (0.74 days) can be easily covered with a few observations. Good orbital coverage is important if emission is only active for certain orbital phases, which is expected. Also, 55 Cnc e is a transiting planet; in the case of detected radio emission, secondary transits are a good way to check a planetary origin of the radio emission. In the case of a detection, flares from the host star (55 Cnc A) and its binary companion (55 Cnc B) will have to be ruled out as potential causes. 

\paragraph{$\upsilon$ Andromedae ($\upsilon$~And)} 
$\upsilon$ Andromedae A is a F9V star with a stellar binary companion, Andromedae B, at a projected distance of $\sim$700 AU from the primary star.
The binary companion $\upsilon$ Andromedae B is of spectral type M4.5V and has been detected in X-rays \citep{Poppenhaeger14}. For model R, the estimated emission frequency is below the ionospheric cutoff, but in model NR, the expected planetary radio emission extends to detectable frequencies. The expected flux density is above the LOFAR detection threshold. Flares from the host star and its binary companion will have to be ruled out as potential causes in the case of a detection.

\paragraph{$\tau$~Bo\"{o}tis ($\tau$~Boo)} 
$\tau$~Bo\"{o}tis A is a hot and young F7V star. Its binary companion $\tau$~Bo\"{o}tis B is of spectral type M3V; it is on a highly eccentric orbit ($e=0.87$) with a semimajor axis of $\sim$ 220 AU \citep{Justesen19}. For model R, the expected planetary radio emission only slightly exceeds the ionospheric cutoff limit. Nominally, the flux exceeds the detection threshold, but the estimate of 50 mJy does not take into account the reduced sensitivity of LOFAR at 15 MHz. However, for model NR the emission extends to detectable frequencies, and the expected flux density is above the detection threshold. As for the two other systems, flares from the host star and its binary companion will have to be ruled out as potential causes in the case of a detection.

Figure \ref{fig:radioprediction:tauBootis} compares the predicted radio flux for $\tau$ Boo b to the sensitivity achieved in previous observations. This figure makes the motivation behind our observations tangible: radio fluxes predicted by at least some of the models are compatible with the theoretical sensitivity limits of several radio-telescopes. In particular, according to the model NR, 
radio emission from $\tau$~Boo~b should be detectable by UTR-2 and LOFAR (low band), and possibly by the VLA.

\section{Observations} \label{sec:obs}

\begin{table}[!tb]
\begin{threeparttable}
\centering
\caption{Setup of the LOFAR-LBA beamformed observations}
\begin{tabular}{ccc}
\hline 
\hline
Parameter   & Value  & Units   \\ 
\hline
\hline 
Array Setup                     & LOFAR Core                &     \\
Number of Stations              & 24                        &\\
Beams                           & ON $\&$ 2 OFF                & \\ 
Configuration                   & LBA outer antennas        &    \\
Antennas per Station            & 48                        &            \\
Minimum Frequency               & 26\tnote{a}\hspace{1ex} or 15\tnote{b}      & MHz   \\
Maximum Frequency               & 74\tnote{a}\hspace{1ex} or 62\tnote{b}    & MHz   \\
Subbands Recorded               & 244                        &    \\
Subband Width                    & 195                       &kHz \\
Channels per Subband            & 64                        &   \\
Frequency Resolution  ($b$)     & 3.05                      & kHz   \\
Time Resolution ($\tau$)        & 10.5                      & msec  \\
Beam Diameter\tnote{c}          & 13.8                      & arcmin    \\
Raw Sensitivity\tnote{d}\hspace{1ex}($\Delta S$) & 208                       & Jy    \\ 
Stokes Parameters               & IQUV                      &   \\
\hline
\end{tabular}
\begin{tablenotes}
\item[a]    Frequency range for the 55 Cnc observations.
\item[b]    Frequency range for the $\upsilon$ And and $\tau$~Boo observations.
\item[c]    Calculated at 30 MHz (\citealt{vanHaarlem2013}).
\item[d]    Raw sensitivity of a pixel in the dynamic spectrum, calculated using $\Delta S= S_{sys}/N\sqrt{n_{pol}\tau b }$, with $S_{sys}$ the system equivalent flux density (SEFD) of one LOFAR Core station (40 kJy; \citealt{vanHaarlem2013}), $N$ the number of stations summed, $n_{pol}$ is the number of polarizations (2), $b$ the frequency resolution, and $\tau$ the time resolution.
\end{tablenotes}
\label{tb:setup}
\end{threeparttable}
\end{table}

Our observations were taken with the LOw Frequency ARray (LOFAR; \citealt{vanHaarlem2013}), using its Low Band Antenna (LBA, 10-90 MHz) in beamformed mode (\citealt{Stappers2011}). The setup used for our observations can be found in Table \ref{tb:setup}. The exact dates and times of each observation can be found in \editsjmg{Appendix \ref{app:Summary_obs} (Table \ref{tb:obs}), 
 which also gives the coordinates used for the ON and OFF beams (Table \ref{tb:beam})}.

The data streams from all 24 \edits{core} stations were summed together during the observations. All the processing steps described in this paper are performed on this summed data product. This means it is not possible to flag out the contribution of any of these stations a posteriori if any station does not operate optimally, \editsjmg{The stations were phased up} to produce multiple simultaneous \edits{digital beams within the broader individual station beams}: one \editsjmg{pointing at the target, and two pointing at different OFF sky positions (used for comparison in the data processing pipeline, see below).} 

We focus on the Stokes-V data.
\editsjmg{\citetalias{Turner2019} have shown that this allows one to detect signal that are 1 to 1.5 orders of magnitude weaker than when using Stokes-I data (provided that the signal is circularly polarized, which is expected here).} The observations were taken at high temporal and spectral resolutions 
\edits{($\tau$ and $b$)} in order to allow proper RFI mitigation. 

 As in our previous studies \citep{Turner2017pre8,Turner2019}, we compare our on-target beam (``ON-beam'') to several simultaneous beams pointing to a nearby location in the sky (``OFF-beam 1'' and ``OFF-beam 2''). A fundamental assumption 
 \editsjmg{of this method} is that the OFF beams provide a good characterization of the ionospheric fluctuations, RFI, and any systematics present in the ON-beam. \edits{For that purpose, the OFF beams are well within the station beams (10$^{\circ}$ FWHM at 30 MHz), far enough from the ON-beam to not overlap within the beam diameter (13.8\arcmin at 30 MHz), and also close enough to the ON-beam to be within the}
 \edits{ionospheric isoplanatic patch (7\degr at 30 MHz; \citealt{Intema2009} and personal communication from C. Vogt).}
 
 \edits{For this project}, 22 exoplanet observations were taken with a total of 89 hours, plus 16 observations of B0809+74 totaling 197 minutes. The observations were all obtained during night-time to avoid \editsjmg{contamination by strong RFI}. During the observing period May 19--November 22, 2016 the LBA calibration tables were unreadable, resulting in 20$\%$ higher noise (communication from ASTRON Radio Observatory staff). Both the 55 Cnc and $\upsilon$ And data are affected by this calibration error. 

\begin{figure*}[!htb]
    \centering
    \begin{tabular}{cc}
      \begin{subfigure}[c]{0.4\textwidth}
        \centering
        \caption{}
       \includegraphics[width=\textwidth]{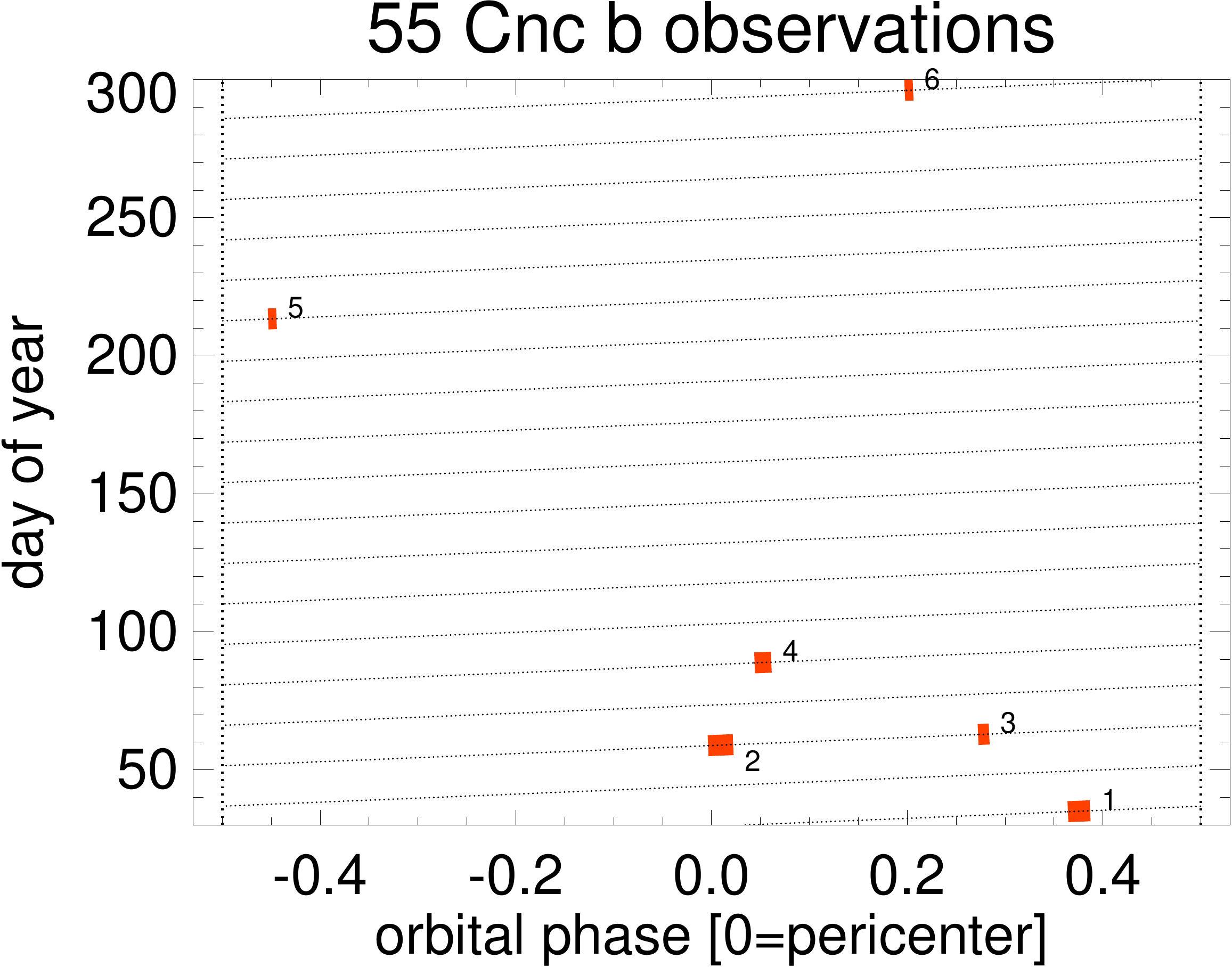}
        \label{}
    \end{subfigure}%
    & 
         \begin{subfigure}[c]{0.4\textwidth}
        \centering
        \caption{}
       \includegraphics[width=\textwidth]{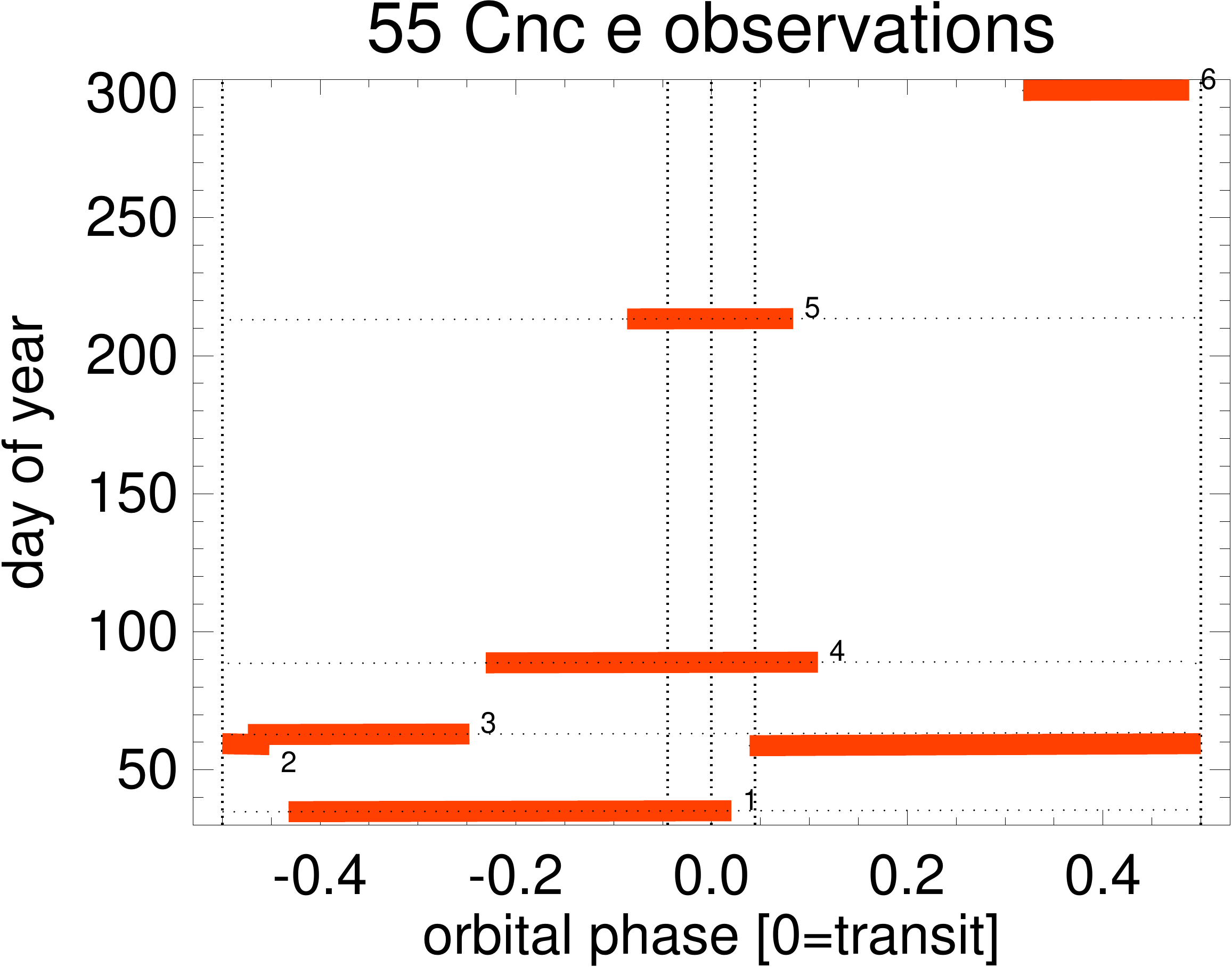}
        \label{}
    \end{subfigure}%
    \\
        \begin{subfigure}[c]{0.4\textwidth}
        \centering
        \caption{}
       \includegraphics[width=\textwidth]{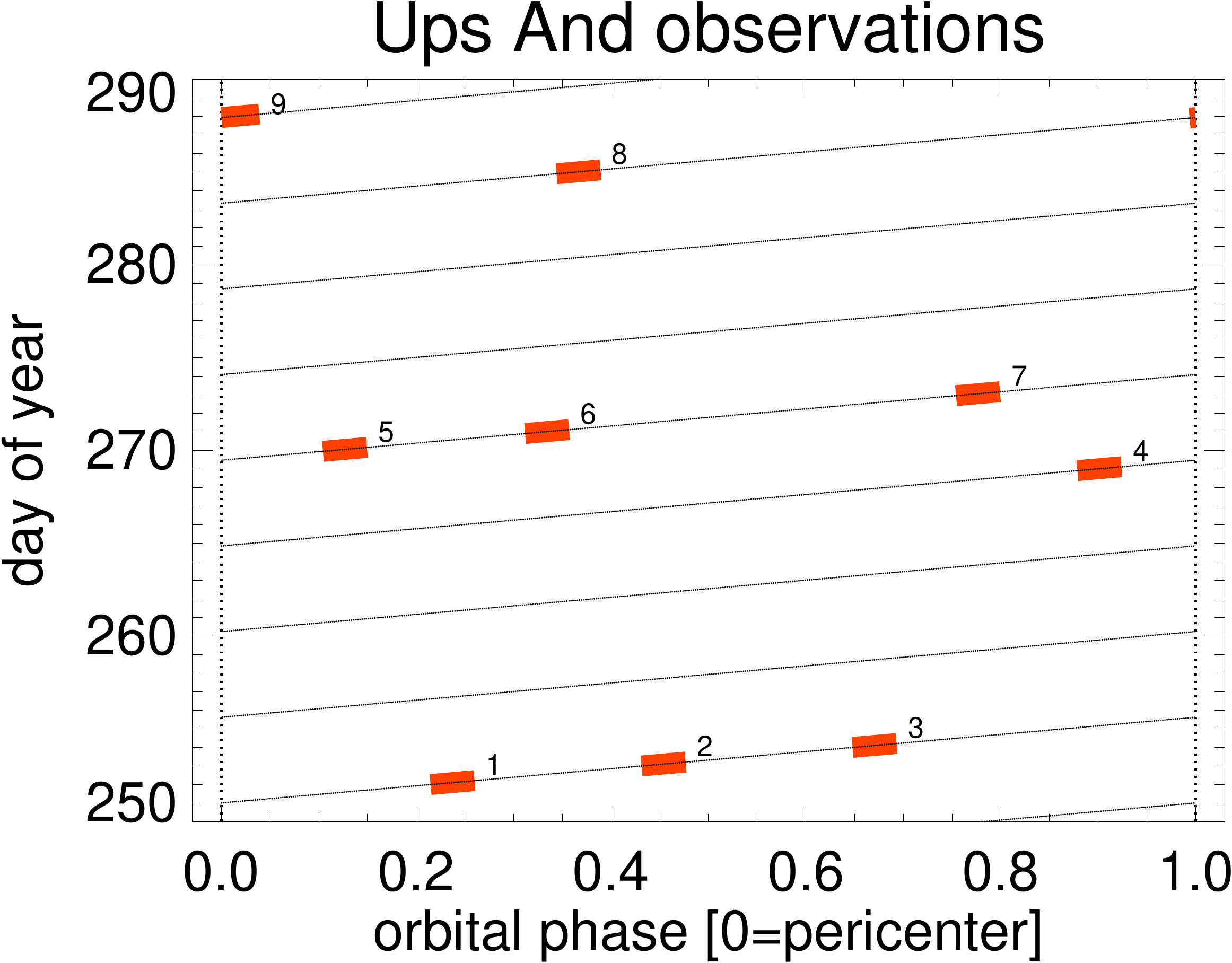}
        \label{}
    \end{subfigure}%
    & 
   \begin{subfigure}[c]{0.4\textwidth}
        \centering
        \caption{}
       \includegraphics[width=\textwidth]{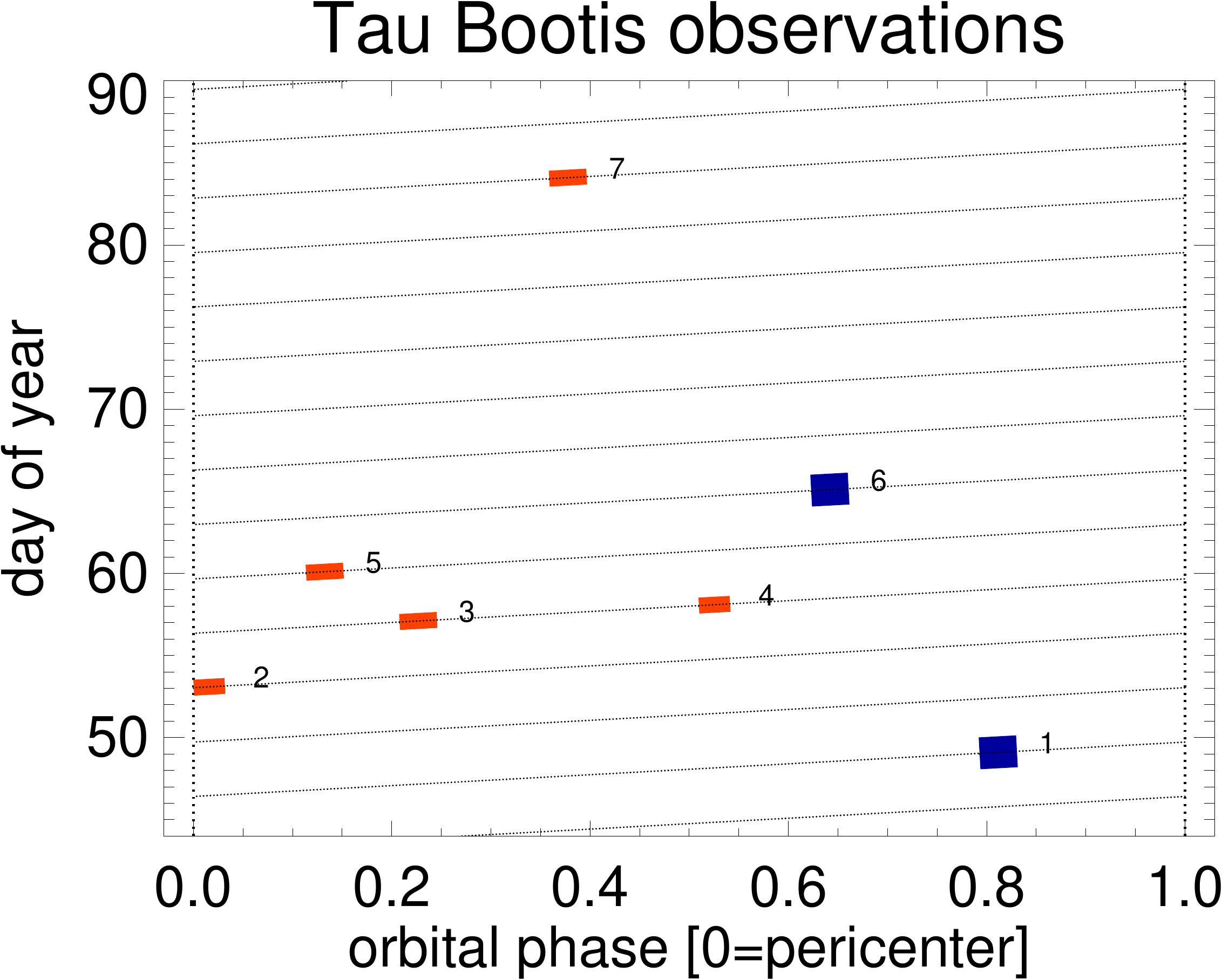}
        \label{}
    \end{subfigure}%
    \\
    \end{tabular}
    \caption{Orbital phase coverage for the observations of 55 Cnc b (\textit{panel a}), 55 Cnc e (\textit{panel b}), $\upsilon$~And~b (\textit{panel c}), and $\tau$~Boo~b (\textit{panel d}). The numbers representing each date are indicted in column 1 of Table \ref{tb:obs}.  We note that 55 Cnc e is a transiting exoplanet. For 55 Cnc e, only a few planetary rotations are shown. For 55 Cne e, the entire orbit is covered by the observations, whereas the orbital coverage is low for 55 Cnc b. Approximately 40$\%$ and 25$\%$ of the orbits of $\upsilon$~And~b and $\tau$~Boo~b  are covered, respectively. The $\tau$~Boo observations with the tentative detections (Section \ref{sec:tauboo_detection}; obs $\#$1 and $\#$6) are displayed as large dark-blue rectangles in \textit{panel d}.
    }
    \label{fig:phase}
\end{figure*}  

For geometrical reasons, we expect the emission to be beamed at Earth only for a fraction of the time (e.g., \citealt{Hess2011}). Assuming the anisotropic beaming is similar to that of Jupiter (\citealt{Zarka2004}), we would expect the emission to be visible from Earth only $\sim$10\% of the time. Also, the intensity of the radio emission is expected to vary as the planet encounters different plasma conditions (particle density, magnetic field) along the orbit (e.g., \editsjmg{\citealt{Griessmeier05AA,Griessmeier07PSS,Fares2009,Vidotto2012}}).
For these reasons, the observation windows were chosen such that the orbital coverage is as wide as possible in order to maximize the chances to detect beamed emission. The orbital coverage of our observations is shown in Figure \ref{fig:phase}. For the observations of 55 Cnc (with a total observing time of 24h), the observation windows can be folded at the orbital period of either 55 Cnc b (Figure \ref{fig:phase}a) or 55 Cnc e (Figure \ref{fig:phase}b). For the short-period \edits{transiting} planet 55 Cnc e, the full orbit is covered \edits{more than once}, increasing the chances of catching beamed emission. Also, secondary eclipses constitute a powerful \editsjmg{tool: If} radio emission is detected, but vanishes during the secondary eclipse, this can be taken as a very strong indication for a planetary origin of the radio emission (e.g., \citealt{Griessmeier05AA,Smith2009,Lecavelier2013}) For 55 Cnc b, the orbital period is longer, and the phase coverage is low (9\%). For $\upsilon$ And (Figure \ref{fig:phase}c), the phase coverage is 40\% with a total observing time of 45h. Finally, for $\tau$ Boo (Figure \ref{fig:phase}d), the phase coverage is 25\% of the orbit with a total observing time of 20h; we cannot exclude that we have missed radio emission concentrated at specific orbital phases. 

\section{Data pipeline } \label{sec:pipeline}
\editsjmg{Based on previous work (\citealt{Vasylieva2015,Turner2017pre8,Turner2019}),
we have developed the BeamfOrmed Radio Emission AnaLysIS  (\texttt{BOREALIS}) pipeline for our exoplanetary beam-formed data. It was applied to the both Stokes-I and Stokes-V data-sets of the observations presented in Section \ref{sec:obs}.}

This pipeline performs RFI mitigation, empirically determines the time-frequency ($t$-$f$) response of the telescope (i.e. the gain), normalizes the data by this $t$-$f$ function, and rebins the data into larger time and frequency bins. The RFI mitigation combines four different techniques (\citealt{Offringa2010,Offringa2012PhD,Offringa2012AA,Zakharenko2013,Vasylieva2015}; and references therein) for optimal efficiency and processing time, following which the data is normalized and rebinned to time--frequency bins of 1 second $\times$ 45 kHz. The expected thermal noise level for the
\edits{rebinned} data is 6 Jy for each measurement. 
The resulting corrected Stokes-V dynamic spectrum ($V^{'}$, defined in equations 9--11 of Section 4.2.1 in \citetalias{Turner2019}) is the input for the post-processing part of the \texttt{BOREALIS} pipeline.

\begin{figure}[!thb]
\centering
    \begin{flushleft} \textbf{(a)} \end{flushleft} \vspace{\baselineskip}
      \centering
      \subfloat{\includegraphics[page=1,width=0.50\textwidth]{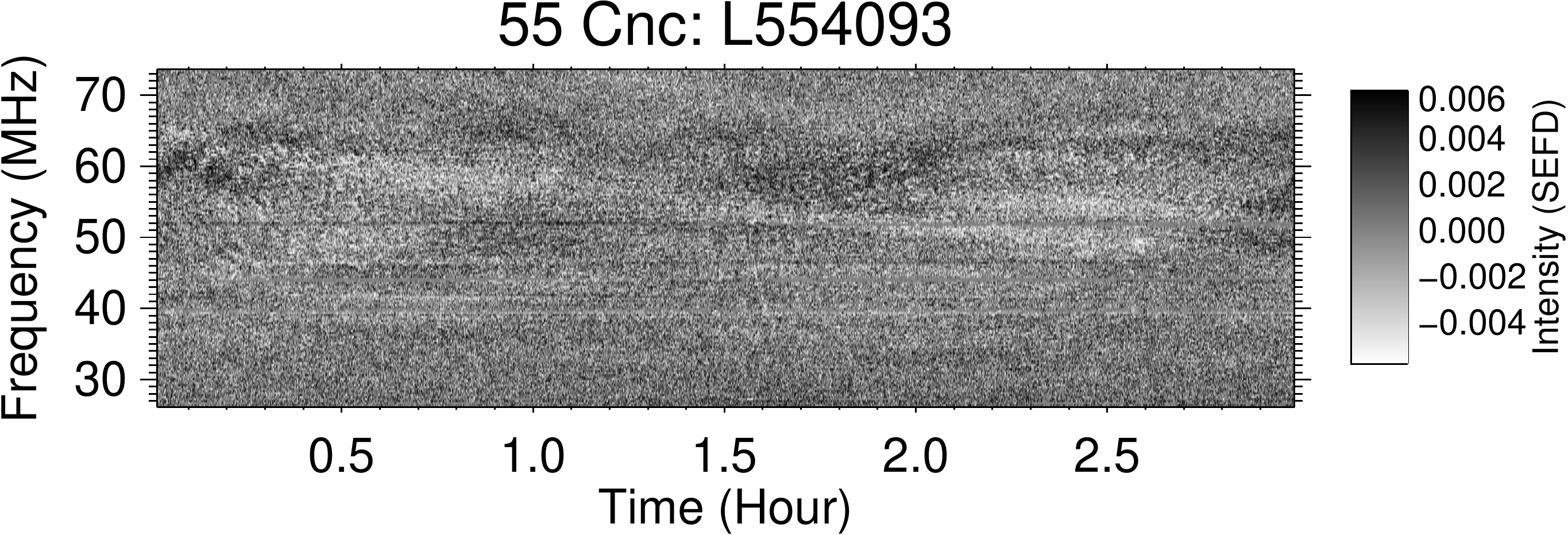}}  \\ 
      \begin{flushleft} \textbf{(b)} \end{flushleft} \vspace{\baselineskip}
      \centering
      \subfloat{\includegraphics[page=1,width=0.50\textwidth]{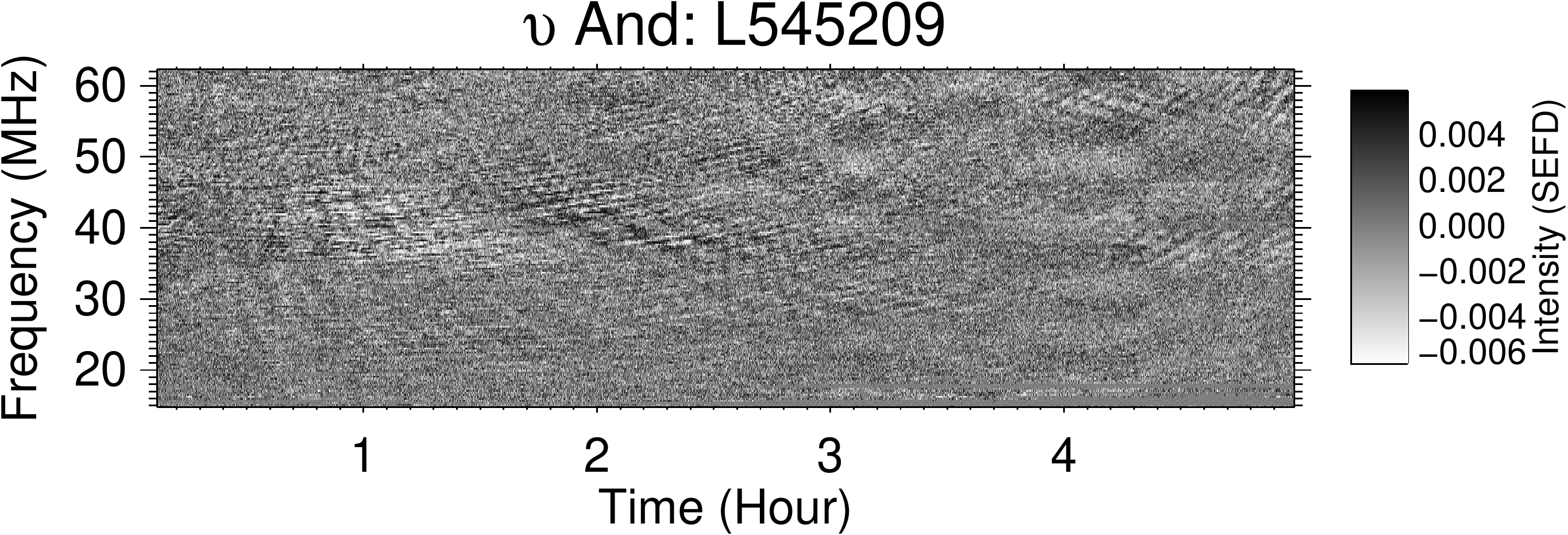}}  \\ 
      \begin{flushleft} \textbf{(c)} \end{flushleft} \vspace{\baselineskip}
      \centering
      \subfloat{\includegraphics[page=1,width=0.50\textwidth]{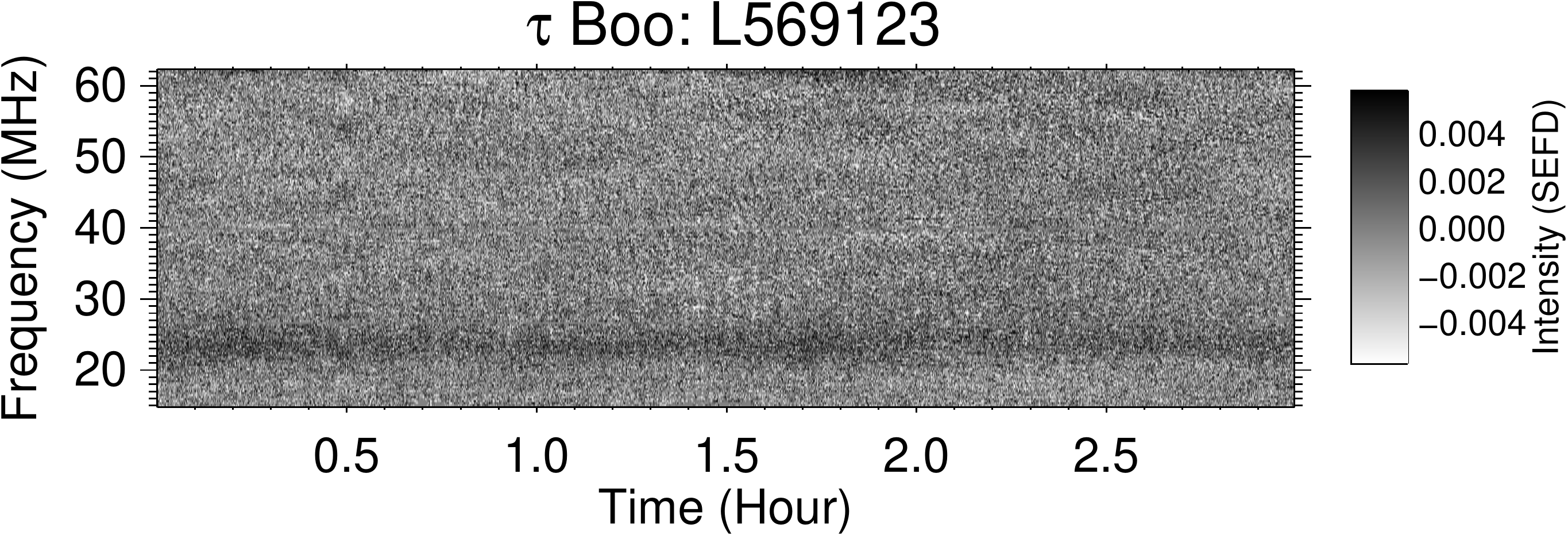}}  \\ 
  \caption{Examples of processed dynamic spectra 
  \edits{in Stokes V}
  (see Section \ref{sec:pipeline}) for \edits{OFF beam 2} of the 55 Cnc observation L554093 (\textit{panel a}), the $\upsilon$ And observation L545209 (\textit{panel b}), and the $\tau$~Boo observation L569123 (\textit{panel c}). The dynamic spectra presented are in units of the system equivalent flux density (SEFD) since they have been normalized by the $t$-$f$ response function of the telescope. 
  }
  \label{fig:Dynspec_non-detection}
\end{figure}

An example of a processed dynamic spectrum for each target is displayed in Figure \ref{fig:Dynspec_non-detection}. The fine structure seen in \editsjmg{Figure \ref{fig:Dynspec_non-detection}} can be caused by ionospheric scintillation on the Galactic background and some small residual instrumental effects. Additionally, the variable patches seen above 40 MHz are likely produced by LOFAR-core's grating lobes. At lower frequencies the grating lobes are below the horizon. The fluctuations \editsjmg{in the flux} of the OFF-beams are less than 1$\%$ \editsjmg{which is} \edits{consistent with the \editsjmg{expected thermal noise error} of $\sim$0.5$\%$}. 
We note that the dynamic spectra presented throughout this paper are in units of the system equivalent flux density (SEFD) since \editsjmg{the data has}
been normalized by the $t$-$f$ response function of the telescope. The SEFD include contributions from both the sky and instrument. The SEFD of LOFAR does change slightly with frequency (see \citealt{vanHaarlem2013}). \edits{In} this paper we use an \edits{average} value of 40 kJy for the 
\editsjmg{SEFD of a single LOFAR station, that is to say ~1.7 kJy for the full LOFAR core (i.e.~the sum of 24 LOFAR stations).} We use these units to simplify the display and analysis of the dynamic spectra and convert to Jy when necessary. 
The post-processing part of the \texttt{BOREALIS} pipeline is based on a series of observable quantities (labeled Q1a, Q1b, Q2, and Q4a-f). These quantities were introduced in \citetalias{Turner2019}; they are also recalled in Appendix \ref{app:obs_quant}. \editsjmg{The quantities Q1a and Q1b represent integrated quantities, namely the frequency-integrated time-series (Q1a) and the time-integrated frequency spectrum (Q1b). Both Q1a and Q1b are designed to find slowly variable emission (with time scales of minutes to hours).} \edits{Q2 represents the high pass filtered frequency-integrated time-series. \editsjmg{It can be displayed as a ``scatter plot'' comparing a pair of beams (e.g., the ON and OFF beam) and is} designed to find bursty emission (\editsjmg{with} time scales $<\sim$ 1 minute).
\editsjmg{The quantities Q4a to Q4f provide} statistical measures of the bursts identified by the Q2 quantity.}

\edits{When examining Q4a-f, the ON and OFF time series are compared to each other;}
\editsjmg{for this, we introduce the difference curve Q4f$_{\text{Diff}}$=Q4f(ON)-Q4f(OFF). We then plot this curve}
\edits{against a reference curve computed from 10000 draws of purely Gaussian noise.}
A \editsjmg{more} detailed description of each observable can be found in
\editsjmg{Appendix \ref{app:obs_quant}, or in \citetalias{Turner2019}}.\\
\indent \editsjmg{The efficiency of the quantities Q4a to Q4f for the detection of faint bursty emission was compared in \citetalias{Turner2019}. This benchmark analysis was based on artificially
attenuated radio emission of Jupiter and led to the conclusion that} Q4f is the most sensitive observable quantity for faint burst emission. \editsjmg{However, exoplanetary radio emission} 
may be different \edits{from} Jupiter radio emission. Therefore, we examined all the observable quantities in our analysis and not just \editsjmg{the most sensitive one indicated by the Jupiter benchmark study (Q4f)}.\\
\indent The post-processing is initially performed on the absolute value of the corrected Stokes-V data ($|V^{'}|$). In case of a tentative detection, information on whether the polarization of the emission is right-handed or left-handed is obtained by analysing $V^{'+}$ and $V^{'-}$ where,
\vspace{-0.9em}
\begin{align}
    V^{'-} =& \begin{cases}
    -V^{'}, & \text{if } V^{'} < 0\\
    0,              & \text{otherwise}
    \end{cases},
\end{align}
\vspace{1em}
\begin{eqnarray}    
        V^{'+} =& 
    \begin{cases}
    V^{'}, & \text{if } V^{'} > 0\\
    0,              & \text{otherwise} \label{eq:Vplus}
    \end{cases}.
\end{eqnarray} 

The post-processing was performed separately over 3 different frequency ranges depending on the observational setup and target. A summary of the parameters used in the post-processing can be found in Table \ref{tb:PPsetup}. While the nominal rebin time was 1 second, we also reanalyzed the dataset with a rebin time of 10 seconds to search for longer and weaker bursts of emission. The RFI mitigation step creates a high-resolution mask (at scales $b$ and $\tau$ \editsjmg{of} Table \ref{tb:setup}) of flags (0 for polluted pixels, 1 for clean ones). Since the processed dynamic spectrum is rebinned, we also have to rebin the flag mask to the same time and frequency resolution. Consequently, the rebinned flag mask contains fractional values between 0.0 and 1.0. In the post-processing we applied a threshold of 90$\%$ on the mask, meaning that 
\editsjmg{that we only retain a bin if $\ge90$\% of its constitutive pixels were clean.} We note also that the elliptical correction described in \citetalias{Turner2019} and \edits{summarised in Appendix \ref{app:obs_quant} was systematically applied to the analysis of Q2 and Q4 as this correction allows one to detect fainter signals.}

\begin{table}[!tbh]
\centering
\begin{threeparttable}
\caption{Parameters for the post-processing pipeline.}
\begin{tabular}{ccc}
\hline 
\hline
Parameter & Value   & Units  \\ 
\hline
\hline 
Frequency ranges            & 26-74, 26-50, 50-74\tnote{a} & MHz \\
                            & 15-62, 15-38, 38-62\tnote{b} & MHz \\
Q1 Time bins ($\delta T$)     & 2                         & minutes \\
Q1 Frequency Bins ($\delta F$)    & 0.5                   & MHz \\
Q2-Q4 rebin times ($\delta\tau$)  & 1, 10                 & sec \\
Mask threshold              & 90                        & \% \\
Smoothing window            &                           & \\
for high-pass filtering     & 10                        & $\delta\tau$ \\
Threshold ($\eta$) range    & 1 - 6                     & $\sigma$ \\
\hline
\end{tabular}
\begin{tablenotes}
    \item[a] Frequency ranges for the 55 Cnc observations.
    \item[b] Frequency ranges for the $\upsilon$ And and $\tau$~Boo observations.
\end{tablenotes}
\label{tb:PPsetup}
\end{threeparttable}
\end{table}

\indent The basic assumption of our analysis is that the OFF beams characterize the noise and systematics that affect the ON-beam well enough to search for a faint signal by comparison (ON vs.~OFF) or difference (ON-OFF). As Q4 difference curves often show large fluctuations, we found it absolutely necessary to record and analyze two OFF beams simultaneously with the ON beam. The analysis of one observation with a given set of processing parameters
includes the statistical comparison of the three different combination of beams: 
ON vs. OFF1, ON vs. OFF2, and OFF1 vs. OFF2. 
\editsjmg{We label a signal as a tentative dection only} if an excess appears in the former 2 pairs and not in the latter one. \editsjmg{If the OFF1 vs. OFF2 analysis shows an excess (with any sign) then a simultaneous tentative signal in ON vs. OFF is considered as a false positive, and discarded.
All the conclusions in this paper are the same when comparing the ON-beam to either of the OFF beams (OFF1 or OFF2). For this reason, we simply refer to the ``ON vs. OFF2'' comparison as ``ON vs. OFF'' throughout the rest of the paper.}\\
\indent In total, $\sim$13000 summary plots were produced, summarizing the analyze of all observations (Table \ref{tb:obs}) using the post-processing parameters \editsjmg{given in} Table \ref{tb:PPsetup}. 
\editsjmg{We applied} the following criteria defined in \citetalias{Turner2019} for a possible detection: (1) one or several Q4 curves show an excess $\ge 2\sigma$ statistical significance for the ON-OFF statistics, (2) the same Q4 curves have a shape clearly different from the OFF1-OFF2 curve, and (3) the detection curve remains positive over a large interval of thresholds $\eta >1.5\sigma$.
\editsjmg{The way these criteria are implemented in an automated search algorithm is described in Appendix \ref{app:summary_numbers}.}

\section{Data analysis and results}  \label{sec:dataAnalysis}

For most of the \editsjmg{observations} and frequency ranges explored we did not find any excess signal in the ON-beam when compared to the OFF beams. 
\editsjmg{The observables Q1a and Q1b do show some structure, but those structures are identical between the ON and OFF beam (examples for Q1b are shown in panels b and d of Figure \ref{fig:Q1b_all} in Appendix \ref{App:Q1ball}).}
\editsjmg{Indeed, for all observations (except for observation L570725 of $\tau$~Boo, see Section \ref{sec:L570725}), the curves for the ON and OFF beams are equivalent} within 1$\sigma$. We also tested the presence of red noise in the difference time-series (Q1a) for all our non-detections using the time-averaging method (\citealt{Pont2006,Turner2016b,Turner2017transit}), and found none. Thus the majority of the difference in the signal between the two beams can be explained purely by Gaussian white noise. \editsjmg{We also searched for burst emission using Q4a-f; for almost all observations, the burst statistics were similar between the ON and OFF beams.}

\editsjmg{For $\tau$~Boo} 
we found \edits{one} tentative detection of faint burst emission in observation L569131 and \edits{one tentative detection of} slowly variable emission in observation L570725, as summarised in Table \ref{tb:detections}. These tentative detections are examined in detail in Section \ref{sec:tauboo_detection}. We also find a marginal signal of faint burst emission for $\upsilon$~And in observation L545197, which is explored more in detail in Section \ref{sec:upsandr_detection}.

\subsection{$\tau$~Bo\"{o}tis} \label{sec:tauboo_detection}
\subsubsection{Observation L569131 (2017-02-18)} \label{sec:tauboo_L569131}
\begin{figure*}[p]
\centering
  \begin{tabular}{cc}
     \begin{subfigure}[c]{0.45\textwidth}
        \centering
        \caption{}
       \includegraphics[width=0.8\textwidth,page=1]{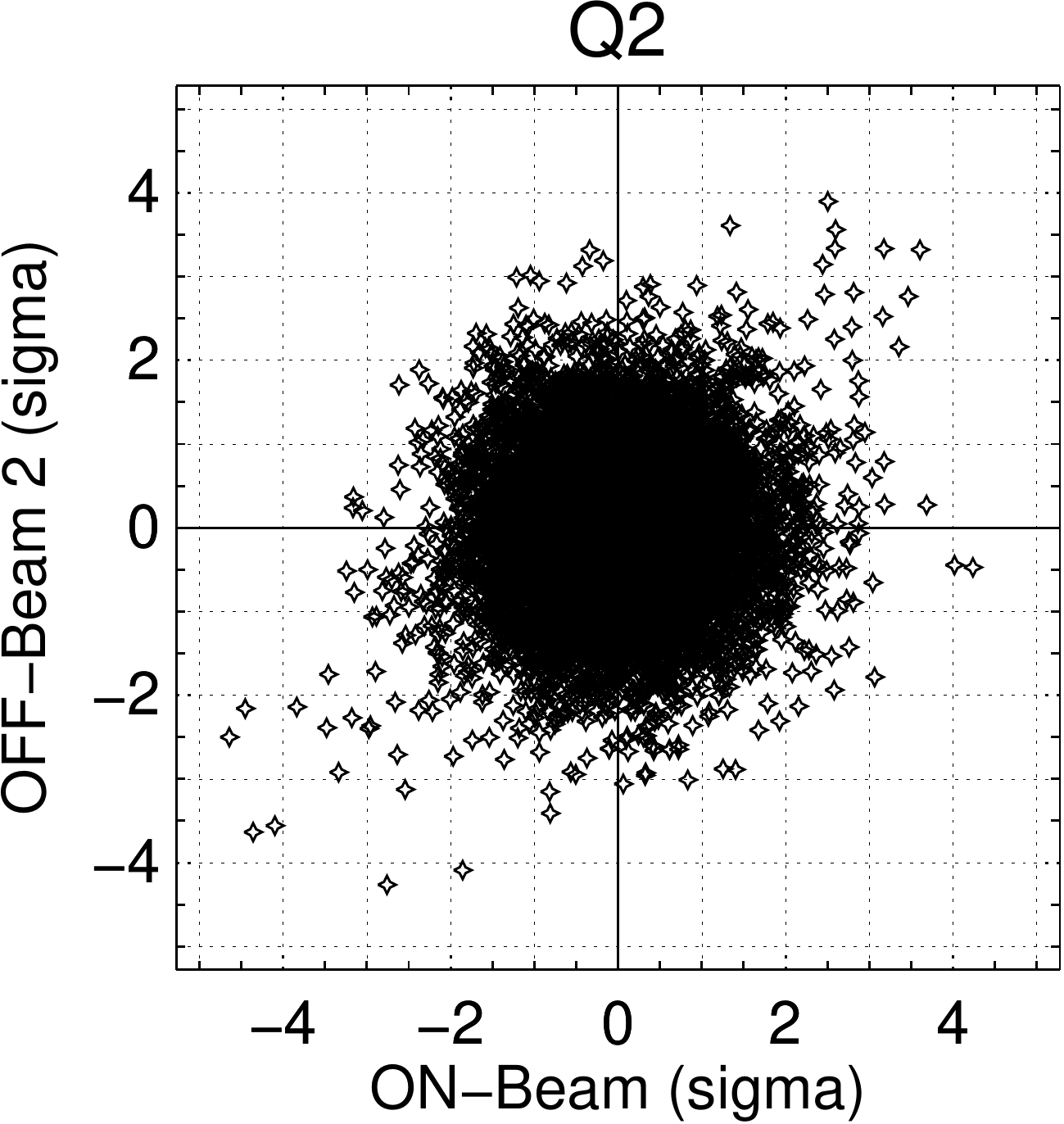}
        \label{}
    \end{subfigure}%
         \begin{subfigure}[c]{0.45\textwidth}
        \centering
        \caption{}
      \includegraphics[width=0.8\textwidth,page=2]{L569131_detection.pdf} 
              \label{}
    \end{subfigure}%
     \\
       \begin{subfigure}[c]{0.45\textwidth}
        \centering
        \caption{}
       \includegraphics[width=\textwidth,page=4]{L569131_detection.pdf} 
        \label{}
    \end{subfigure}%
\hspace{5mm}
     \begin{subfigure}[c]{0.45\textwidth}
        \centering
        \caption{}
     \includegraphics[width=\textwidth,page=6]{L569131_detection.pdf} 
              \label{}
    \end{subfigure}%
     \\
            \begin{subfigure}[c]{0.45\textwidth}
        \centering
        \caption{}
       \includegraphics[width=\textwidth,page=8]{L569131_detection.pdf} 
        \label{}
    \end{subfigure}%
\hspace{5mm}
     \begin{subfigure}[c]{0.45\textwidth}
        \centering
        \caption{}
     \includegraphics[width=\textwidth,page=10]{L569131_detection.pdf} 
              \label{}
    \end{subfigure}%
\end{tabular} 
  \caption{Q2 (\textit{panels a and b}) and beam differences for Q4a (\textit{panel c}), Q4b (\textit{panel d}), Q4e (\textit{panel e}), and Q4f (\textit{panel f}) for $\tau$~Boo in observation L569131 in the range 14-38 MHz in Stokes-V ($|V^{'}|$).
  \textit{Panel a:} Q2 for the ON-beam vs the OFF beam 2. 
  \textit{Panel b:} Q2 for the OFF beam 1 vs the OFF beam 2.
  \textit{Panel c:} Q4a (number of peaks). \textit{Panel d:} Q4b (power of peaks). \textit{Panel e:} Q4e (peak 
  \edits{number}
  offset). \textit{Panel f:} Q4f (peak \edits{power} offset). For \textit{panels c to f} the black lines are the ON-beam difference with the OFF beam 2 and the red lines are the OFF beam difference. The dashed lines are statistical limits (1, 2, 3$\sigma$) of the difference between all the Q4 values derived \edits{from 10000 runs} using two different Gaussian distributions. In all panels the ON-beam shows an excess above 2 $\sigma$ statistical significance and is distinctly different from the OFF-beams. \edits{The probability to obtain the OFF beam curve in Q4f (\textit{Panel f}) by chance is $\sim$81$\%$, whereas it is $7\times10^{-4}$ for the ON-beam curve}, corresponding to a 3.2$\sigma$ detection. \edits{A} Kolmogorov–Smirnov statistical test on the two curves for Q4f in \textit{panel f} \edits{conclude} that the probability to reject the null hypothesis (that the two curves are drawn from the same parent distribution) is 98$\%$. }
  \label{fig:L569131_detection}
\end{figure*}

For $\tau$~Boo, we detect Stokes-V burst emission in the frequency band 15-38 MHz in observation L569131  \edits{from the Q4} observables (Figure \ref{fig:L569131_detection}; Table \ref{tb:detections}). \editsjmg{In particular, Figure \ref{fig:L569131_detection} shows an excess of bursts in the ON-beam in Q4f (panel f, curve ON-OFF), whereas both OFF beams show comparable statistics (the curve OFF1-OFF2 is almost flat). The burst emission is also seen in $V^{'-}$ \editsjmg{(defined in equation \ref{eq:Vplus}),} suggesting that the emission is left-hand polarized.}

To isolate the frequency range of the 
\editsjmg{signal,}
we processed a finer grid of frequency ranges (\edits{6 MHz in bandwith}) within 15-38 MHz. Performing this test, we find that the majority of the signal is from the 15-21 MHz range. Therefore, we 
\editsjmg{use}
this frequency range to \editsjmg{tentatively deduce} physical information about 
\editsjmg{the planet (see Section \ref{sec:tauBooPhysics})}.

\indent We estimate the \edits{flux density} of this signal using the sensitivity limit derived in \citetalias{Turner2019}. In that work, this
limit was found
\editsjmg{to be}
equal to 1.3 times the LOFAR theoretical sensitivity ($\sigma_{LOFAR}$)\footnote{\edits{The 1.3$\sigma_{LOFAR}$ sensitivity limit in Stokes-V is also consistent with the recent finding by \citet{Mertens2020} \editsjmg{who use imaging observations}.}} and consisted of 30 data points above $\eta$ = 2$\sigma$ in Q4f. 
The signal in Q4f 
\editsjmg{for observation}
L569131 consists of 21 data-points (each binned to a resolution of 1 sec and 24 MHz) above $\eta$ = 2$\sigma$ level\footnote{\editsjmg{The number of 21 points is not directly visible in the figures; it relates to the underlying Q2 distribution that was used to calculate the observable quantity Q4f.}}. 
\editsjmg{These points result}
\edits{in a Q4f$_{\text{Diff}}$ 
curve above the 2 sigma Gaussian reference \editsjmg{curve.} }
The flux in the tentative detection should be near the sensitivity limit since the number of detected points between the \editsjmg{Jupiter observation (\citetalias{Turner2019}) and the $\tau$ Boo observation (the present work)} are \edits{comparable}%
\footnote{\editsjmg{The relationship between the signal intensity and the number of points is nonlinear, see Fig.~2c in \citetalias{Turner2019}.}}.
\editsjmg{Also, the} 
uncertainty \editsjmg{is} similar to that derived in \citetalias{Turner2019}. 

 \edits{The 
 \editsjmg{theoretical sensitivity limit}
 of LOFAR for broadband bursts} is 
\begin{align}
    \sigma_{LOFAR} =& \frac{S_{Sys}}{N\sqrt{b\tau}}, \label{eq:sigmaLOFAR_nonumbers}\\
    \sigma_{LOFAR} =& \frac{40 kJy}{24\sqrt{1 \ sec \times 6\ MHz}} = 680 \ mJy \label{eq:sigmaLOFAR} 
\end{align}
where $N$ is the number of stations (24), $b$ is the bandwidth (\edits{6 MHz}), $\tau$ is the timing resolution (1 sec), and $S_{Sys}$ is the station SEFD with a value of 40 kJy (\citealt{vanHaarlem2013}). Using the sensitivity limit of $1.3\times\sigma_{LOFAR}$ (\citetalias{Turner2019}) and equation \eqref{eq:sigmaLOFAR}, we find \edits{a flux density estimate} for the burst emission of \edits{890}$^{+690}_{-500}$ mJy. 
\editsjmg{In this, we assumed an uncertainty of a}
\edits{quarter of an order of magnitude (10$^{0.25}$) 
\editsjmg{for the sensitivity limit \citepalias[see][]{Turner2019}.}
The \editsjmg{uncertainties given here are merely rough estimates; the real values} 
depend on the time and frequency scale of the emission.}

To quantity the statistical significance of the detected signal we use the method outlined in \citetalias{Turner2019}. \edits{The difference of the ON-OFF beam in the observable Q4f (Q4f$_{\text{Diff}}$ the solid black line in Fig.~\ref{fig:L569131_detection}f) is normalized by the 1$\sigma$ Gaussian statistical limit (the lower-positive dashed-line in Fig.~\ref{fig:L569131_detection}f) and we calculate its average value ($<$Q4f$_{\text{Diff}}>$). $<$Q4f$_{\text{Diff}}>$ is then compared to the case when the ON and the OFF beams only contain random noise. Using this procedure, we find that the probability of a false positive for obtaining the ON-beam signal is $7\times10^{-4}$, equivalent to} a 3.2$\sigma$ detection. Finally, we compare these values to those of the two OFF beams (the solid red line in Fig.~\ref{fig:L569131_detection}f). For the OFF beams we find a false positive rate of 19$\%$ indicating clearly a non-detection.

\edits{To further assess the statistical significance of the detected signal we perform a two-sided Kolmogorov-Smirnov (K-S) test in which we compare the Q4f ON-OFF and OFF1-OFF2 curves. The null hypothesis is that the two curves are drawn from the same parent distribution. Performing the K-S test we find that the probability to reject the null hypothesis (that the two curves are drawn from the same parent distribution) is high with a value of 98$\%$. This test adds to the evidence that the two curves are distinctly different and that \editsjmg{this} signal is real. }   

\editsjmg{To summarize this analysis: We detect a bursty signal in this observation. This signal is statistically significant. Its potential origin is discussed in Section \ref{sec:discussion}.}

 \begin{figure}[!thb]
     \begin{tabular}{c}
     \begin{subfigure}[c]{0.47\textwidth}
        \centering
        \caption{}
     \includegraphics[width=\textwidth,page=1]{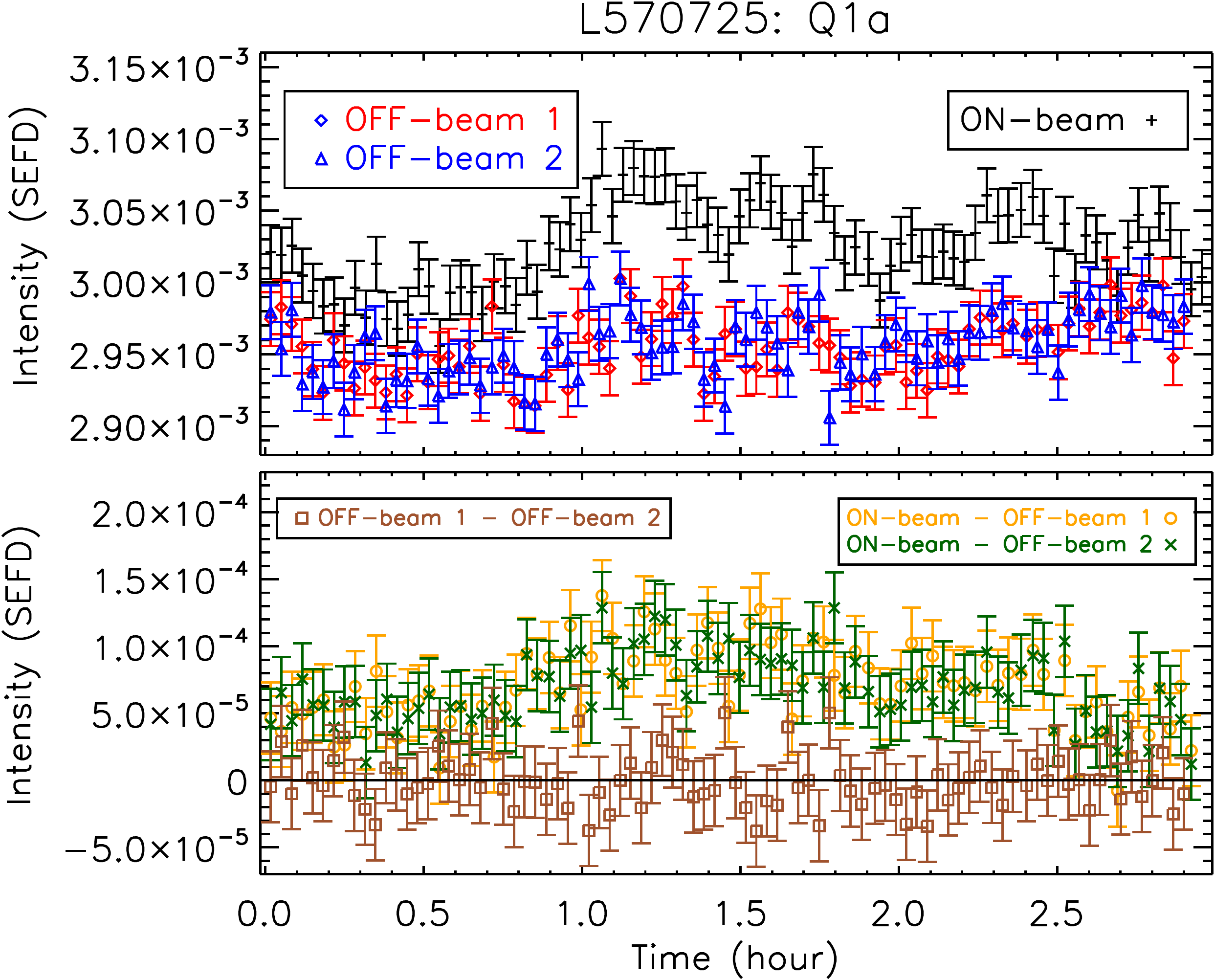} 
             \label{}
    \end{subfigure}%
    \\
      \begin{subfigure}[c]{0.47\textwidth}
        \centering
        \caption{}
     \includegraphics[width=\textwidth,page=2]{L570725_detection.pdf} \\
             \label{}
    \end{subfigure}%
          \end{tabular}
     \caption{Time-series (Q1a; \textit{panel a}) and integrated spectrum (Q1b; \textit{panel b}) in Stokes-V (using $|V'|$ as defined in equations 9--11 of \citetalias{Turner2019})
     for $\tau$~Boo during the observation L570725. In \textit{panel a}, the ON-beam shows excess signal above both OFF beams at all times. In \textit{panel b}, the signal in the ON-beam is concentrated between 21 and 30 MHz and is distinctly different than in both OFF-beams. In both panels, 
     the two OFF beams are equivalent 
     within the error bars, calculated assuming pure Gaussian noise ($\sigma=1/\sqrt{b\tau}$). The dynamic spectrum for this signal can be found in Figure \ref{fig:Dynspec_Detection}. We find by performing Gaussian simulations that the probability to randomly reproduce the signal in the ON-beam curve in Q1a is 2.1$\times10^{-12}$ and 1$\times10^{-18}$ for Q1b. This probability corresponds to a statistically significant detection of 6.9$\sigma$ and 8.6$\sigma$, respectively. 
     }
     \label{fig:L570725_detection}
 \end{figure}

\begin{figure*}
\centering
\begin{tabular}{cc}
     \begin{subfigure}[c]{0.50\textwidth}
        \centering
        \caption{}
    \includegraphics[page=1,width=\textwidth]{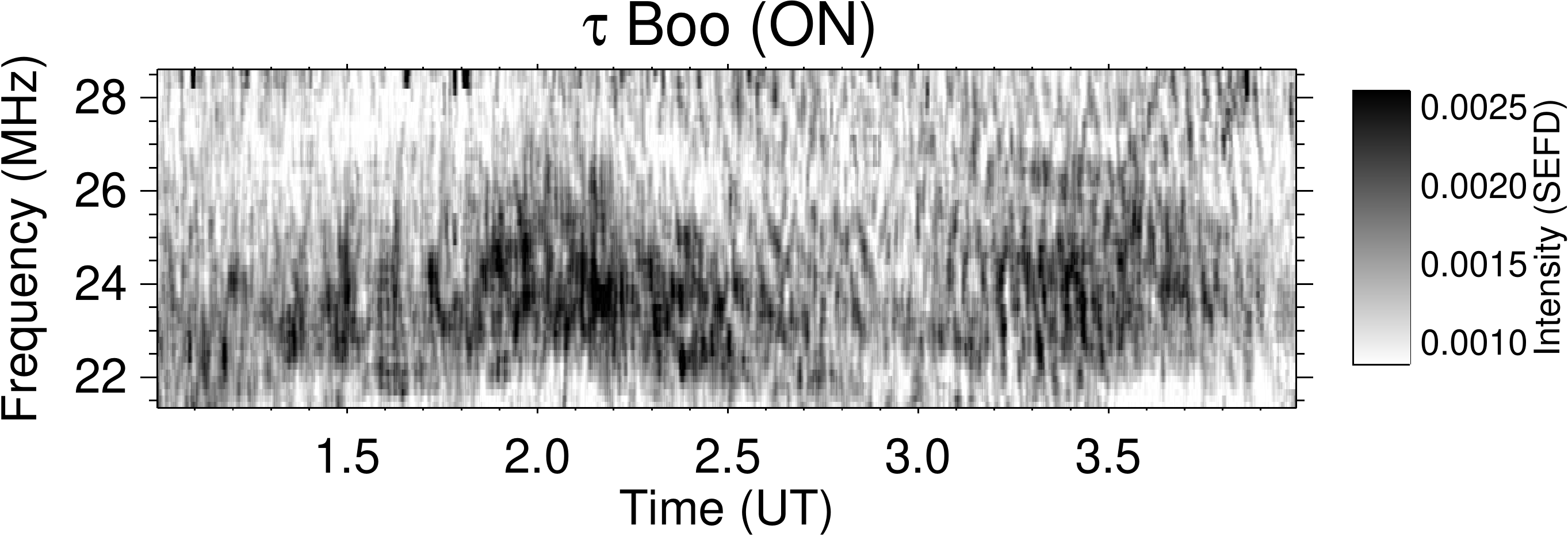} 
             \label{} 
    \end{subfigure}%
    \hspace{5mm}

         \begin{subfigure}[c]{0.50\textwidth}
        \centering
        \caption{}
   \includegraphics[page=2,width=\textwidth]{TauBoo_L570725_DynSpec_Diff.pdf}
             \label{}
    \end{subfigure}%
    \\
    \begin{subfigure}[c]{0.50\textwidth}
        \centering
        \caption{}
    \includegraphics[page=4,width=\textwidth]{TauBoo_L570725_DynSpec_Diff.pdf}
             \label{}
    \end{subfigure}%
    \hspace{5mm}

     \begin{subfigure}[c]{0.50\textwidth}
        \centering
        \caption{}
   \includegraphics[page=5,width=\textwidth]{TauBoo_L570725_DynSpec_Diff.pdf}
             \label{}
    \end{subfigure}%
    \\

    \begin{subfigure}[c]{0.50\textwidth}
        \centering
        \caption{}
        \includegraphics[page=2,width=\textwidth]{TauBoo_L570725_DynSpec_Ratio.pdf}
         \label{}
    \end{subfigure}%
    \hspace{5mm}
     \begin{subfigure}[c]{0.50\textwidth}
        \centering
        \caption{}
        \includegraphics[page=3,width=\textwidth]{TauBoo_L570725_DynSpec_Ratio.pdf}
         \label{}
    \end{subfigure}%
\end{tabular}
  \caption{Dynamic spectra of the $\tau$~Boo~b observation L570725 in Stokes-V ($|V^{'}|$). \textit{Panel a:} ON-beam. \textit{Panel b:} OFF-beam 2. \textit{Panel c:} ON-beam minus OFF-beam 2. \textit{Panel d:} OFF-beam 1 minus OFF-beam 2. \textit{Panel e:}  ON-beam divided by the OFF-beam 2. \textit{Panel e:}  OFF-beam 1 divided by the OFF-beam 2.  Data from the range 21-29 MHz is shown. The ON-beam signal is clearly seen as structured excess emission in \textit{panels a}, \textit{c}, and \textit{e}. A faint signal is seen in the OFF beams (e.g., \textit{panel c}) likely due to ionospheric refraction of the ON-beam signal (also see Appendix \ref{App:pulsar}). However, there is no visible excess difference when the two OFF-beams are subtracted by each other (\textit{panel d}) suggesting that whatever is causing the faint OFF-beam signal is the same for both beams.  }
  \label{fig:Dynspec_Detection}
\end{figure*}

\subsubsection{Observation L570725 (2017-03-06)}  \label{sec:L570725}

\edits{For} $\tau$~Bo\"{o}tis, we \edits{also} detected slowly variable emission in the range 14-38 MHz in observation L570725.
\edits{This} emission was detected using \editstwo{the dynamic spectra integrated over frequency (time-series, Q1a) and time (integrated spectrum, Q1b)} in Stokes-V (Figure \ref{fig:L570725_detection}; Table \ref{tb:detections}). 
The signal is also an outlier when compared to all $N2$ numbers for all observations (Appendix \ref{App:run_numbers_compare}, Figure \ref{fig:PP_stats}). 
The OFF beams are equivalent
within $1\sigma$ for both Q1a and Q1b 
suggesting that there is an excess signal in the ON beam. Further examination of Q1b determined that the bulk of the ON-beam emission is coming from the range 21-30 MHz. We did not find any short-term (1--10 secs) burst emission using the observable Q4 for this observation (Appendix \ref{App:L570725_Nondetection}, Figure \ref{fig:L570725_Nondetection}). \edits{The slow emission feature is seen in $V^{'+}$ but not in $V^{'-}$ \editsjmg{(not shown),} suggesting that the emission \editsjmg{---if physical---} is right-hand \editsjmg{circularly} polarized.}

The flux density of the signal
can be estimated from the ON-OFF beam difference (bottom \editsjmg{panels} in Figures \ref{fig:L570725_detection}a,b). We find a mean signal of $1.1\pm0.2\times10^{-4}$ of the theoretical SEFD from Q1a. Since all the emission is coming from the range 21-30 MHz (Figure \ref{fig:L570725_detection}b), the calculation of Q1a was limited to this range.  
\editsjmg{We find a value of 190$\pm$30 mJy for the time-averaged signal.}
The quoted error above is the standard deviation of the difference of Q1a for the OFF beams. We find a maximum emitted flux of 430$\pm$30 mJy using the peak spectral flux ($\sim 2.5\times10^{-4}$ of the SEFD) observed in the difference of Q1b. Again, the error is the standard deviation of the difference of Q1b for the OFF beams.

\edits{To quantify the statistical significance of this signal, we use a method similar to the technique outlined in \citetalias{Turner2019} and described thoroughly in Appendix \ref{App:stats}. We find that the probability of a false positive for obtaining a signal like the one we observe is 2.1$\times10^{-12}$ for the Q1a(ON)-Q1a(OFF2) (bottom panel in Figure \ref{fig:L570725_detection}a) and 1$\times10^{-18}$ for Q1b(ON)-Q1b(OFF2) (bottom panel in Figure \ref{fig:L570725_detection}b). This 
corresponds to a statistically significant signal of 6.9$\sigma$ and 8.6$\sigma$, respectively. For the OFF-beams we find that the false positive rate is $\sim$90$\%$ for Q1a(OFF1)-Q1a(OFF2) and $\sim$100$\%$ for Q1b(OFF1)-Q1b(OFF2), clearly corresponding to a non-detection. }

We examine the dynamic spectrum of the ON-beam and OFF-beams to determine the time-frequency structure of the emission in observation L570725. In Figure \ref{fig:Dynspec_Detection}, we show the dynamic spectrum of \edits{the beams ON (\textit{panel a}), OFF 2 (\textit{panel b}), ON - OFF 2 (\textit{panel c}), OFF 1 - OFF 2 (\textit{panel d}), ON/OFF 2 (\textit{panel e}), and OFF 1/OFF 2 (\textit{panel f})}. 
The processed dynamic spectrum (with an original time and frequency resolution of 1 sec and 45 kHz, respectively) has been binned to 6 seconds and a boxcar 
window 3 pixels across (135 kHz) has been applied along the frequency direction to 
\editsjmg{reduce the noise}.
The ON-beam signal is clearly seen as structured excess emission in \textit{panels a} and \textit{c}. These structured features are very similar to the observed radio dynamic spectrum of Jupiter (e.g., \citealt{Zarka1998}; \citealt{Marques2017}). 
\editsjmg{There is no equivalent large scale structure in the OFF beam difference plot (\textit{panel d}), confirming that these two beams are very similar to each other.}

\begin{figure}[!htb]
    \centering
    \includegraphics[page=1,width=0.50\textwidth]{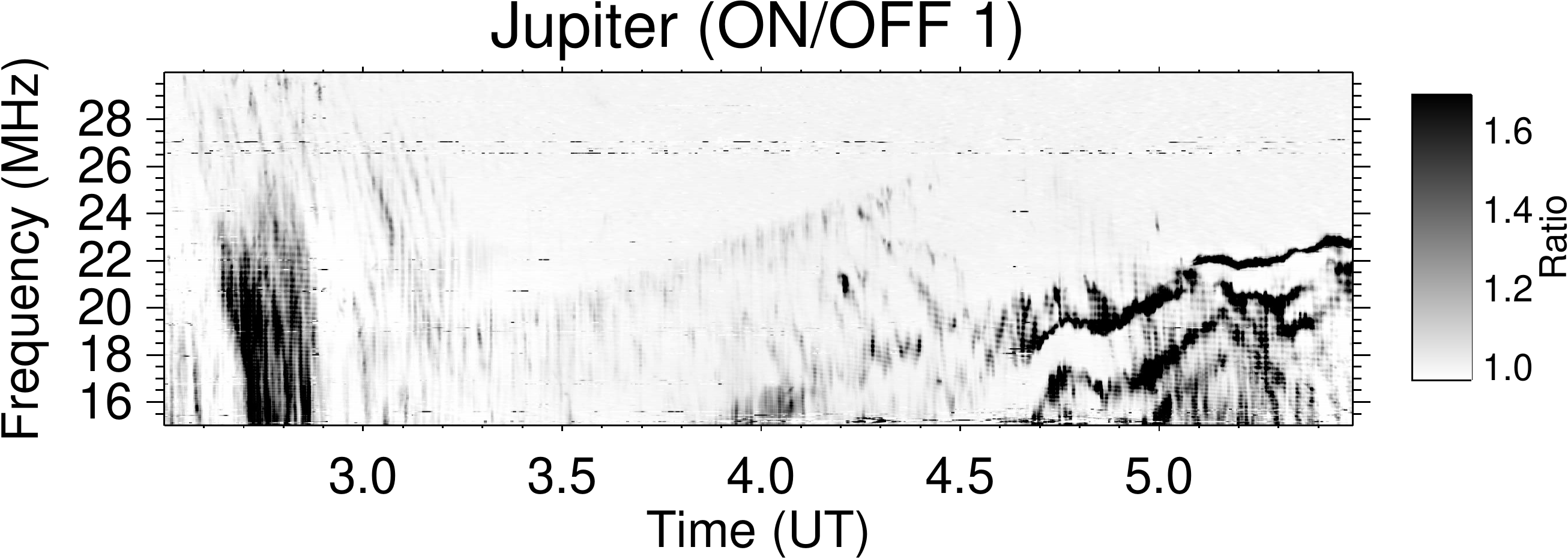}\\ 
  \caption{Dynamic \editsjmg{spectrum} of the Jupiter observation L568467 (\citealt{Turner2019}) in Stokes-V ($|V^{'}|$). 
  \editsjmg{Shown: ON-beam divided by OFF-beam 1.}}
  \label{fig:Dynspec_Jupiter}
\end{figure}

\editsjmg{\textit{Panel b} shows that the OFF beams contain a}
\edits{replica of the ON signal.} \editsjmg{The signal is $\sim$ 1.6 times fainter in the OFF beam dynamic spectrum when compared to the ON-beam (\textit{panel e}).}
\edits{We found convincing evidence that the
replica signal in the OFF beam originates from the ON beam. Using the pulsar B0809+74 observation L570723 (taken 15 minutes before observation L570725), we detect the pulsar in the ON-beam but also in the two OFF beams (Appendix \ref{App:pulsar}, Figure \ref{fig:OFFbeams_pulsar}). 
Additionally, we examine the Jupiter observation L568467 (data taken from \citetalias{Turner2019}) and find that the ratio of the ON to OFF beam flux is $S(ON)/S(OFF)\sim 1.6$ 
(Figure \ref{fig:Dynspec_Jupiter}). 
\editsjmg{The flux density ratios (ON/OFF) for the 
Jupiter observation is}
consistent with the value found for the $\tau$~Boo observations. The}
\edits{replica signal is}
\edits{likely caused by imperfect phasing of LOFAR}
\edits{at the epoch of the observations,}
\editsjmg{which}
\edits{leads to strong 
side lobes within the station beam (personal communication from M. A. Brentjens, ASTRON).}

\edits{We explored several 
non-astronomical sources (e.g., instrumental systematics) for the signal in the ON-beam in observation L570725.} 
\editsjmg{First, during each of our observations, 1-2 LOFAR stations did not operate optimally (cf.~Table\ref{tb:obs}, second-to-last column). Close inspection showed that the ON-beam signal in this observation was not caused by this (see Appendix \ref{App:staion_inspection_plots} for details).}
\editsjmg{Next, if}
\edits{
the ON beam signal was} 
\edits{a permanent
instrumental effect, it should also appear in \editsjmg{the observation of the pulsar B0809+74 (observation L570723, preceding the $\tau$ Boo observation by 16 minutes)}. We find no }
\edits{such large scale features in the range 20-30 MHz in}
\editsjmg{that observation}
\edits{(Appendix \ref{App:pulsar},} \editsjmg{Figure \ref{fig:OFFbeams_pulsar})}.
\editsjmg{This suggests that the ON-beam signal is either}
\edits{real excess flux in 
that beam or a 
time-variable instrumental effect. }
\editsjmg{Finally,} 
\edits{low-level features in the integrated spectrum (Q1b) are seen for all ON and OFF beams} \editsjmg{for} 
\edits{all targets (Appendix \ref{App:Q1ball}, Figure \ref{fig:Q1b_all}). 
For each observation \editsjmg{except} L570725}
\editsjmg{these features do not change in time 
and are similar for the ON and OFF beam within the error bars. }
\edits{The ON-beam feature in observation L570725}
\editsjmg{(Figure \ref{fig:Q1b_all}f, orange line)}
\edits{has an amplitude comparable to the features seen in}
\edits{ON and
OFF beams in all other $\tau$~Boo observations (Figure} \editsjmg{\ref{fig:Q1b_all}e-f). However,}
\edits{both OFF-beams in observation L570725 have a lower amplitude, \editsjmg{leading to the} detection of an excess \editsjmg{signal (ON-OFF)}. 
\editsjmg{The fact that the ON-OFF excess is caused by a lower level in the OFF beam rather than a higher level in the ON-beam casts some doubt}
about the astrophysical origin of the detected signal}. 
\edits{However, the
\edits{time-frequency} structure in the dynamic spectrum of observation L570725 is not the same as}
\editsjmg{for}
\edits{
the other} 
\editsjmg{observations.}
\editsjmg{The differences between the dynamic spectra (Appendix \ref{App:dynspecdiff_tauboo}, Figure \ref{fig:Dynspec_OtherDiff}) show large scale systematics, which suggests a different source for the emission in this observation.}

\edits{Summarizing the detailed arguments presented \editsjmg{above and} in the Appendices, we could not identify an instrumental origin or systematic error for the excess ON-beam signal detected in observation L570725. The amplitudes of ON and OFF signal on that day compared to the other dates encourage us to skepticism, but the time-frequency structure of the signal leaves some room for an astrophysical origin. In all cases, follow-up observations with several radio telescopes are necessary to confirm \editsjmg{or invalidate this potential signal.}}

\begin{table*}[!thb]
\centering
    \caption{Summary of detected signals and derived planetary parameters assuming that the emission originates from the planet (see Section \ref{sec:discussion} for a discussion of the arguments for and against a planetary origin for the signals).
    Column 1: planet name. Column 2: LOFAR observation ID. Column 3: frequency range of the detection. Column 4: polarization signature of the emission. Column 5: observable quantity that found the detection. Column 6: statistical significance derived using Gaussian simulations. Column 7: respective figures showing the detected signals. Column 8: estimated flux density of the emission. Column 9: estimated power of the emission. Column 10: brightness temperature of the tentative emission. Column 11: 
    \edits{maximum} polar surface magnetic field of the planet
    derived using the maximum frequency, assuming CMI emission. 
    }
    \begin{tabular}{ccccccc|cccc}
    \hline  
      Planet  & LOFAR   & $\nu$            & Polarization      &Q$\#$  & Signifi- & Figure &  Flux      & Power & $T_{B}$           &  Magnetic   \\
      Name        & Obs   ID    & (MHz)            & ($|V|$, V+, V-)   &       & cance ($\sigma$)    & & (mJy)    & (W)   & (K)&  Field (G)                    \\
      \hline
      \hline
     $\tau$~Boo~b    & L570725 & 21-30      &  $|V|$, V+                     & Q1a  & 6.9 & \ref{fig:L570725_detection} & 190$\pm$30 & $\edits{6.3}\times10^{14-15}$                  & $\edits{4.2}\times10^{17}$ &  7.5-10.7      \\
     $\tau$~Boo~b    & L570725 & 21-30  &$|V|$, V+ &  Q1b  &  8.6 & \ref{fig:L570725_detection} & 430$\pm$30 & $\edits{1.4}\times10^{15-16}$        & $\edits{1.0}\times10^{18}$& 7.5-10.7                \\
     $\tau$~Boo~b   & L569131  & 15-21 &$|V|$, V- & Q4f        &3.2   & \ref{fig:L569131_detection} & \edits{890}$^{+690}_{-500}$  & $\edits{2.0}\times10^{15-16}$ & $\edits{2.0}\times10^{18}$&5.4-7.5   \\
      \hline
    \end{tabular}
    \label{tb:detections}
\end{table*}

\subsection{$\upsilon$ Andromedae} \label{sec:upsandr_detection}

For $\upsilon$ And, we detected burst emission in the frequency band 14-38 MHz in observation L545197 (2016-09-08)
using the Q4f observable (Figure \ref{fig:L545197_detection}). 
The Q4f signal consists of 8 data-points greater than $\eta$ = $2\sigma$ and 
\edits{a Q4f$_{\text{Diff}}$ curve above the}
2 sigma Gaussian reference curve (Figure \ref{fig:L545197_detection}c).  

Similar to Section \ref{sec:tauboo_L569131},
we find an approximate emission flux
of 1.3$\times\sigma_{LOFAR}$(1+0.20)$\sim$ 540$^{+460}_{-240}$ mJy using equation \ref{eq:sigmaLOFAR} and taking into account the 20$\%$ extra noise mentioned in Section \ref{sec:obs}. The probability that the ON$-$OFF Q4f curve is a false positive is 1.3\% (assuming Gaussian statistics), equivalent to a $2.2\sigma$ detection. By contrast, the probability of obtaining the OFF1$-$OFF2 curve by chance is 41\%. Additionally, we used the K-S statistical test to assess whether the ON-OFF and OFF1-OFF Q4f curves in Figure \ref{fig:L545197_detection}f could have been drawn from the same distribution. We find that the probability to reject the null hypothesis (that the two curves are drawn from the same parent distribution) is marginal with a value of 76$\%$. The signal we find for $\upsilon$~And is not highly statistically significant, so the possibility for a false-positive is high. Observations (L545197) also marginally stands out when compared to other observations (detection criterion N6, Appendix \ref{App:run_numbers_compare}, Figure \ref{fig:PP_stats}).
In rare cases (1$\%$), OFF-beams reach comparable or higher values, adding to the uncertainty on the detection. \editsjmg{We} present this marginal detection \edits{for reference} in hope to facilitate and guide future observations of this system.      

\section{Discussion} \label{sec:discussion} 

\subsection{Origin of the detected signals}

We first examine arguments for and against a celestial origin of the signals in the $\tau$ Boo observations (Section \ref{sec:origin_a} for observations L569131, Section \ref{sec:origin_b} for observation L570725). Then, assuming that the bursty emission (L569131) and the slowly varying signal (L570725) are of celestial nature, we explore in greater detail whether those signals could originate from the $\tau$~Boo system (Sections \ref{sec:dis_tauboo})
and from the exoplanet $\tau$~Boo b itself (Sections \ref{sec:dis_planet}).

\label{sec:origin}
\subsubsection{Arguments on whether the detected signal in observation L569131 is of a celestial origin}
\label{sec:origin_a}

\editsjmg{The
arguments for
a celestial origin of}
the bursty signal in observation L569131 of $\tau$~Boo 
(i.e.~it not being an instrumental effect or RFI) 
\editsjmg{are the following:}
\begin{enumerate}
 \item We find a signal in Q4f equivalent to a 3.2$\sigma$ detection (Figure \ref{fig:L569131_detection}).
 \editsjmg{The ON-OFF difference (black curve) is distinctly different from the OFF1-OFF2 difference (red curve red).}
 The K-S test finds that the probability to reject the null hypothesis (that the two curves are drawn from the same parent distribution) is 98$\%$.  
 \item The signal is circularly polarized, as expected for CMI emission.  
 \item There is no evidence \editsjmg{for} low-level \editsjmg{residual} RFI. 
 \editsjmg{Indeed, residual RFI would cause flux to appear simultaneously in the ON- and OFF-beams, and would be seen as points with high values close to the main diagonal in the Q2 plot (Figure \ref{fig:L569131_detection}.a).}
Also, the main diagonal is excluded in the calculation of the Q4f observable quantity.
\end{enumerate}
An argument against an celestial origin is:
\begin{enumerate}
 \item There is a possibility that low-level RFI could cause the signal in the event that the RFI conditions between the ON and OFF beams are significantly different. However, there is no evidence that the RFI is different between the beams. \edits{Also, 
 we have demonstrated that the RFI mitigation step
 in the pipeline is efficient (\citealt{Turner2017pre8,Turner2019}). }
\end{enumerate}

\editsjmg{Since we do not find} potential false positives for the bursty signal, \editsjmg{this} is likely a real detection of celestial emission and not an instrumental effect. 
\editsjmg{However, a detection at 3.2$\sigma$ level is not highly significant}
\editsjmg{and calls for confirmation via follow-up observations.}
 
\subsubsection{Arguments on whether the detected signal in observation L570725 is of a celestial origin}

\label{sec:origin_b} 

Likewise, we now summarize the arguments for and against the possibility that the slowly varying signal in observation L570725 of $\tau$~Boo is indeed real celestial emission. The arguments for a real detection are:
\begin{enumerate}
 \item There is a clear signal in Q1a equivalent to 6.9$\sigma$ and Q1b equivalent to 8.5$\sigma$ (Figure \ref{fig:L570725_detection}).
\item The signal is found to be circularly polarized, as expected for CMI emission.
\item The signal has a complex structure in time-frequency (Figure \ref{fig:Dynspec_Detection}) reminiscent of Jupiter radio arcs (e.g., \citealt{Marques2017}). 
\item The structures seen in the dynamic spectra of the ON-beam are distinctly different than all OFF-beams for every $\tau$~Boo observation (Figure \ref{fig:Dynspec_OtherDiff}). 
\item All observations in this study were processed identically with the \texttt{BOREALIS} pipeline (\citealt{Vasylieva2015}; \citealt{Turner2017pre8}; \citetalias{Turner2019}). The L570725 signal is the only one among many observations and runs with similar observing conditions (Table \ref{tb:obs}). 
\item All OFF-beams in each individual observation are consistent with each other \edits{within expected} Gaussian error bars (Figure \ref{fig:Q1b_all}). There is no indication that any of the beams (ON or OFF) in any observation has abnormal behavior. 
\item All ON and OFF beams in each individual observation except L570725 are also consistent with each other assuming Gaussian error bars (Figure \ref{fig:Q1b_all}). 
\item The RFI mitigation in the pipeline is efficient (\citealt{Turner2017pre8,Turner2019}) and there is no evidence that RFI remains in the data. Also, the signal does not look like RFI. 
\item The ON and OFF beams are close enough to each other (Table \ref{tb:beam}; 2-3$\degree$) so that the ionospheric fluctuations should be quite similar in all three beams.
\end{enumerate}

\edits{
The arguments against a celestial signal
\editsjmg{and potential explanations}} are as follows:
\begin{enumerate}
    \item In observation L570725, the two OFF-beams contain weak replicas of the ON \editsjmg{beam} signal (Figure \ref{fig:Dynspec_Detection}).
    \editsjmg{We believe this replica signal can be explained by imperfect phasing of the LOFAR core.}
    \editsjmg{This assessment is based on the analysis of observations of Jupiter (Figure \ref{fig:Dynspec_Jupiter})
    and the pulsar B0809+74
    (Appendix \ref{App:pulsar}, Figure \ref{fig:FFT_pulsar}),  
    taken in the same mode as our exoplanet observations.}
    \item 
    \editsjmg{During each of our observations, 1-2 LOFAR stations did not operate optimally and showed some spurious behaviour in the frequency range 21-30 MHz (Appendix \ref{App:staion_inspection_plots}, Figure \ref{fig:stations_new}).}
    A quantitative analysis suggests that this is not the origin of the apparent signal (Appendix \ref{App:staion_inspection_plots}, \editsjmg{Figure \ref{fig:stations_new}).}
    \item
    \editsjmg{The ON-beam excess in observation L570725 is caused by an unusually low signal in both OFF-beams rather than an unusually high high signal in the ON-beam. 
    Indeed, the ON-beam feature at 21-30 MHz has an amplitude comparable to the features seen in 
    \edits{ON and} OFF beams in all other $\tau$~Boo observations (Figure \ref{fig:Q1b_all}e).}    
    \edits{However, the structure in the dynamic spectrum of the observation L570725 is not the same as in}
    \edits{all other} \editsjmg{observations}\edits{, especially those with large-scale systematics (Appendix \ref{App:dynspecdiff_tauboo}, Figure \ref{fig:Dynspec_OtherDiff}).}
    \editsjmg{It is unclear} why observation L570725 is different from \editsjmg{our} other $\tau$~Boo \editsjmg{observations.}
    \editsjmg{We cannot rule out a small temporary variation within the telescope.}
\end{enumerate}

\editsjmg{
We do have some doubt about the slowly varying signal. Further observations may show whether such a signal is confirmed; if not, it may have been caused by an instrumental artifact.}

\subsubsection{Arguments for the detected signals originating from the $\tau$~Boo system} \label{sec:dis_tauboo} \label{sec:dis_tauboo}

\editsjmg{Assuming the bursty and slowly variable signal are real and of celestial origin, 
do these signals originate from the $\tau$~Boo system? 
Or could other possible radio sources} be at the origin of the detected \editsjmg{circularly polarized} signals? 

In general, most astrophysical \editsjmg{radio} sources display little circular polarization. 
\editsjmg{There is only one known object detected in Stokes-I at 150 MHz in the TGSS survey (\citealt{Intema2017}) within the 13.8 arcmin $\tau$~Boo ON-beam of the LOFAR core.} This same object is also detected in the 1.4\,GHz VLA FIRST Survey Catalog (\citealt{Helfand2015}). This object is probably a background radio-loud quasar since it produces emission over a large frequency range. This quasar is likely not the origin of our detected signals since quasars 
\editsjmg{generally} \editsjmg{have a small degree of circular polarization ($<2\%$, e.g., \citealt{Bower2002}).}
During all our observations, simultaneous observations of Jupiter between 10 and 40 MHz were taken using the Nan\c{c}ay Decameter Array (NDA; \citealt{Boischot1980}; \citealt{Lamy2017}). No emission was seen during the time period of the observation L570725. Therefore, it is very unlikely that the ON-beam signal is caused by Jupiter decameter emission in a side lobe. Radio signals from artificial satellites are also sometimes polarized. However, the detected signals are not from a low Earth orbit (LEO) spacecraft because a LEO satellite would pass from horizon to horizon in several minutes; \editsjmg{in our data, the detected signals are visible for much longer timescales.} \edits{It is also unlikely that the signal is a geostationary (GEO) spacecraft since all three beams are close together on the sky and the same beams were used for all observations \editsjmg{of each planet (cf.~Table \ref{tb:beam})}. 
Besides, the time-frequency structure of the signal does not resemble a satellite beacon.} \editsjmg{As we do not see any other compelling explanation, we conclude it is indeed 
likely that the source of the detected signals is located within the $\tau$~Boo system.}

\subsubsection{Arguments for the detected signals originating from the exoplanet $\tau$~Boo b} \label{sec:dis_planet}

Several physical arguments suggest that the detected signals originate from the planet $\tau$~Boo b rather than its host star. The planetary emission is expected to be much stronger than the stellar emission for hot Jupiters (e.g., \citealt{Griessmeier05AA}; \citealt{Zarka1997pre4, Zarka11PREVII}). \citet{Griessmeier05AA} estimated that the radio emission from the planet $\tau$~Boo~b would be stronger by \edits{several orders of magnitude} than the galactic background, the quiet and quiescent stellar emission, and stellar noise storms. While stellar radio bursts can in principle be as intense as the planetary emission, there is no evidence that $\tau$~Boo undergoes large flare events\edits{; no such flares have been seen in} two years of long-term monitoring of the system with the \texttt{MOST} satellite (\citealt{Walker2008}) and X-ray observations from XMM-Newton (\citealt{Mittag2017}). 

The Radio--Magnetic Bode's law (\citealt{Zarka2001,Zarka2007,Gr2007,Zarka2018,Zarka2018haex}) predicts that for close-in planets an emission flux up to 10$^{6-7}$ times Jupiter's radio flux should be possible. \citet{Zarka2010} tested this scaling law by examining the radio emission from magnetized binary stars. They found that the 
\edits{radio emission (likely not due to CMI) from}
the RS CVn stellar system V711 Tauri matched approximately the 
\edits{extrapolation of the law}
derived for the Solar System (\citealt{Mottez2014}). Their result suggests that the radio--Magnetic Bode's law could hold for 10 orders of magnitude above the range of solar system planets. The recent detection of Ganymede-induced radio emission confirms the Radio--Magnetic Bode's law (\citealt{Zarka2018}). Several models suggest that there could be deviations of one order of magnitude from this scaling law (\citealt{Nichols2011}; \citealt{Saur2013}; \citealt{Nichols2016}),
\edits{but the deviation remains modest compared to the large-scale tendency.}

\edits{ Stellar and planetary radio emissions are also expected to have different polarization properties
(\citealt{Zarka1998, Griessmeier05AA}). The \editsjmg{detected signal is} circularly polarized, which is expected for planetary CMI emission. Quiet and quiescent stellar radio emission 
usually have a low polarization degree
\citep[quiescent emission of M dwarfs has occasionally been observed to reach a polarization degree of 50\%,][and references therein]{Guedel02}.
Stellar radio bursts can be circularly polarized if they are caused by the CMI. 
\editsjmg{However, there is no previous evidence of radio flares from the $\tau$~Boo system. More importantly,}
a star needs to be strongly magnetized (B$_{star}>$10--100\,G) to produce CMI emission that is not quenched by the 
coronal plasma (\citealt{Zarka2007}).
For $\tau$~Boo, the 
mean stellar magnetic field has been measured
to be 
1.7--3.9\,G (\citealt{Catala2007}; \citealt{Donati2008}; \citealt{Fares2009,Fares2013}; \citealt{Mengel2016}; \citealt{Jeffers2018}).
CMI emission from the star over the frequency range of the detected signals is thus unlikely. Still,} 
\editsjmg{since the surface magnetic field maps (e.g., \citealt{Vidotto2012}) do not cover small-scale magnetic structures (\citealt{Mittag2017}), CMI emission from small active regions cannot be not ruled out.}
\editsjmg{Also, the magnetic field of the M-dwarf companion star $\tau$~Boo B is unknown.}
\editsjmg{For this reason, stellar flares cannot be ruled out and could potentially be the cause for the detected radio signal.}

\editsjmg{CMI emission from the planet $\tau$~Boo b remains a  possible cause for the detected circularly polarized radio signal. It it not the only possible source, the other being radio emission by stellar flares. A major argument in favor of planetary radio emission would be the detection of a radio signal compatible with the planetary rotation period. Follow-up observations are required to confirm the presence of this faint signal, and subsequently verify its origin.}


\subsection{Physical constraints on the planetary systems}

\subsubsection{$\tau$~Bo\"{o}tis}\label{sec:tauBooPhysics}


\edits{Assuming that at least one of the two detections (L569131 and L570725) is real and due to CMI emission from the planet (Table \ref{tb:detections}), we can} constrain the 
\edits{maximum surface}
magnetic field of the planet $\tau$~Boo~b to be in the range $\sim$5-11 G.
CMI emission is produced at the local gyrofrequency, $\nu_{g} (MHz) =  2.8 \times B_{p} (G)$, and we use the full frequency range of the detected signals (15-30 MHz) to constrain the magnetic field range for $\tau$~Boo~b. \edits{ This value is slightly smaller than Jupiter's maximum surface polar magnetic field of 14 G (\citealt{Acuna1976,Connerney1993}).
The magnetic moment ($\mathcal{M}_{p}$) of the planet that can be expressed as:
\begin{equation}
    \mathcal{M}_{p} = \mathcal{M}_{Jup}  \left(\frac{B_{p}}{B_{Jup}}\right) \label{eq:mag_moment} \left(\frac{R_{p}}{R_{Jup}}\right)^3,
\end{equation}
where B$_{p}$ and B$_{Jup}$ are the exoplanet's and Jupiter's polar magnetic field strength, and $R_{p}$ and R$_{Jup}$ are the radius of the planet and Jupiter. $R_{p}$ is estimated to 1.06 R$_{Jup}$ using a parametric equation for the radius of an irradiated planet (\citealt{Wang2011}). We find a magnetic moment $\mathcal{M}_{p}$ of 0.94 $\mathcal{M}_{Jup}$ for Tau $\tau$~Boo~b.} \edits{Our tentative} $\tau$~Boo detections are labeled as observation 9 in Figure \ref{fig:radioprediction:tauBootis}. The figure shows that the magnetic field and emission strengths derived for $\tau$~Boo~b are consistent with the predictions by \citet{Gr2007} and \citet{Griessmeier17PREVIII} (in particular with the NR model). Our derived magnetic field strengths could place constraints on the dynamo theory. The field strength estimated for the planet should allow for a sustained planetary magnetosphere, thus protecting it from the stellar wind (\citealt{Nicholson2016}). The detection of radio emission from $\tau$~Boo~b is also interesting in view of atmospheric simulations. For a typical hot Jupiter, CMI quenching by a high plasma frequency in the planetary ionosphere can potentially prevent radio emission; for $\tau$~Boo~b, however, this problem is alleviated by the high planetary mass \edits{and thus low electron density in the ionosphere} \citep{Weber2017,Weber2017pre8,Weber2018}.

We can estimate the power and brightness temperature of the observed emission in both detections (L569131 and L570725). The power of the polarized emission can be estimated by:
\begin{eqnarray}
    P =& S \Omega d^2 \Delta\nu, \label{eq:Power} 
\end{eqnarray}
where $S$ is the observed polarized flux density, $\Omega$ is the solid angle filled by the CMI emission beam (assumed to be in the range \edits{0.16--1.6 sr} similar to Jupiter's decameter emission; \citealt{Zarka2004}), $d$ is the distance to the planet (15.6 pc for $\tau$~Boo~b), and $\Delta\nu$ is the frequency range 
\edits{(column 3 of Table \ref{tb:detections}).} 
Using equation \eqref{eq:Power} and the maximum observed flux, we find a power of \edits{6.3$\times10^{14}$--2.0$\times10^{16}$ W} for $\tau$~Boo~b (Table \ref{tb:detections}). This derived power is $10^{4-5}$ greater than Jupiter's maximum decametric emission ($4.5\times10^{11}$ W; \citealt{Zarka2004}).
\edits{The derived power is consistent with theoretical predictions (\citealt{Gr2007}, \citealt{Griessmeier17PREVIII}) and} compatible with our tests on LOFAR beam-formed data (\citetalias{Turner2019}) which showed that 
\edits{such a large power is needed to be detectable from $\sim$15 pc distance.}
The brightness temperature (T$_{B}$) of the radio source is:
\begin{equation}
    T_{B}  = \frac{S}{\omega k} \lambda^{2}, \label{eq:TB}
\end{equation}
where $\omega$ is the angular size of the emission source, k is the Boltzmann constant, and $\lambda$ is the wavelength. With the power of the exoplanetary radio emission being much higher than that of Jupiter, equation \eqref{eq:TB} means that either the brightness temperature of the emission from $\tau$~Boo~b is much higher than that of Jupiter, or the source region is much more extended. A source size of 1 $R_{Jup}$ implies a brightness temperature of 4.2$\times10^{17}$ to $2.0\times10^{18}$ K (Table \ref{tb:detections}), similar to the brightness temperature of Jovian radio bursts (2$\times10^{17}$ K; \citealt{Zarka1992}), but the latter is reached in sources of size 10-100 km only (\citealt{Zarka1996GeoRL}). These flux densities require a very high Poynting flux due to the proximity of the planet to the star, following the radio magnetic scaling law (e.g., \citealt{Zarka2001,Gr2007,Zarka2018}). Once the planetary origin of the signal is established, modeling of the dynamic spectrum with a software like \texttt{ExPRES} (\citealt{Hess2011}; \citealt{Louis2019}) will provide more information about the emission source and magnetic field structure.

\subsubsection{$\upsilon$ Andromedae}
\edits{The only radio signal tentatively seen from the $\upsilon$~And system is at 2.2$\sigma$ 
\edits{level}
in observation L54519.} If this marginal detection is real, the flux from the system is 540$^{+460}_{-240}$ mJy. If it is a false-positive, we can derive 
a 3$\sigma$ upper limit of 124 mJy from the range 26-73 MHz using the Q1a observable for slowly varying emission. For the moment, we classify this as a non-detection, implying that either the observations were not sensitive enough, the planetary magnetic field is too weak to emit at the observed frequencies, or that Earth was outside the beaming pattern of the radio emission at the time the observations were carried out. 

\subsubsection{55 Cancri}

We can estimate an upper limit on the radio emission from our non-detection of the 55 Cnc system. We find a 3$\sigma$ upper limit of 73 mJy from the range 26-73 MHz using the Q1a observable for slowly varying emission.
Using the attenuated Jupiter modeling done in \citetalias{Turner2019}, this is equivalent to a 
\editsjmg{flux density}
less than $10^{5}$ times the peak flux of Jupiter's decametric burst emission ($\sim5\times10^{6}$ Jy; \citealt{Zarka2004}). \edits{Due to our full orbital coverage of 55 Cnc e, we can rule out beaming effects (Earth outside the beaming pattern) as the cause of our non-detection. Therefore, our non-detection of 55 Cnc implies that either the planetary magnetic field is too weak to emit at the observed frequencies or that the emission is too weak.}


\subsection{Limitations of beamformed observations }

Beamformed observations provide very good resolution in time and frequency (\editsjmg{in the present study, we used} 10 ms and 3 kHz). They are extremely powerful for studying strong (e.g., Jupiter; \citealt{Marques2017}, the Sun; \citealt{Pick2008}), periodic (e.g., pulsars; \citealt{Pilia2016}), or dispersed bursts (e.g., RRATs; \citealt{Karako2015}, FRBs; \citealt{CHIME2019}), as well as spectral lines (e.g., RRLs; \citealt{Asgekar2013}). The situation is much less favorable for weak non-periodic broadband bursts without a clear dispersion signature such as radio bursts expected from exoplanets (e.g., \citealt{Zarka1997pre4}). \edits{This emission} is the most difficult kind of emission to detect at low-frequencies in beamformed mode. 

\edits{Imaging observations excel at detecting continuous and moderately bursty signals. Progress in sky imaging at low-frequencies has been huge recently (e.g., LoTSS, \citealt{Shimwell2019}). For low-frequency \editsjmg{radio telescopes} like LOFAR-LBA (\citealt{vanHaarlem2013}) and NenuFAR \citep{Zarka2012,Zarka2020} imaging is 
\edits{still difficult, computationally expensive, limited by RFI and ionospheric effects, and sometimes impossible due to the lack of} good calibrators, thus beamformed observations are still relevant. There has been recent development of a software package called \texttt{DynSpecMS}\footnote{https://github.com/cyriltasse/DynSpecMS} (Tasse et al., in prep.) that produces low resolution ($\sim$ 8 secs $\times$ $\sim$10 kHz in LoTSS) dynamic spectra from the calibrated visibilities of imaging data. 
\editsjmg{\texttt{DynSpecMS} will provide as many OFF-beams as image pixels, and thus} may be a good alternative to beamformed observations when high-quality imaging is possible (e.g., LOFAR-HBA).}


Beamformed observations retain several advantages over imaging data. They have higher time resolution (in our case 10 msec, compared to several seconds), which can be used to mitigate RFI \editsjmg{on shorter time scales, although not on individual station level}. Beamformed observations also excel at the detection of short bursty signals. The computational cost of their processing is significantly less than for imaging observations since only a handful of pixels have to be analyzed. As done in this paper, a typical observation in beamformed mode should \edits{involve 1 ON-beam and 2} to 3 simultaneous OFF-beams (e.g \citealt{Zarka1997pre4,Turner2017pre8}).

\editsjmg{However,}
\edits{beamformed observations also suffer from several drawbacks.}
Any spurious emission in the side lobes or from instrumental origin is difficult to distinguish from real emission. \editsjmg{Having several simultaneous OFF beams is} absolutely crucial in that case. We found that two OFF beams are a critical minimum to be able to compare statistics as done in this paper.  Future observations may need to have more OFF beams (3 or 4 surrounding the target), even at the expense of the frequency bandwidth (or increasing the data volume in the case of the LOFAR core), \editsjmg{to better evaluate} the background and side lobe effects. Also, if a variable strong low-frequency source is in the sky (e.g., Jupiter or the Sun), it should be monitored (e.g., with a 
\editsjmg{dedicated}
beam) in order to identify any emission that it could contribute to the ON-beam. 

\editsjmg{
It is currently not possible to conclude whether the imaging or beamformed approach is better adapted for the study of exoplanetary radio emission. As long as this question is unanswered, both types of observations should be pursued. In the ideal case, observations should be executed in both modes simultaneously whenever possible.}
\editsjmg{
This paper has demonstrated that beamformed observations can provide important information; as a consequence, they will certainly continue to play a useful role in studying exoplanetary radio emission.}

\section{Conclusions}  \label{sec:conclusion}

In this study, we obtained and analyzed LOFAR-LBA beamformed circularly polarized (Stokes-V) observations of the exoplanetary systems 55~Cancri, $\upsilon$~Andromedae, and $\tau$~Bo\"{o}tis. For the $\tau$~Boo system, we tentatively detect circularly polarized burst emission in the range 14-21 MHz with a statistical significance of 3.2$\sigma$ (Table \ref{tb:detections}; Figure \ref{fig:L569131_detection}). 
\editsjmg{We cannot rule out stellar flares as the source of the emission and emission from the planet $\tau$~Boo remains a possible cause (see Section \ref{sec:dis_planet}); follow-up observations are required (see next Section)}.\editsjmg{For $\tau$~Boo, we also detect} slowly variable emission in the range 21-30 MHz with a significance of 8.6$\sigma$ (Table \ref{tb:detections}; \editsjmg{Figure} \ref{fig:L570725_detection}). A thorough analysis did not allow any firm conclusion: the existence of this signal can neither be confirmed with certainty, nor can it fully be refuted
(Section \ref{sec:origin_b}). Our observations of the 55 Cnc system cover twice the full orbit of the inner planet (Figure \ref{fig:phase}b). \editsjmg{This} is the first time an exoplanetary radio search \editsjmg{project} has \editsjmg{full} orbital coverage of an exoplanetary target. \edits{No} emission is seen from 55 Cnc and we placed a 3$\sigma$ upper limit of 73 mJy on the system from our observations. For the $\upsilon$~And system, we found burst emission in the range 14-38 MHz with a marginal statistical significance of 2.2$\sigma$. 
\editsjmg{We classify this as a non-detection.}
For $\tau$ Boo~b and $\upsilon$~And~b the phase coverage is 25$\%$ (Figure \ref{fig:phase}c) and 40$\%$ (Figure \ref{fig:phase}d), respectively. \editsjmg{For this reason,} we cannot rule out that we have missed radio emission concentrated at specific orbital phases not covered.\\
\indent Using the $\tau$~Boo detections (L569131 and L570725) from the range 15-30 MHz and assuming the emission is from the planet and generated by the CMI mechanism (Section \ref{sec:discussion}), we 
\edits{derived a maximum surface polar} magnetic field for $\tau$~Boo~b between $\sim$5-11 G. The signals for $\tau$~Boo~b range from 190 to \edits{890} mJy with an emitted power of \edits{6.3$\times10^{14}$--2.0$\times10^{16}$ W} and a brightness temperature of 0.42--2.0$\times10^{18}$ K (Table \ref{tb:detections}). The magnetic field and emission strengths derived for $\tau$~Boo~b are consistent with the predictions (Table \ref{tb:predictions}; Figure \ref{fig:radioprediction:tauBootis}) by \citet{Gr2007} and \citet{Griessmeier17PREVIII}.

Follow-up low-frequency radio observations \edits{(e.g., LOFAR, UTR-2, LWA-OLWA, NenuFAR)} are needed to confirm our tentative detections from $\tau$~Boo and the marginal detection from $\upsilon$~And. 
Searching for periodicity in the detected signals will be crucial in confirming their origin and nature. Simultaneous observations between two facilities (e.g LOFAR and NenuFAR) is highly encouraged to rule out possible false-positives due to instrumental effects. \edits{We also hope to incorporate machine learning techniques (e.g., \citealt{Baron2019}) into \texttt{BOREALIS} in the future to more efficiently search through the post-processing outputs.} \\
\indent Ancillary data on these targets are also critical. The planetary mass and planetary inclination are well constrained from high-resolution spectroscopy observations (\citealt{Brogi2012}; \citealt{Rodler2012};  \citealt{Piskorz2017}). The planetary rotation period could also be constrained using high-resolution data (e.g., \citealt{Brogi2016}). Continued or simultaneous monitoring of stellar lightcurves for stellar flares can be done with the northern version of Evryscope that is currently undergoing commission (\citealt{Law2015}; \citealt{Ratzloff2019}; \citealt{Howard2019}). X-ray monitoring is also useful as an indicator to discriminate the planetary signal from that of the star. Follow-up and if possible simultaneous stellar magnetic field maps (e.g., obtained by Zeeman-Doppler Imaging) and wind measurements (e.g., astrospheric absorption) would also be extremely beneficial for \editsjmg{the interpretation of} future observations. 

\editsjmg{
Beamformed observations are notoriously difficult to exploit for exoplanetary radio emission.
Despite these drawbacks, they can provide useful information, as demonstrated in this paper.
We expect them to continue to play an important role in the future alongside other observing modes such as imaging observations.}

\section*{Acknowledgements} 
J.D. Turner was partially funded by the National Science Foundation Graduate Research Fellowship under Grant No. DGE-1315231. J.D. Turner was also funded by a postdoctoral research position at Cornell University. Most of the work on this paper was done during J.D. Turner's PhD at the University of Virginia under the supervision of Robert E. Johnson. Part of this research was carried out at the Jet Propulsion Laboratory, California Institute of Technology, under a contract with the National Aeronautics and Space Administration. This work was supported by the ``Programme National de Plan\'{e}tologie'' (PNP) of CNRS/INSU co-funded by CNES \editsjmg{and} by the ``Programme National de Physique Stellaire'' (PNPS) of CNRS/INSU co-funded by CEA and CNES. 

This research has made use of the Extrasolar Planet Encyclopaedia (exoplanet.eu) maintained by J. Schneider (\citealt{Schneider2011}), the NASA Exoplanet Archive, which is operated by the California Institute of Technology, under contract with the National Aeronautics and Space Administration under the Exoplanet Exploration Program, and NASA's Astrophysics Data System Bibliographic Services. This research has also made use of Aladin sky atlas developed at CDS, Strasbourg Observatory, France (\citealt{Bonnarel2000}; \citealt{Boch2014}). In this paper, all the physical characteristics for the pulsar B0809+74 were taken from the ATNF Pulsar Catalogue \citep{Manchester2005} located at http://www.atnf.csiro.au/research/pulsar/psrcat. 

This paper is based on data obtained with the International LOFAR Telescope (ILT) under project codes LC2$\_$018, LC5$\_$DDT$\_$002, LC6$\_$010, and LC7$\_$013. LOFAR (\citealt{vanHaarlem2013}) is the Low Frequency Array designed and constructed by ASTRON. It has observing, data processing, and data storage facilities in several countries, that are owned by various parties (each with their own funding sources), and that are collectively operated by the ILT foundation under a joint scientific policy. The ILT resources have benefited from the following recent major funding sources: CNRS-INSU, Observatoire de Paris and Universit\'{e} d'Orl\'{e}ans, France; BMBF, MIWF-NRW, MPG, Germany; Science Foundation Ireland (SFI), Department of Business, Enterprise and Innovation (DBEI), Ireland; NWO, The Netherlands; The Science and Technology Facilities Council, UK. 

We acknowledge the use of the Nan\c{c}ay Data Center computing facility (CDN - Centre de Donn\'{e}es de Nan\c{c}ay). The CDN is hosted by the Station de Radioastronomie de Nan\c{c}ay in partnership with Observatoire de Paris, Universit\'{e} d'Orl\'{e}ans, OSUC and the CNRS. The CDN is supported by the Region Centre Val de Loire, d\'{e}partement du Cher.

We thank the ASTRON staff for their help with these observations (pre- and post observing) and understanding where the systematics originate. 

We also thank the following people who were \edits{co-I's} on the LOFAR observing proposals (LC2$\_$018, LC5$\_$DDT$\_$002, LC6$\_$010, and LC7$\_$013) and helped us obtain the data presented in this paper: Alain Lecavelier des Etangs, Alexander Konovalenko, Artie Hatzes, Cyril Tasse, Gregg Hallinan, Iaroslavna Vasylieva, Robert E. Johnson, Sander ter Veen, Vladimir Ryabov, and Walid Majid. 

This paper uses data from the TGSS survey (\citealt{Intema2017}) and we thank the staff of the GMRT that made this survey possible. GMRT is run by the National Centre for Radio Astrophysics of the Tata Institute of Fundamental Research. NDA observations of Jupiter were used to monitor its emission. The NDA is hosted by the Nan\c{c}ay Radio Observatory/ Unit\'{e} Scientique de Nan\c{c}ay of the Observatoire de Paris (USR 704-CNRS, supported by Universit\'{e} d’Orléans, OSUC, and Region Centre in France).

\edits{We thank the anonymous referee for their useful and thoughtful comments. }

\textbf{Facilities:} \texttt{LOFAR} (\citealt{vanHaarlem2013})

\textbf{Software:} \texttt{BOREALIS} (\citealt*{Vasylieva2015}; \citealt*{Turner2017pre8}; \citealt*{Turner2019}); \texttt{IDL Astronomy Users Library} \citep{Landsman1995}; \texttt{Coyote IDL} created by David Fanning and now maintained by Paulo Penteado (JPL).

\bibliographystyle{aasjournal} 
\bibliography{reference.bib} 

\begin{appendix}


\section{Observational setup}\label{app:Summary_obs}

\editsjmg{
Table \ref{tb:obs} gives the exact dates, times and observation IDs for each one of our observations. It also includes the number of stations that were found to behave non-optimally during the observation, and the an assessment of the data quality (percentage of data masked during RFI flagging).}

\editsjmg{
Table \ref{tb:obs} shows that we observed the pulsar B0809+74 for 10 or 15 minutes
either before or after each of the $\upsilon$ And and $\tau$~Boo observations.
For these observations, we used the same settings as the exoplanet observation. 
These pulsar observations were taken to test and calibrate the processing pipeline. Some of these tests are shown in Appendix \ref{App:pulsar}.}

\editsjmg{
For all observations, the beam coordinates of the ON and OFF beams are given in Table \ref{tb:beam}. For the observations of 55 Cnc, the positions of the two OFF beams correspond to the position of  
the pulsar B0823+26 (OFF beam 1) and at a nearby \edits{``empty''} sky region (OFF beam 2). 
For the $\upsilon$ And and $\tau$~Boo observations, the two OFF beams correspond to ``empty'' sky patches, i.e. positions without point sources at a level $\geq$ 100 mJy according to the LOFAR MSSS-HBA survey \citep[150 MHz,][]{Heald2015} and without point sources at $\geq 5$ mJy in the TGSS survey \citep[150 MHz,][]{Intema2017}.}

\begin{table*}[!th]
\begin{threeparttable}
\centering
\caption{Summary of the observations. Column 1: observation number. Column 2: LOFAR cycle. Column 3: LOFAR observation ID. Column 4: date and start time of the observation (UTC). Column 5: duration of the observation. 
Column 6: target elevation range.  
Column 7: number of bad stations present during the observation found by examining the station inspection plots. 
Column 8: amount of RFI masked by the \texttt{BOREALIS} pipeline.
}
\begin{tabular}{cccccccc}
\hline 
\hline
Obs $\#$ & LOFAR Cycle &  LOFAR ID  &  Date $\&$ Time  & Duration  &Elevation    & Bad Stations       & RFI Masked     \\
         &             &            & (UTC)          & (hours)       &$(\degree)$          & ($\#$)                   &($\%$)       \\
\hline
                \multicolumn{8}{c}{55 Cnc [24 hours]}   \\ 
\hline
 1  & 5 & L429868   & 2016-02-03 20:00          & 8 &  37--65  & 2 & 4.5  \\  
 -  & 5&L432752\tnote{1}   & 2016-02-27 18:03      & 9 &  30--65  & 2  &100\\  
 2  & 5&L433872   & 2016-03-02 19:00            & 4 & 52--65   & 2  &3.4\\  
 3  & 5&L441630   & 2016-03-28 18:00            & 6 &  40--65  & 2  & 5.9 \\  
 4  & 5&L527649   & 2016-07-31 08:00             & 3 &  40--60  & 2   & 7.1\\  
 5  & 5&L554093   & 2016-10-22 03:00            & 3 & 45--65   & 1  &5.7\\  
\hline
                \multicolumn{8}{c}{$\upsilon$ And [45 hours]}   \\
\hline
 1  &6 &L545197   & 2016-09-08 00:00    & 5 & 58--78&2   & 4.7\\  
 2  &6 &L545213   & 2016-09-09 00:00    & 5 &58--78 &1  & 2.9 \\  
 3  &6 &L545209   & 2016-09-10 00:00    & 5 &58--78 &1   &5.2  \\  
 4 &6 &L547657   & 2016-09-24 22:00     & 5 & 58--78&1  & 3.4\\  
 5 &6 &L547653   & 2016-09-25 23:00     & 5 & 55--78&1  & 3.4 \\  
 6 &6 &L547649   & 2016-09-26 22:00     & 5 & 58--78&1 &4.1 \\  
 7 &6 &L547645   & 2016-09-28 23:00     & 5 & 55--78&1  & 4.0   \\  
 8 &6 &L551195   & 2016-10-10 22:02     & 5 & 55--78&2 &  4.3  \\  
 9 &6 &L552145   & 2016-10-13 22:00     & 5 & 55--78&2 &   5.3  \\  
\hline
                  \multicolumn{8}{c}{$\tau$~Boo [20 hours]}   \\
\hline
 1 & 7&L569131   & 2017-02-18 01:12     & 3    & 45--53&2 &  6.3 \\  
 2 &7 &L569127   & 2017-02-22 01:00     & 2.5  & 45--55&2  &  4.6 \\  
 3 & 7&L569123   & 2017-02-26 01:16     & 3    & 48--55&2  &  5.6  \\  
 4 & 7&L569119   & 2017-02-27 01:16     & 2.5  & 48--55&2  &  6.2\\  
 5 &7 &L570729   & 2017-03-01 01:16     & 3    & 50--55&2  & 6.4 \\  
6 & 7 &L570725   & 2017-03-06 01:00     & 3    & 50--55&1  & 4.7\\  
 7 &7 &L581807   & 2017-03-25 01:00     & 3    & 40--55&2  & 4.8 \\  
\hline
                \multicolumn{8}{c}{B0809+74 [197 min.]} \\  
\hline
 1 & 6&L545199   & 2016-09-07 23:49   & 0.17 &38 &2 &2.5 \\  
 2 &6 &L545215   & 2016-09-08 23:49   & 0.17 & 38&1 & 3.1 \\  
 3 & 6&L545211   & 2016-09-09 23:49   & 0.17 &38 &1 & 4.5\\  
 4 &6 &L547659   & 2016-09-24 21:49  & 0.17 & 38&1& 3.2   \\  
 5 &6 &L547655   & 2016-09-25 22:49   & 0.17 &39 &1&2.8 \\  
 6 & 6&L547651   & 2016-09-26 21:49  & 0.17 &38 &1& 6.1     \\  
 7 &6 &L547647   & 2016-09-28 22:49  & 0.17 &39 &1& 2.8 \\  
 8 & 6&L551197   & 2016-10-11 03:03   & 0.17 &52 &2& 6.2  \\  
 9 &6 &L552143   & 2016-10-13 21:49  & 0.17 & 39&2&  5.9 \\   
 
 10 &7 &L569129   & 2017-02-18 00:56  & 0.25 &67 &2 &5.9 \\  
 11 &7 &L569125   & 2017-02-22 00:44   & 0.25 &67 &2 & 2.9\\  
 12 &7 &L569121   & 2017-02-26 01:00   & 0.25 & 65&2& 4.4 \\  
 13 & 7&L569117   & 2017-02-27 01:00   & 0.25 &65 &2&4.9\\  
 14 &7 &L570727   & 2017-03-01 01:00   & 0.25 &65 &2& 3.9 \\  
 15 &7 &L570723   & 2017-03-06 00:44   & 0.25 &64 &1 & 3.5\\  
 16 &7 &L581805   & 2017-03-25 00:44   & 0.25 &60 &2&  4.1 \\  
\hline
\hline
\end{tabular}
\label{tb:obs}
\begin{tablenotes}
\item[1] This observation was not use in the analysis due to near-continuous RFI at all frequencies for the entire observing period. \editsjmg{In addition,during}
this observation, there was also an antenna cable delay compensation issue (communication from ASTRON Radio Observatory staff).
\end{tablenotes}
\end{threeparttable}
\end{table*}

\begin{table}[!htb]
\centering
\caption{Beam coordinates used for the observations. Column 1: beam name. Column 2: right ascension (RA) of the beam. Column 3: declination (DEC) of the beam. Column 4: distance of the beam from the ON-beam. } 
\begin{tabular}{cccc}
\hline 
\hline
Beam   & RA (J2000)  & DEC (J2000)  & Distance \\
       &  (h:m:s)    & ($^\circ$:':") & ($\degree$) \\
\hline
\hline
\multicolumn{4}{c}{55 Cnc} \\
\hline
ON      & 08:52:34.81   &   +28:19:51.00    & ---\\
OFF 1   & 08:26:51.38   &   +26:37:23.80    & 6.0\\
OFF 2   & 08:58:10.14   &   +27:50:53.09    & 1.3\\
\hline
\multicolumn{4}{c}{ $\upsilon$ And} \\
\hline
ON      & 01:36:47.84   &   +41:24:19.60    & --- \\
OFF 1   & 01:40:00      &   +38:00:00       & 3.4 \\
OFF 2   & 01:30:00      &   +48:00:00       & 6.7 \\
\hline
\multicolumn{4}{c}{$\tau$~Boo} \\
\hline 
ON      & 13:47:15.74  &    +17:27:24.90    & --- \\
OFF 1   & 13:54:44.953 &    +16:49:29.20    & 1.9      \\
OFF 2   & 13:58:10.366 &    +19:00:01.37    & 3.0   \\
\hline
\multicolumn{4}{c}{ B0809+74} \\
\hline 
ON      & 08:14:59.52   &   +74:29:06.00    & ---\\
OFF 1   & 08:23:24.48   &   +71:52:58.80    & 2.7\\
OFF 2   & 09:03:31.92   &   +74:52:58.80    & 3.2\\
\hline
\end{tabular}
\label{tb:beam}
\end{table}

\section{Observable quantities} \label{app:obs_quant}
The post-processing section of the \texttt{BOREALIS} pipeline computes a series of observable quantities, Q1 to Q4. 
\editsjmg{As defined in \citetalias{Turner2019},} 
Q1 is designed to search for slowly variable emission on the order of minutes to hours. Q2 is the frequency-integrated time-series designed to find bursty emission and is created by integrating the processed data over a selected frequency range (several trial values were used, see Table \ref{tb:PPsetup}), followed by high-pass filtering using a smoothing window of 10 time bins (e.g 10 seconds for a rebin time of 1 second) and normalization by the standard deviation. Q3 is designed to localize the bursts (Q2) in time with a resolution $\delta T$ (e.g., 2 mins) for one selected threshold $\eta$ (in units of standard deviations). Q4 is the statistical analysis of the broadband burst emission (Q2), searched over an entire observation and a selected frequency range, and plotted against the threshold $\eta$. 

\editsjmg{For convenience, the definition of those each observable quantities that are used in the present work is summarised below:}
\begin{itemize}
    \item Q1: Slowly variable emission observables.
     \begin{itemize}
         \item Q1a (Time-series): Dynamic spectrum integrated over all frequencies and rebinned to a time resolution of $\delta T$ (e.g., 2 minutes)
         
         \item      Q1b (Integrated spectrum): 
                        Dynamic spectrum integrated over all times and and rebinned to a frequency resolution of $\delta F$ (e.g., 0.5 MHz)
     \end{itemize}  
    \item Q2 : Time series obtained by integrating the processed data over a selected frequency range (several trial values were used, see Table \ref{tb:PPsetup}), followed by high-pass filtering created using a smoothing window of 10 time bins (e.g., 10 seconds for a rebin time of 1 sec) and normalization by the standard deviation.

    \item Q4: 
    Statistical analysis of the broadband burst emission (Q2), searched over an entire observation and a selected frequency range, and plotted against the threshold $\eta$ (in units of standard deviations).
      \begin{itemize}
         \item Q4a (Number of Peaks): Number of peaks where Q2$ \ge \eta$. 
         \item Q4b (Power of Peaks): Sum of the power of peaks where Q2 $\ge \eta$.
         \item Q4c (Peak Asymmetry): Number of peaks where Q2$ \ge \eta$ subtracted by the number of peaks where Q2$ \le - \eta$.
         \item Q4d (Power Asymmetry): Sum of the power of peaks where Q2$ \ge \eta$ subtracted by the sum of the $|\text{power}|$ of peaks where~Q2$~\le~-\eta$.
         \item Q4e (Peak Offset): Number of peaks where Q2$ \ge \eta$ for the ON-beam and exceeding the corresponding OFF values by a factor $\ge 2$.
         \item Q4f (Power Offset): Sum of the power of peaks where Q2$ \ge \eta$ for the ON-beam and exceeding the corresponding OFF values by a factor $\ge 2$.
      \end{itemize}
\end{itemize} 
\editsjmg{When examining Q4a-f, we compare two beam against each other 
in the form of difference curves, e.g.~Q4f$_{\text{Diff}}$=
Q4f(ON)-Q4f(OFF) or Q4f$_{\text{Diff}}$=Q4f(OFF1)-Q4f(OFF2).} 

The post-processing pipeline produces Q1-Q4 synthesis plots, for each observation of Table \ref{tb:obs}, pair of beams, and set of parameters of Table \ref{tb:PPsetup}. Example Q1, Q2, and Q4 plots can be found in Figures \ref{fig:L569131_detection} and \ref{fig:L570725_detection}. 
\editsjmg{Figure \ref{fig:L570725_detection} shows the Q1 observables
(Q1a in Figure  \ref{fig:L570725_detection}.a and Q1b in Figure  \ref{fig:L570725_detection}.b).}
\editsjmg{
The}
\editsjmg{ON}
beam is displayed in black, and the OFF-beams are \editsjmg{displayed} in red \editsjmg{and blue}. 
\editsjmg{The lower panel show the differences between two beams (orange, green, and brown curves)}.
\editsjmg{Figure \ref{fig:L569131_detection}.a and  Figure \ref{fig:L569131_detection}.b display}
the \editsjmg{observable quantity Q2 in the form of scatter plots}.
\editsjmg{These scatter plots allow to compare two beams by}
showing 
the normalized signal in one beam versus the normalized signal in the other. Panels c-f in Figure \ref{fig:L569131_detection} display the Q4a, Q4b, Q4e, and Q4f observables for the ON-OFF and OFF1-OFF2 difference curves. In these plots, the 3 pairs of dashed curves surrounding the horizontal axis indicate the $\pm$1, $\pm$2 and $\pm$3$\sigma$ levels reached when computing these Q4 observables from 10000 pairs of Gaussian noise time series, as a function of the threshold on normalized signal intensity. Q4 difference curves exceeding the 2$\sigma$ to 3$\sigma$ reference curves over a significant range of thresholds indicate \editsjmg{a} possible detection. 
\editsjmg{We consider the observables Q4a-f separately.}

\editsjmg{In most observations,}
\edits{
the distribution of points in the Q2 scatter plot \editsjmg{is} not} 
\edits{as ``circular'' as \editsjmg{it} should be if both signals \editsjmg{were} Gaussian. 
Ionospheric fluctuations affecting both the ON- and the OFF-beam simultaneously}
\edits{may elongate the scatter-plot along
the main diagonal and thus cause}
\editsjmg{cause a non-circular distribution.}
\editsjmg{To remedy this problem, an elliptical correction was designed \citepalias{Turner2019}. It circularizes}
\edits{the distribution of points in the Q2 scatter plot (i.e. \editsjmg{mitigates} the ionospheric fluctuations
), \editsjmg{thus allowing} to detect \editsjmg{more easily} outliers \editsjmg{which appear} only in the ON-beam.}

\section{\editsjmg{Detection criteria}} \label{app:summary_numbers}

\editsjmg{
At the end of Section \ref{sec:pipeline}, we showed the criteria that are used to determine a tentative detection. These criteria are based on the observable quantities (Q1a, Q1b, Q2, and Q4a-f, as defined in Appendix \ref{app:obs_quant}).
In this Appendix, we describe how these criteria are implemented in an automated search algorithm.
}

\subsection{\editsjmg{Criteria for slowly varying emission (observable quantities Q1a and Q1b)}}

\editsjmg{For the observables Q1a and Q1b, we define the following detection criteria (sometimes also called ``summary numbers''):}
\begin{itemize}[noitemsep,topsep=0pt]
    \item $N1$: Integral of the ON-OFF difference curve.
    \item $N2$: Percentage of positive values in the ON-OFF difference curve.
    \item $N3$: Standard deviation of the ON-OFF difference curve.
    \item $N4$: Probability of false-detection, comprised between 0 and 1. It describes the likelihood that the ON-OFF difference curve can be reproduced from simulations based on random Gaussian noise. For example, a probability of $6.3\times10^{-5}$ corresponds to a 4$\sigma$ detection.
\end{itemize}
The \editsjmg{detection criteria} $N1$ to $N3$ should increase if there is an excess signal in the ON-beam.  

\subsection{\editsjmg{Criteria for burst emission (observable quantities Q4a to Q4f)}}

\editsjmg{For the observables Q4a to Q4f, we define the following detection criteria:}
\begin{itemize}
    \item $N5$: Integral of the ON-OFF difference curve.
    \item $N6$: Ratio of $N5$ to the integral of the 1$\sigma$ reference Gaussian curve (e.g., dashed lines in Figure \ref{fig:PP_Q4f_diff}). The reference curve was created by running simulations on 10000 pairs of Gaussian noise distributions and calculating the respective observable (i.e. Q4f).
    \item $N7$: Percentage of positive values in the ON-OFF difference curve. 
    \item $N8$: Percentage of ON-OFF values above the 1$\sigma$ reference Gaussian curve.
    \item $N9$: Minimum ($\eta_{min}$) and maximum ($\eta_{max}$) threshold values where ON-OFF $>0$.
    \item $N10$: Percentage of positive values in the ON-OFF difference curve for $\eta_{min} \le \eta \le \eta_{max}$
    \item $N11$: Percentage of negative values in the ON-OFF difference curve for $\eta \ge \eta_{max}$
    \item $N12$: Probability of false-detection, comprised between 0 and 1 (similar to $N4$).
\end{itemize}
If an excess signal is in the ON-beam then \editsjmg{the detection criteria} $N5$ and $N6$ should both be greater than 0. \editsjmg{The detection criteria} $N5$, $N6$, and $N10$ probe the continue nature of the emission which is essential for a detection. \editsjmg{Additionally,} $N11$ can be used to eliminate false-positives because $N11 = 0$ for a detection.  

\subsection{\editsjmg{Test of the detection criteria}}

\begin{figure}[!thb]
\centering
  \begin{tabular}{c}
     \begin{subfigure}[c]{0.48\textwidth}
        \centering
        \caption{}
    \includegraphics[width=\textwidth,page=2]{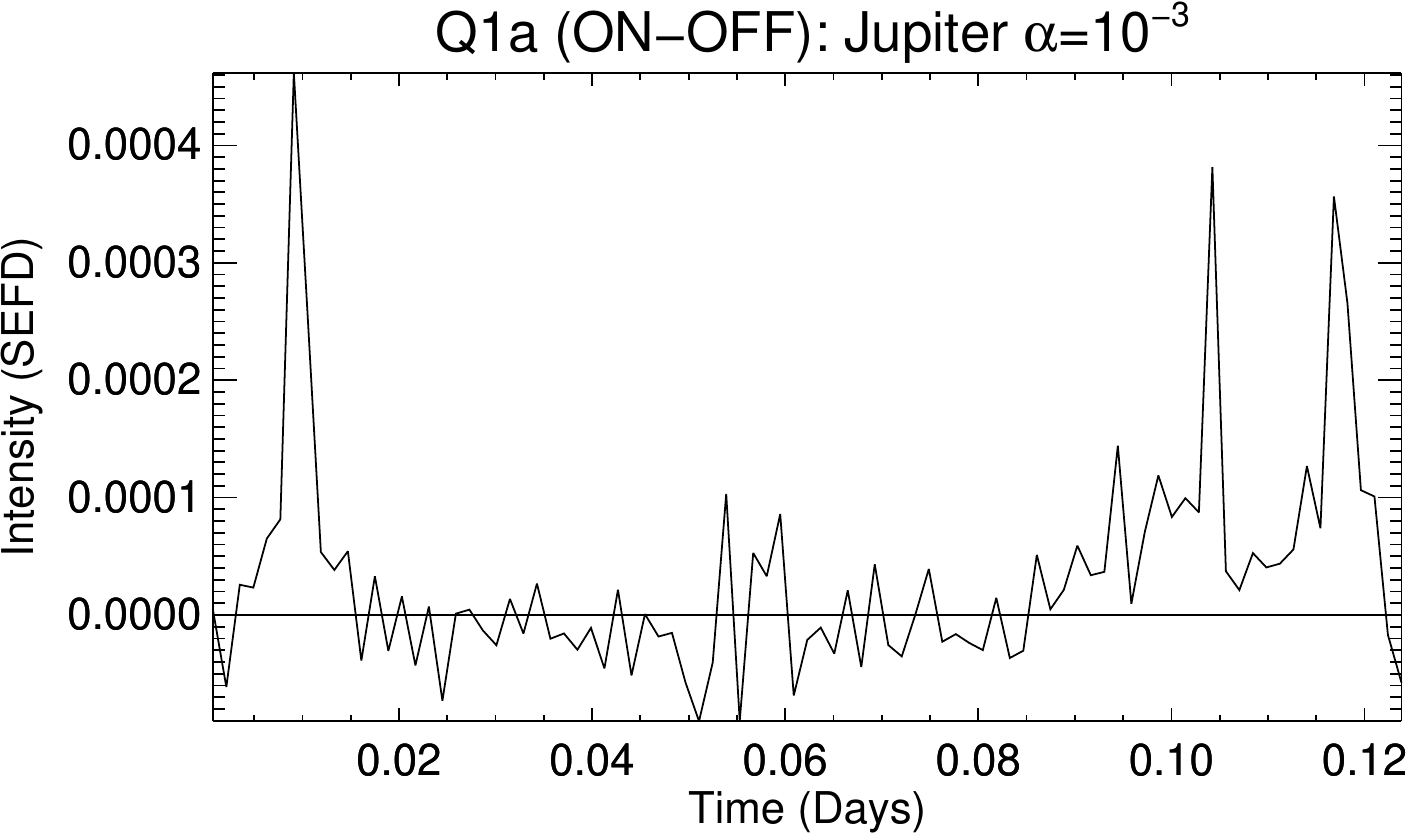}
             \label{}
    \end{subfigure}%
    \\
    \begin{subfigure}[c]{0.48\textwidth}
        \centering
        \caption{}
        \includegraphics[width=\textwidth,page=4]{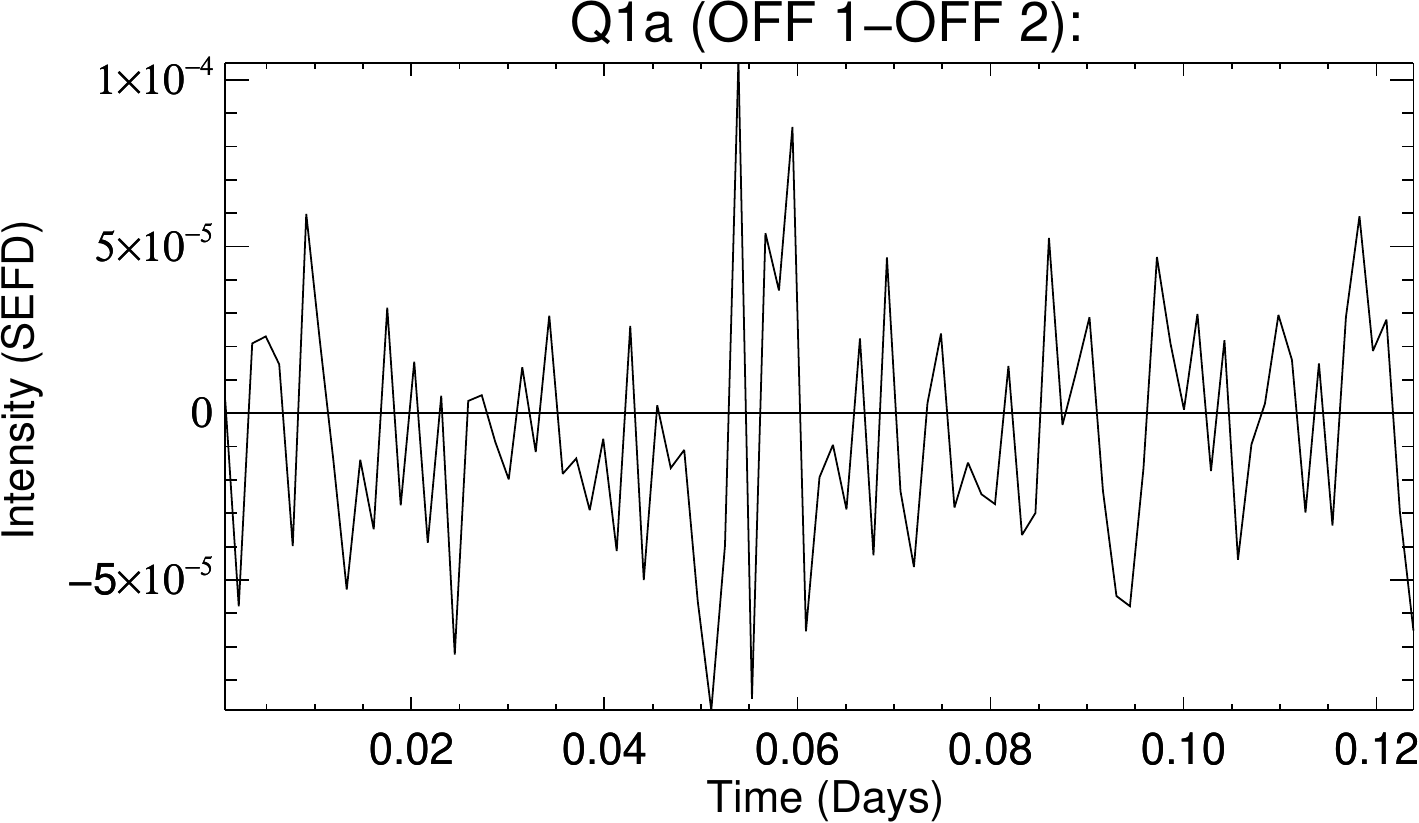} 
         \label{}
    \end{subfigure}%
\end{tabular}
  \caption{Example Q1a difference curves for Jupiter signal attenuated by a factor $\alpha=10^{-3}$ (panel \textit{a}) and for the two OFF beams (panel \textit{b}). }
  \label{fig:PP_Q1_diff}
\end{figure}

In order to determine the reliability of 
\editsjmg{the detection criteria}
in finding a signal, we
tested them on LOFAR data containing \editsjmg{radio emission from Jupiter}.
We used the 
LOFAR dataset of Jupiter radio emission attenuated by a factor $\alpha = 10^{-3}$ from \citetalias{Turner2019}. An example Q1a difference curve can be seen in Figure \ref{fig:PP_Q1_diff} and the corresponding 
\editsjmg{detection criteria}
from these plots are listed in Table \ref{tb:PP_Q1_diff}. All the Q1 
\editsjmg{detection criteria}
show an excess in the ON-beam. The excess is moderate because the emission essentially consisted of \editsjmg{two burst, one near the beginning and one ear the end} 
of the interval \citep[see Fig. 1 in][]{Turner2019}. From $N4$, we conclude this is a real detection at 3.8$\sigma$ \editsjmg{significance}.

\begin{table}[!htb]
\centering
\caption{\editsjmg{Values of detection criteria}
for Q1a from Figure \ref{fig:PP_Q1_diff}. 
Column 1: \editsjmg{Detection criterion}.
Column 2: 
\editsjmg{Detection criterion applied to the difference between ON and OFF beam.}
Column 3:  
\editsjmg{Detection criterion applied to the difference between OFF1 and OFF2 beam.}
Column 4: Absolute value of the ratio of ON-OFF (column 2) by OFF1-OFF2 (column 3). }
\begin{tabular}{cccc}
\hline
$\#$             & ON- OFF    & OFF 1 - OFF 2  & $|$Ratio$|$\\
\hline
N1      & $3.3\times10^{-6}$          & -$1.2\times10^{-6}$  & 2.75\\
N2      &  60         &44                                 & 1.35\\
N3      & $9.4\times10^{-5}$         & $3.7\times10^{-5}$ & 2.54\\
N4      &$10^{-4}$, $>3.8\sigma$  & 1, 1$\sigma$ & 3.8$\sigma$\\
\hline
\end{tabular}
\label{tb:PP_Q1_diff}
\end{table}

\begin{figure}[!hbt]
\centering
\begin{tabular}{c}
     \begin{subfigure}[c]{0.45\textwidth}
        \centering
        \caption{}
    \includegraphics[width=\textwidth,page=1]{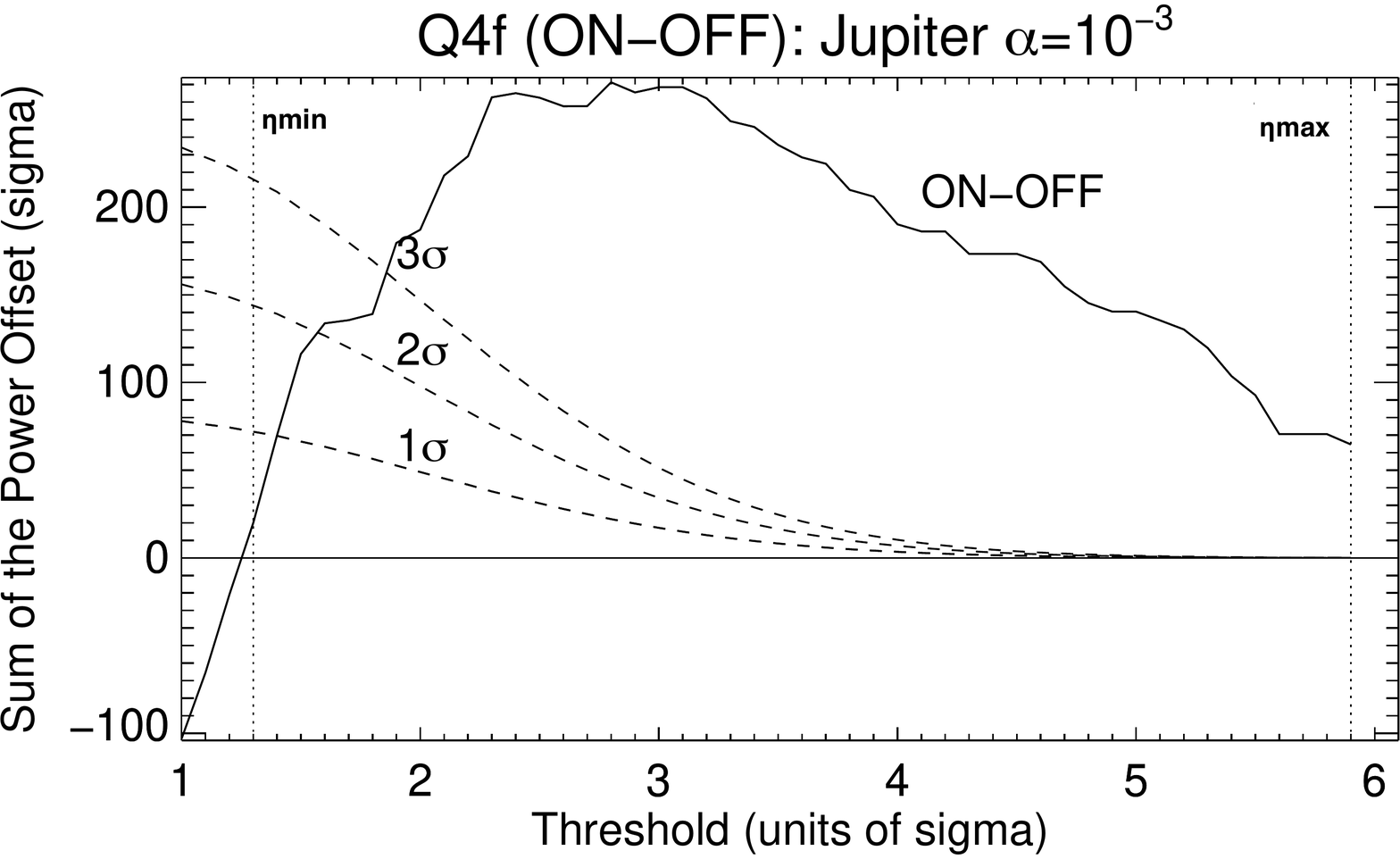} \\
             \label{}
    \end{subfigure}%
    \\
     \begin{subfigure}[c]{0.45\textwidth}
        \centering
        \caption{}
    \includegraphics[width=\textwidth,page=3]{PP_Qs_new2.pdf} 
             \label{}
    \end{subfigure}%
\end{tabular}
  \caption{Example Q4f difference curves for Jupiter signal attenuated by a factor $\alpha=10^{-3}$. \textit{Panel a:} ON vs. OFF beam. \textit{Panel b:} OFF1 vs. OFF2 beam. The dashed lines are the 1, 2, 3$\sigma$ reference curves derived from computing Q4f on 10000 pairs of Gaussian noise distributions. }
  \label{fig:PP_Q4f_diff}
\end{figure}

An example Q4f difference \editsjmg{plot} is shown in Figure \ref{fig:PP_Q4f_diff}
\editsjmg{; the corresponding detection criteria}
are listed in Table \ref{tb:PP_Q4f_diff}. We note that the line in the plot for $\eta_{max}$ and $\eta_{min}$ are slightly offset due to the granularity of the simulations in $\eta$-space. 
\editsjmg{In Table \ref{tb:PP_Q4f_diff}, all}
\editsjmg{detection criteria}
show an excess in the ON-beam with very large ratios for $N5$ and $N6$. $N6$ also probes the continuous nature of the curve which is essential for detection. $N12$ is also very useful because it allows us to check the false positive rate of any tentative detections. From these \editsjmg{detection criteria}.
we conclude a firm detection of bursty emission in this dataset.

\subsection{\editsjmg{Numerical implementation}}

\edits{Based on these results, we can 
\editsjmg{define a set of parameters for an automatic search algorithm.}
The \editsjmg{parameter} for the automated search for detections in Q1a and Q1b was $N2>80$\%, indicating that the ON-beam should contain more slowly variable emission than the OFF beam for most of the observation. For Q4f, $N6$ appears to be the most sensitive \editsjmg{criterion} for bursty signals since its ratio between the ON-OFF and OFF1-OFF2 curves is the highest. Also, $N6$ probes the continuous nature of the Q4 difference curve, which is essential for a detection. Therefore, we \editsjmg{required} for automatic burst detection for Q4f that $N6$(ON-OFF) $>1$ and $N6$(ON-OFF) $>$ $N6$(OFF1-OFF2). \editsjmg{Finally, the criterion} $N12$ is also very useful because it allows us to check the false positive rate of any tentative detection.}

\begin{table}[!htb]
\centering
\caption{
\editsjmg{Detection criteria}
for Q4f from Figure \ref{fig:PP_Q4f_diff}. 
Column 1: 
\editsjmg{Detection criterion.}
\editsjmg{Column 2}: 
\editsjmg{Detection criterion applied to the difference between ON and OFF beam.}
Column 3:  
\editsjmg{Detection criterion applied to the difference between OFF1 and OFF2 beam.}
\editsjmg{Column 4: Absolute value of the ratio of ON-OFF (column 2) by OFF1-OFF2 (column 3).}}
\begin{tabular}{cccc}
\hline
$\#$      &   ON- OFF    &  OFF 1 - OFF 2      & $|$Ratio$|$ \\
\hline
N5 & 657  & 14     &  47\\
N6 & 6.1    & 0.1  &  61              \\ 
N7 & 88    &26     & 3\\
N8 & 82       & 0   & --          \\
N9a &1.3       &   1.3 & 1.2  \\
N9b & 5.9       & 3.2  & 1.7\\
N10         & 84       & 22 & 3.8\\
N11          & 0 & 8 &  --\\
N12         & $10^{-7}$, $5\sigma$  & 0.49, 1$\sigma$  & $5\sigma$ \\
\hline
\end{tabular}
\label{tb:PP_Q4f_diff}
\end{table}

%
\subsection{\editsjmg{Comparison of detection criteria for different post-processing runs}}\label{App:run_numbers_compare}

\begin{figure}[!htb]
\centering
 \begin{tabular}{c}
  \begin{subfigure}[c]{0.45\textwidth}
        \centering
        \caption{}
      \includegraphics[width=\textwidth]{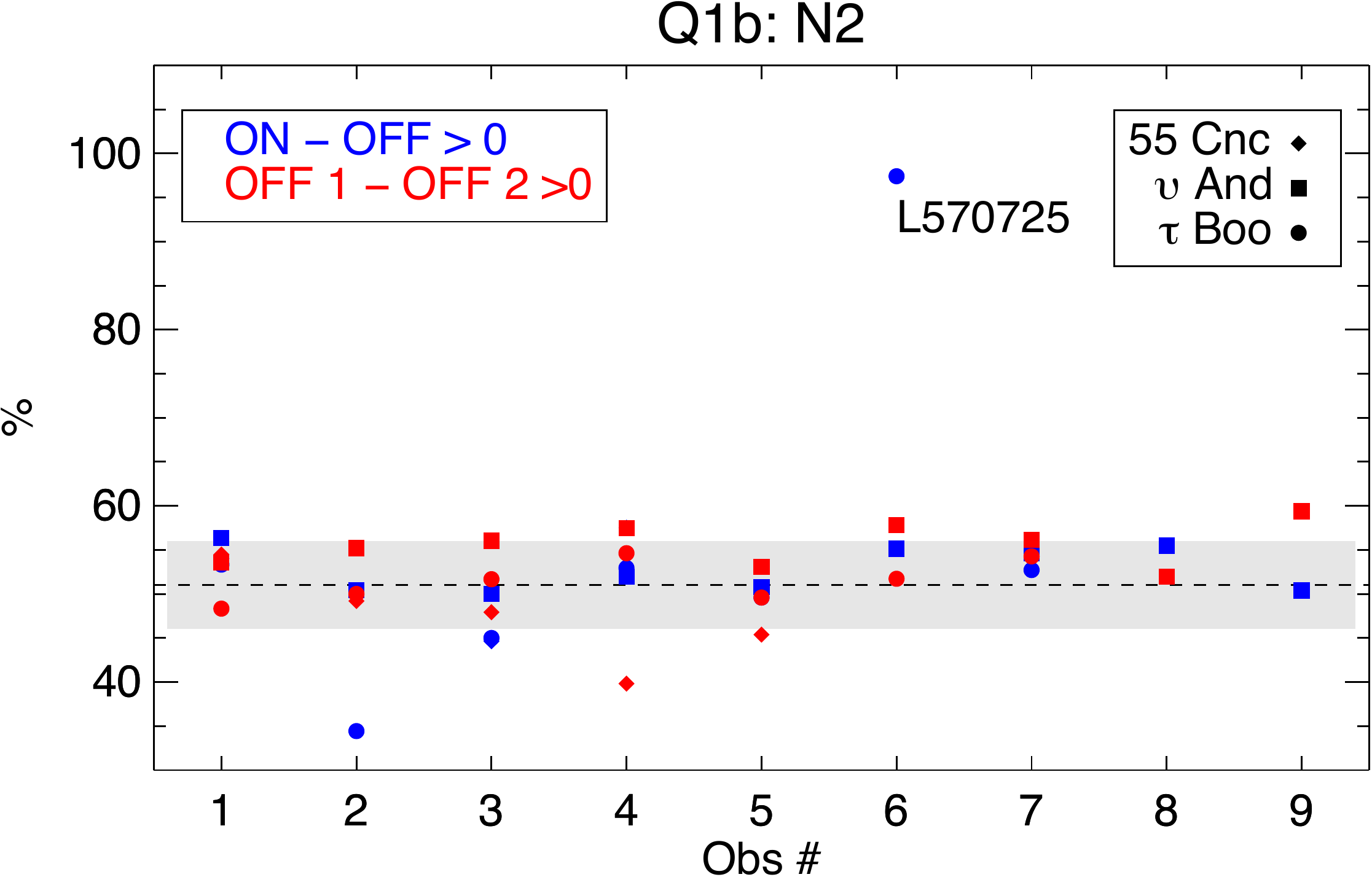}  \\
              \label{}
    \end{subfigure}%
    \\
  \begin{subfigure}[c]{0.45\textwidth}
        \centering
        \caption{}
       \includegraphics[width=\textwidth]{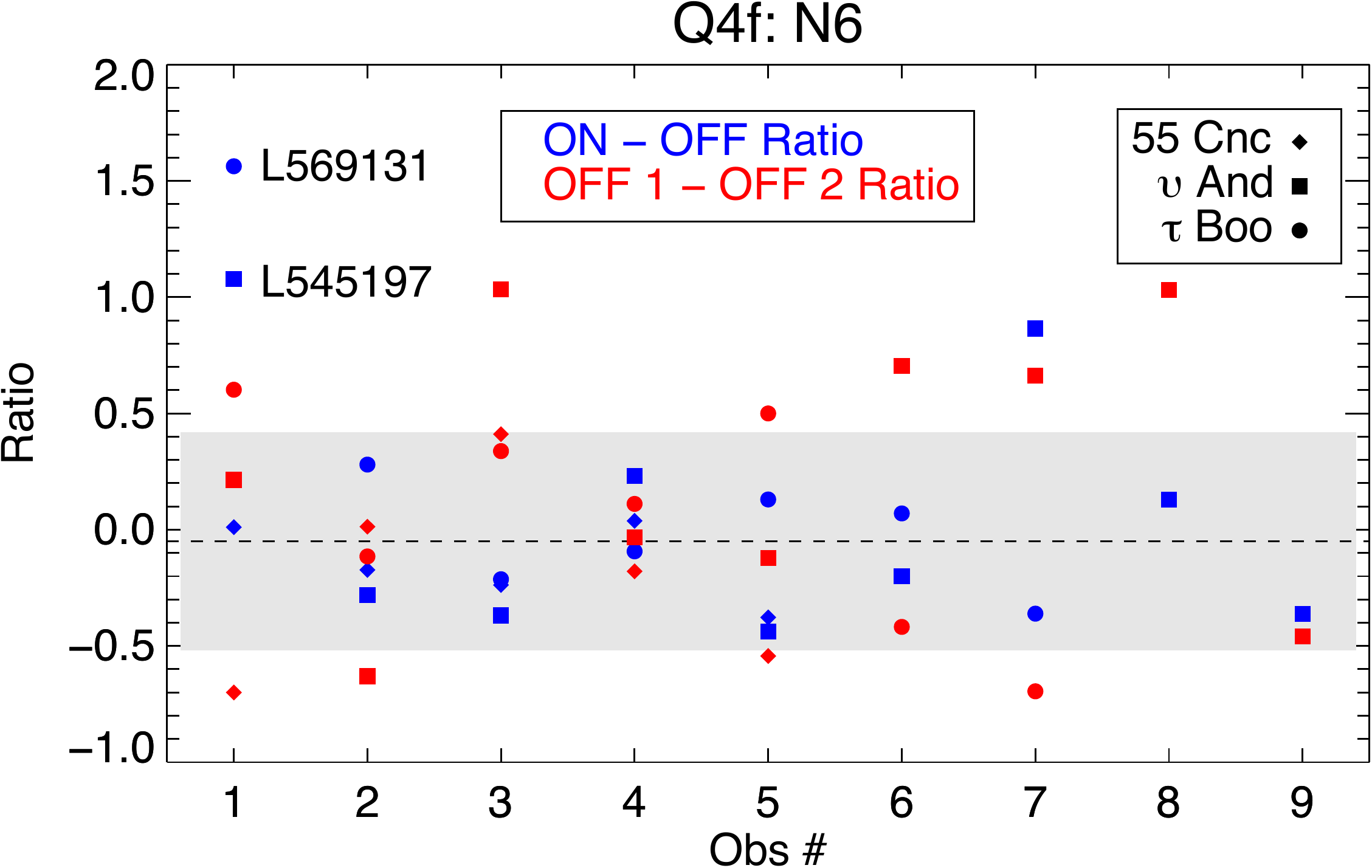}\\
             \label{}
    \end{subfigure}%
    \end{tabular}
 \caption{\editsjmg{
 Detection criteria for all observations.
 \textit{Panel a}: Detection criterion N2 for the observable quantity Q1b.
 \textit{Panel b}: Detection criterion N6 for the observable quantity Q4f.
 Only the lowest frequency band (15--38 MHz for $\tau$ Boo and $\upsilon$~And and 26-50 MHz for 55 Cnc) and |V| is shown. The median value for all post-processing runs is shown as a dashed line. The 1 $\sigma$ standard deviation is shown as the gray area.}
 }
\label{fig:PP_stats} 
\end{figure}
 
We can plot all post-processing runs (each run is referred to as a run number) together to look for outliers and \editsjmg{locate observations or processing runs that require a more detailed inspection.}
A run number is defined by a specific target (e.g., $\tau$~Boo), date and observation (e.g., L569131), frequency range (e.g., 15-38 MHz), polarization (e.g., $|V|$), and rebin time ($\delta \tau$, 1 sec or 10 secs).  An example of this can be found Figure \ref{fig:PP_stats},
\editsjmg{which shows, as a function of observation number, (a) the detection criteria N2 calculated for the observable quantity  Q1b, (b) the detection criterion and N6 for the observable quantity Q4f}. 

\editsjmg{For the slowly varying emission, Figure \ref{fig:PP_stats}.a shows that observation L570725 of $\tau$ Bootis stands out from the other observations and requires a detailed analysis. This detailed analysis is presented in Section \ref{sec:L570725}.}

\editsjmg{For the burst emission, panel b of Figure \ref{fig:PP_stats} shows the detection criterion N6 for the observable quantity Q4f. As for panel a, this is used to identify observations which require a detailed analysis.}
\editsjmg{Figure \ref{fig:PP_stats}.b shows two observations which stand out from the overall distribution, namely L569131 and L545197. L569131 corresponds to the $\tau$~Boo observation in which we tentatively identify burst emission (Section \ref{sec:tauboo_L569131}); L545197 is the observation of $\upsilon$~And in which we see a marginal detection (Section \ref{sec:upsandr_detection}).} 

\section{Integrated spectrum (Q1b) plots for all exoplanet observations}\label{App:Q1ball}

\begin{figure*}[!htb]
\centering
  \begin{tabular}{cc}
    \begin{subfigure}[c]{0.45\textwidth}
        \centering
        \caption{}
        \includegraphics[page=1,width=\textwidth]{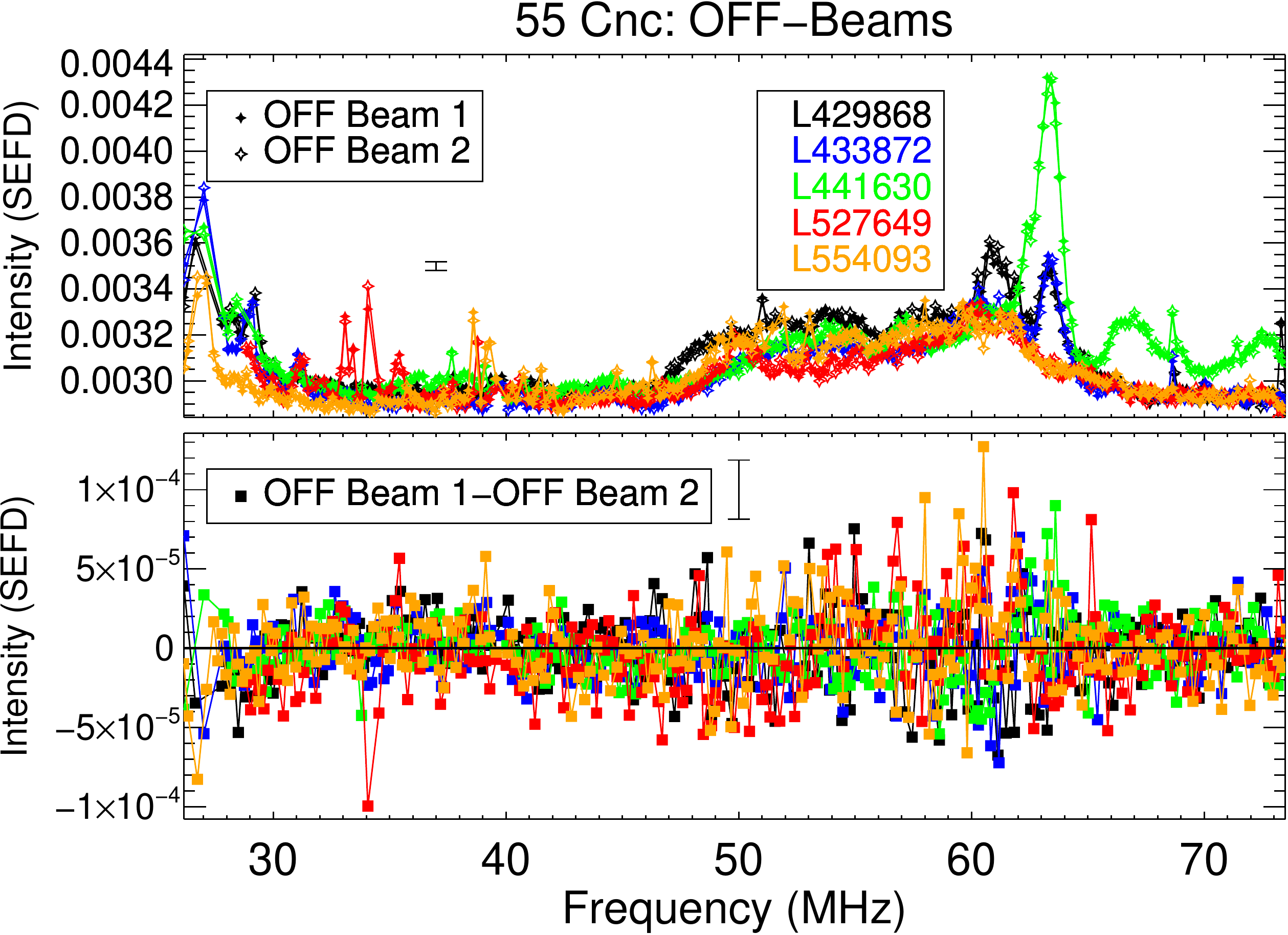}
               \label{}
    \end{subfigure}%
    \hspace{5mm}

        \begin{subfigure}[c]{0.45\textwidth}
        \centering
        \caption{}
      \includegraphics[page=1,width=\textwidth]{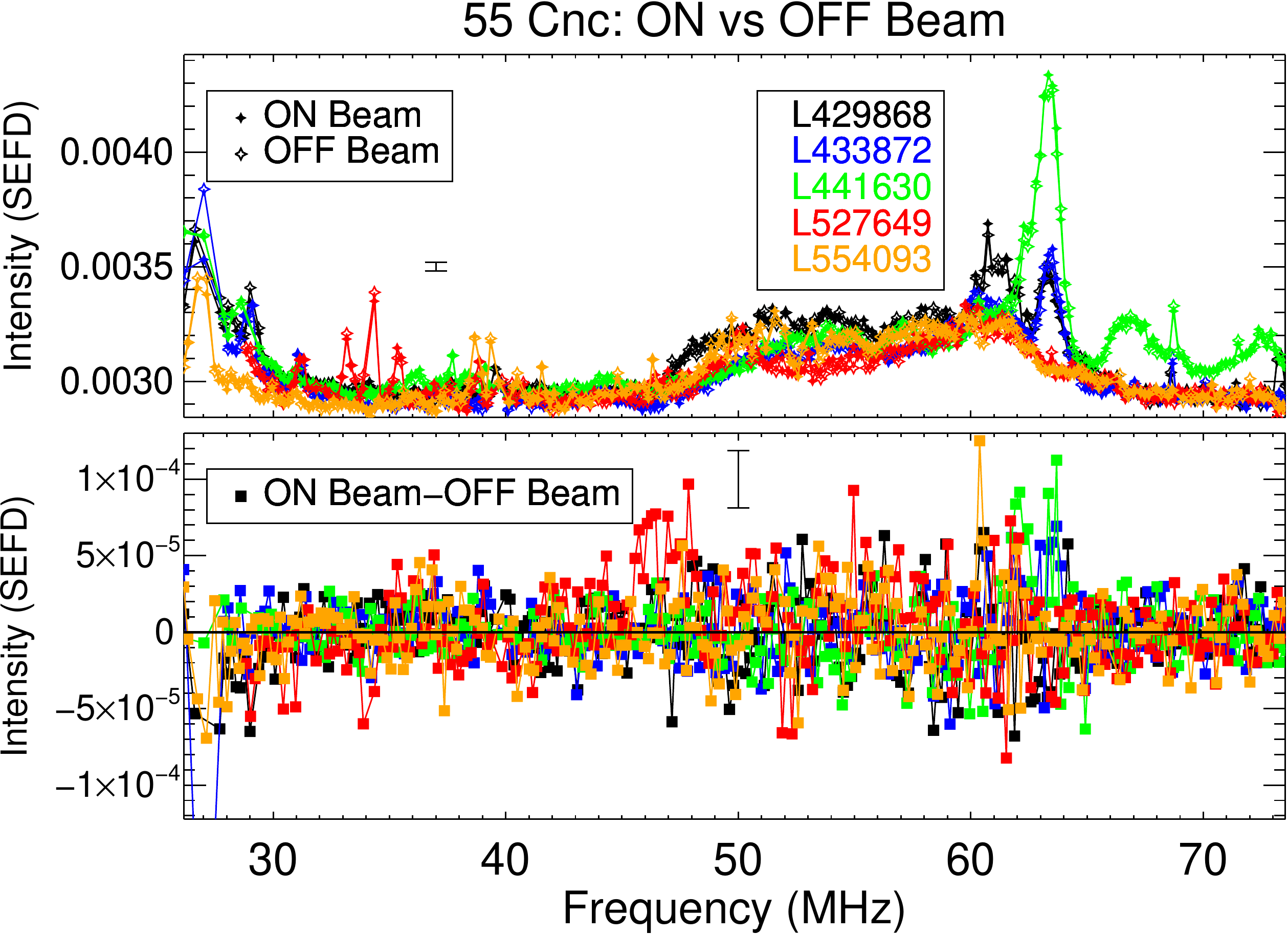}  
              \label{}
    \end{subfigure}%
    \\
     \begin{subfigure}[c]{0.45\textwidth}
        \centering 
        \caption{}
    \includegraphics[page=1,width=\textwidth]{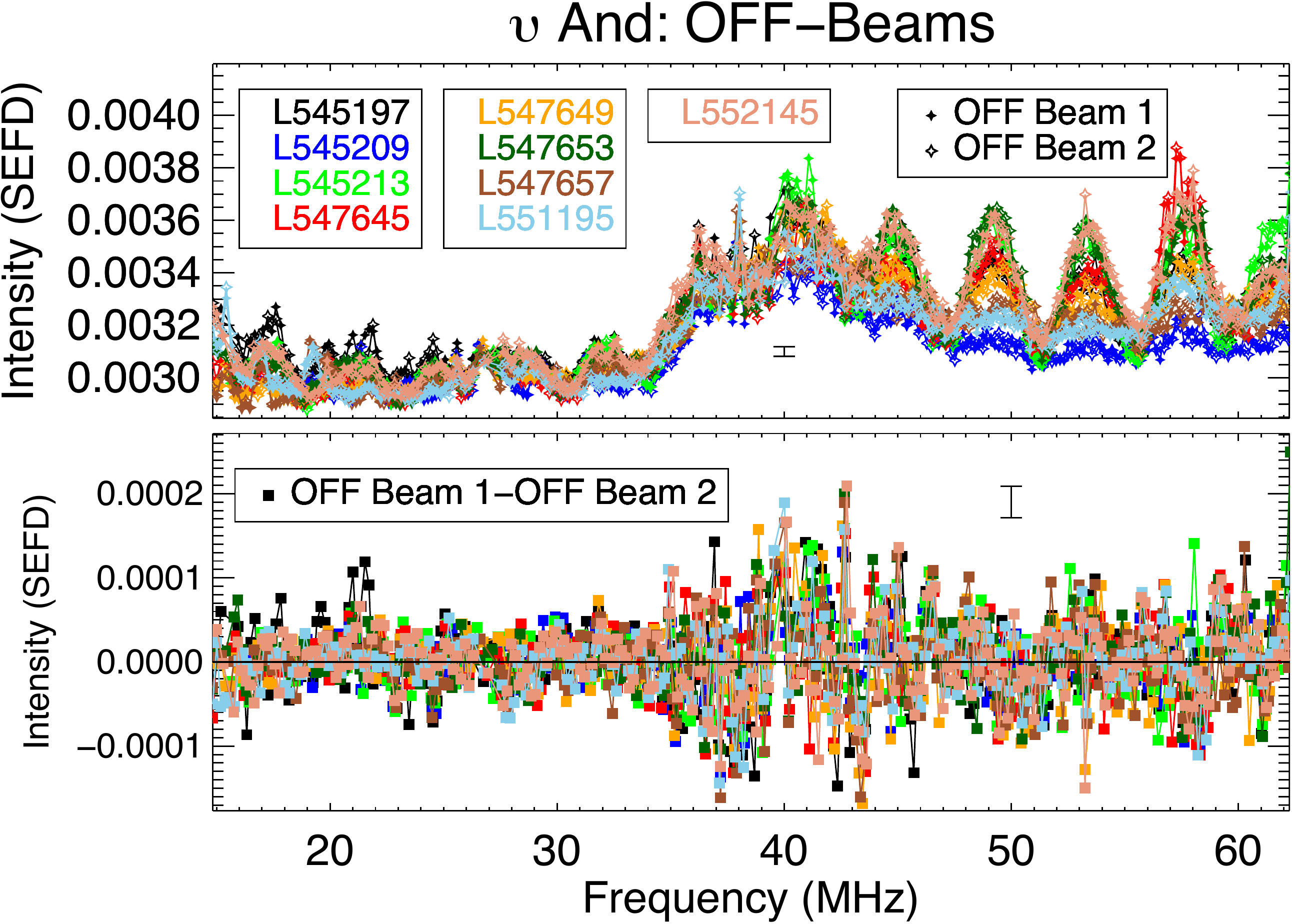} 
               \label{} 
    \end{subfigure}%
    \hspace{5mm}

      \begin{subfigure}[c]{0.45\textwidth}
        \centering
        \caption{}
         \includegraphics[page=1,width=\textwidth]{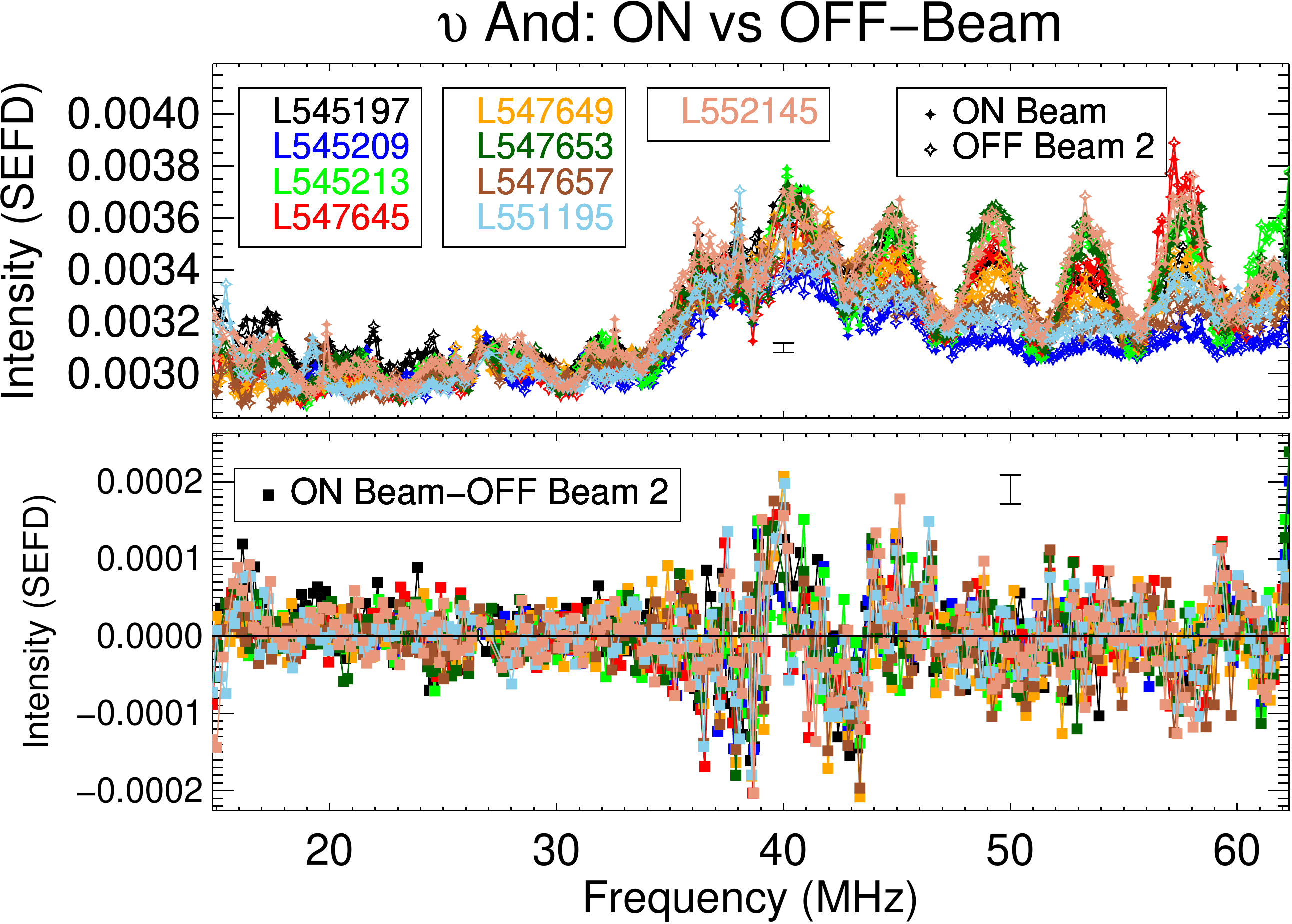} 
              \label{}
    \end{subfigure}%
    \\
     \begin{subfigure}[c]{0.45\textwidth}
        \centering
        \caption{}
     \includegraphics[page=1,width=\textwidth]{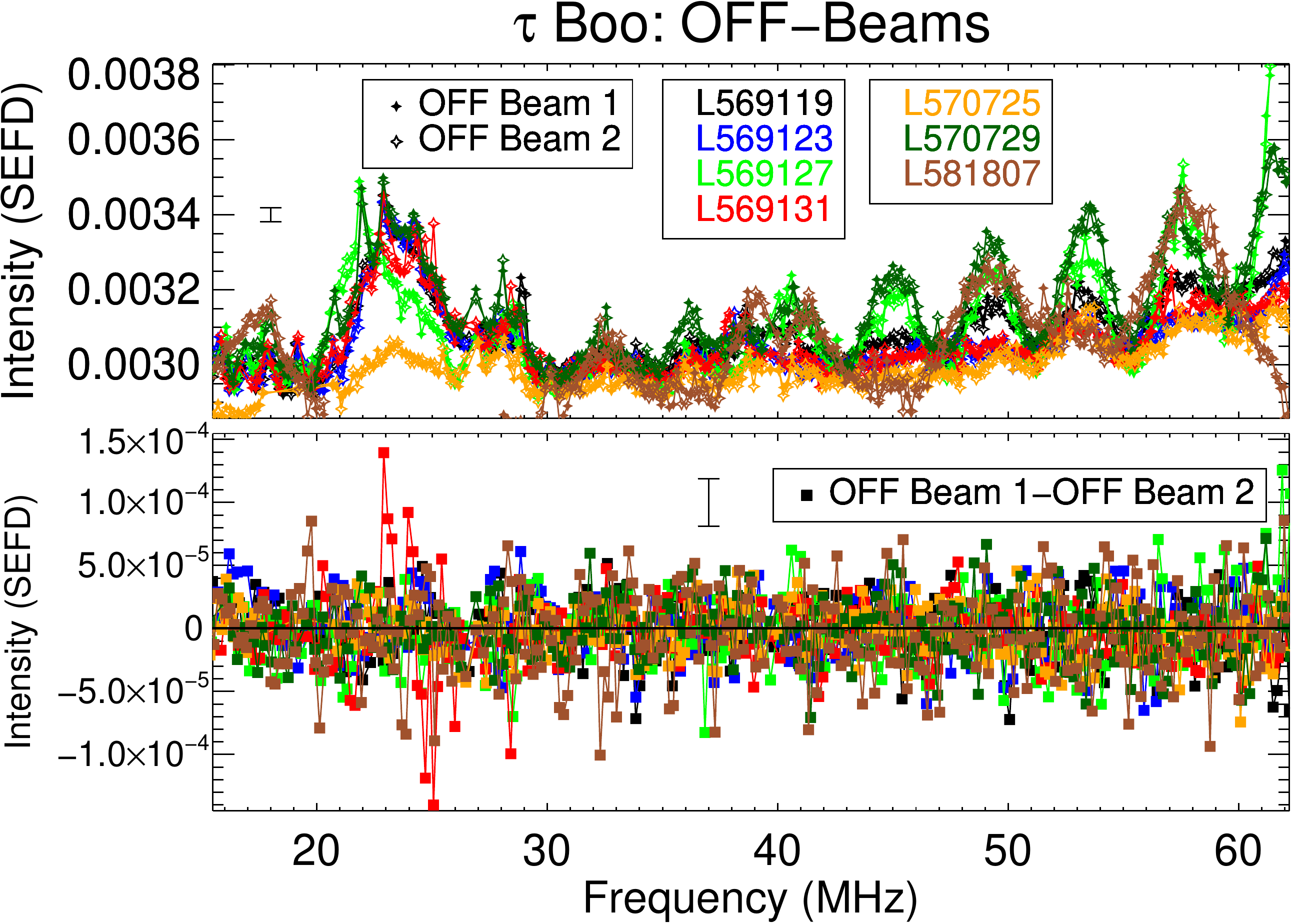} 
               \label{}
    \end{subfigure}%
    \hspace{5mm}

      \begin{subfigure}[c]{0.45\textwidth}
        \centering
        \caption{}
    \includegraphics[page=1,width=\textwidth]{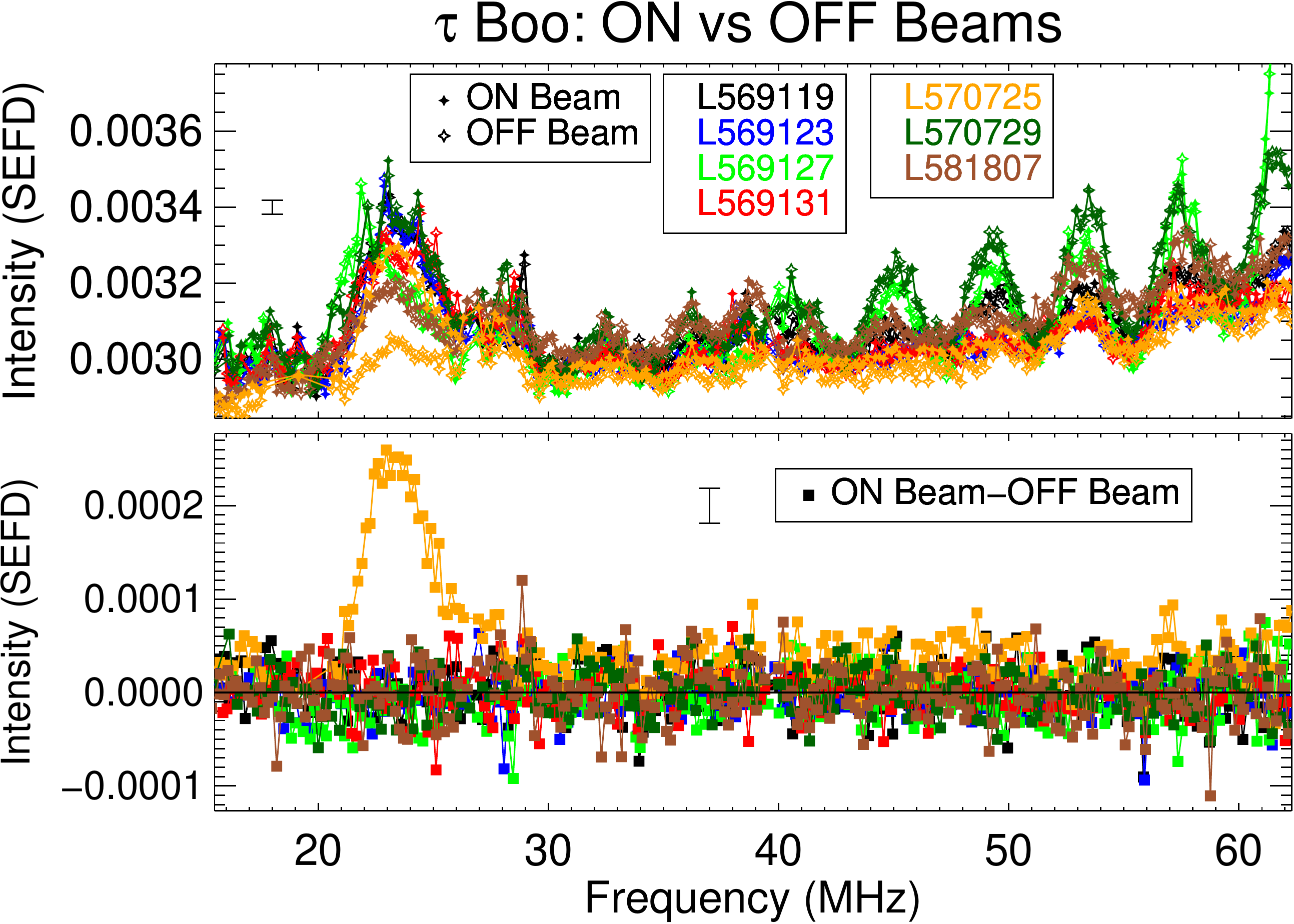}     
              \label{}
    \end{subfigure}%
    \\
     \end{tabular}
  \caption{Integrated spectrum (Q1b) for all beams in all observations for 55 Cnc (\textit{panel a and b}), $\upsilon$ And (\textit{panel c and d}), and $\tau$~Boo (\textit{panel e and f}). Large scale features are seen for all dates, however, they change between observations. In \textit{panels c-f} we see a ripple pattern that is likely due to an instrumental effect (e.g., imperfect phasing). There are no large-scale differences seen between the two OFF beams (bottom plots in \textit{panels a, c, and e}) for any date. In every individual observation, the two OFF-beams and ON-beam (expect L570725) are equivalent with each other within the pure Gaussian noise error bars ($\sigma = 1/\sqrt{b\tau}$). Dynamic spectrum differences of all OFF beams in every $\tau$~Boo observation can be found in Figure \ref{fig:Dynspec_OtherDiff}. The $\tau$~Boo ON-beam signal in observation L570725 (orange curve in \textit{panel f}) is the only large-scale difference between the ON and OFF beams that is seen (bottom plot in all panels). The 20-30 MHz features in many of the $\tau$~Boo~b dates are the same order of magnitude as the ON-beam signal in observation L570725 (\textit{panel e}). The dynamic spectrum of the ON-beam in observation L570725 subtracted by the OFF beams in the other $\tau$~Boo observations can be found in Figure \ref{fig:Dynspec_OtherDiff}.
   }
  \label{fig:Q1b_all}
\end{figure*}

\edits{Low-level systematic features in the integrated spectrum (Q1b) are seen in all beams for all the 55~Cnc, $\upsilon$ And, $\tau$~Boo observations (Figure \ref{fig:Q1b_all}). In the $\tau$~Boo (\textit{panel e}) and $\upsilon$ And (\textit{panel c}) observations we see a ripple pattern that is likely due to an instrumental effect. Since all the observations are processed identically with the \texttt{BOREALIS} pipeline it is unlikely that the data processing created the ripples since they are not the same in the two different data sets and do not do appear in the 55~Cnc observations.} 

\edits{For each individual observation, the features in the OFF1 and OFF2-beam are equivalent within the error bars as can be demonstrated by subtracting the two beams (bottom panels in the left column of Figure \ref{fig:Q1b_all}). We also find for each observation that the ON and OFF-beams are equivalent within the error bars (right column of Figure \ref{fig:Q1b_all}) with the exception of observation L570725 (Figure \ref{fig:Q1b_all}f). Therefore, any variations in the observing conditions (ionospheric and instrumental) are similar between all beams. We do observe large differences between the OFF beams of different dates
\editsjmg{showing}
that the systematics change between observations. 
\editsjmg{However,}
these systematic features do not change over time in one observation, indicating that they may be caused by either an instrumental effect, by a source that is in the side-lobes, or a combination of both effects. For all observations, the lowest intensity values found are similar ($\sim$0.003 of the SEFD or 5.1 Jy) suggesting low-level instrumental noise is always present. The features on top of that are likely caused by a combination of sources in the side lobes and non-uniform instrumental effects.}

\edits{Specifically, for every $\tau$~Boo observation except L570725 there is a bump in the OFF signal at 20-30 MHz. We find that ON-beam feature in observation L570725 has an amplitude comparable to the features seen in both OFF beams in all other $\tau$~Boo observations (Figure \ref{fig:Q1b_all}e -- f). Only for observation L570725 are the 2 OFF beams lower than every other OFF beam (Figure \ref{fig:Q1b_all}e). However, the structure in the dynamic spectrum of the observation L570725 is not the same as in the dates with the large-scale systematics (see Appendix \ref{App:dynspecdiff_tauboo}).  }

\newpage 
\onecolumn

\newpage
\clearpage
\section{Non-detection of burst emission in observation L570725 }\label{App:L570725_Nondetection}
\begin{figure*}[!thb]
\centering
  \begin{tabular}{c c}
       \begin{subfigure}[c]{0.45\textwidth}
        \centering
        \caption{}
    \includegraphics[width=0.8\textwidth,page=1]{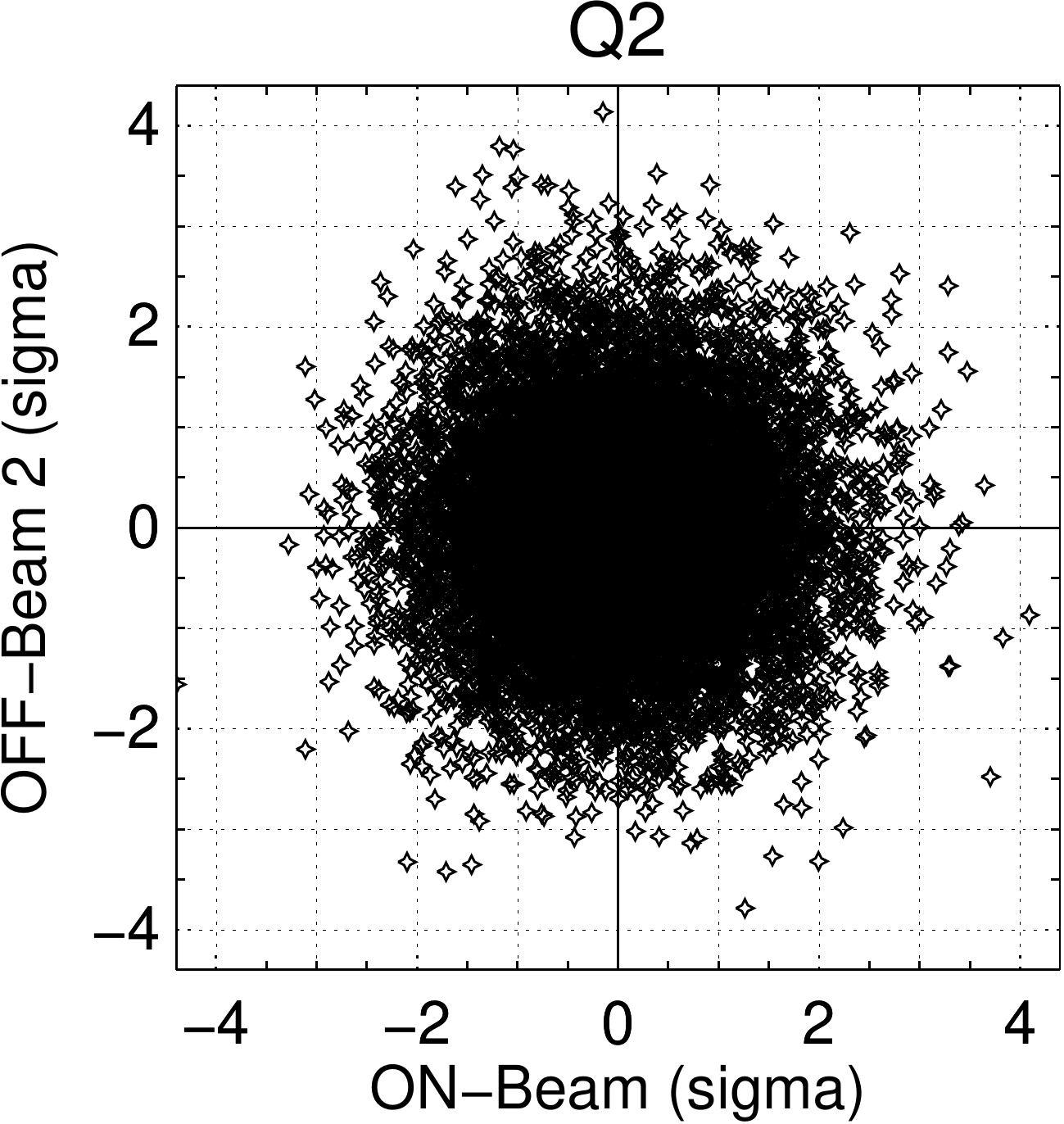} 
               \label{}
    \end{subfigure}%
    \hspace{5mm}

      \begin{subfigure}[c]{0.46\textwidth}
        \centering
        \caption{}
 \includegraphics[width=0.8\textwidth,page=2]{L570725_Nondetection.pdf}            
    \label{}
    \end{subfigure}%
    \\
    \begin{subfigure}[c]{0.45\textwidth}
        \centering
        \caption{}
       \includegraphics[width=\textwidth,page=4]{L570725_Nondetection.pdf}
                 \label{}
    \end{subfigure}%
    \hspace{5mm}

    \begin{subfigure}[c]{0.45\textwidth}
        \centering
        \caption{}
       \includegraphics[width=\textwidth,page=6]{L570725_Nondetection.pdf}       \label{}
    \end{subfigure}%
    \\
     \begin{subfigure}[c]{0.45\textwidth}
        \centering
        \caption{}
    \includegraphics[width=\textwidth,page=8]{L570725_Nondetection.pdf}
                     \label{}
    \end{subfigure}%
    \hspace{5mm}

    \begin{subfigure}[c]{0.45\textwidth}
        \centering
        \caption{}
        \includegraphics[width=\textwidth,page=10]{L570725_Nondetection.pdf} 
    \end{subfigure}%
\end{tabular}
  \caption{Q2 (\textit{panels a and b}) and beam differences for Q4a (\textit{panel c}), Q4b (\textit{panel d}), Q4e (\textit{panel e}), and Q4f (\textit{panel f}) for $\tau$~Boo in observation L570725 from the range 21-30 MHz in Stokes-V ($|V^{'}|$).
  \textit{Panel a:} Q2 for the ON-beam vs the OFF-beam 2. 
  \textit{Panel b:} Q2 for the OFF-beam 1 vs the OFF-beam 2.
  \textit{Panel c:} Q4a (number of peaks). \textit{Panel d:} Q4b (power of peaks). \textit{Panel e:} Q4e (peak offset). \textit{Panel f:} Q4f (peak offset). For \textit{panels c} to \textit{f} the black lines are the ON-beam difference with the OFF beam 2 and the red lines are the OFF beam difference. The dashed lines are statistical limits (1, 2, 3$\sigma$) of the difference between all the Q4 values derived using two different Gaussian distributions (each performed 10000 times). We do not see any excess signal in the ON-beam compared to the OFF-beams. Therefore, this observation is a non-detection for burst emission.
  We find by performing Gaussian simulations that the probability to obtain the OFF beam curve in Q4f (\textit{Panel f}) is $\sim$73$\%$, whereas the probability to randomly reproduce the ON-beam curve is $\sim$82$\%$.
}
  \label{fig:L570725_Nondetection}
\end{figure*}

\newpage 
\twocolumn
\clearpage
\section{Statistical significance of slow emission detection in L570725}\label{App:stats}
\edits{To quantify the statistical significance of}
\editsjmg{the detection of a slowly varying signal in observation L570725}, 
\edits{we use 
a method similar to the technique outlined in \citet{Turner2019} for}
\edits{the significance of the Q4f detections (see also Section \ref{sec:tauboo_L569131}).}

\editsjmg{We first calculate the observable quantity Q1a for the ON and OFF beam. From this, we obtain the difference between both, Q1a(ON)-Q1a(OFF)=Q1a$_{\text{Diff}}$. 
We then normalize Q1a$_{\text{Diff}}$ by its standard deviation and calculate the average value of the normalized curves, which we denote as $<$Q1a$_{\text{Diff}}$$>$. 
For observation L570725, we obtain $<$Q1a$_{\text{Diff}}$$>=2.7$.
In the same way, we define $<$Q1b$_{\text{Diff}}$$>$ and find $<$Q1b$_{\text{Diff}}$$>=8.2$.
}

\edits{These values are compared to those obtained in the case when both 
beams only contain random Gaussian noise. We generated a random distribution of points for the ON and OFF beams \editsjmg{(generating an artificial dynamic spectrum with the same number of points)} and calculated} 
\editsjmg{$<$Q1a$_{\text{Diff}}$$>$ and $<$Q1b$_{\text{Diff}}$$>$.}
\editsjmg{We generated over $10^6$ instances of this artificial data set. In none of these simulated cases, the value of $<$Q1a$_{\text{Diff}}$$>$ and $<$Q1b$_{\text{Diff}}$$>$ reached the values obtained in observation L570725.}
\editsjmg{By interpolating the peak values obtained for $<$Q1a$_{\text{Diff}}$$>$ and $<$Q1b$_{\text{Diff}}$$>$ over a set of $N$ simulations (with $N$ between $10^5$ and $10^6$),}
\edits{we find that the probability 
\editsjmg{to randomly obtain a signal as strong as the observed one} 
is 2.1$\times10^{-12}$ for Q1a$_{\text{Diff}}$ and 1$\times10^{-18}$ for Q1b$_{\text{Diff}}$. This false positive rate corresponds to a 
\editsjmg{statistical significance}
of 6.9$\sigma$ and 8.6$\sigma$, respectively.}

\edits{As a final step, we compare this value to those obtained \editsjmg{for} the OFF beams. In that case, we find that the false positive rate is $\sim$90$\%$ for Q1a(OFF1)-Q1a(OFF2) and $\sim$100$\%$ for Q1b(OFF1)-Q1b(OFF2). Therefore, the OFF beams difference clearly corresponds to a non-detection.}

\edits{We note that this procedure assumes all points in the Q1a and Q1b curves are uncorrelated.}
\editsjmg{In the case of instrumental effects, sources in the side lobe or other systematic errors this assumption does not hold.}

\section{Pulsar B0809+74 observation L570723}\label{App:pulsar}

\subsection{\editsjmg{Replica signal in OFF-beam}}
\begin{figure}[!tbh]
    \centering
    \begin{tabular}{c}
    \includegraphics[width=0.45\textwidth]{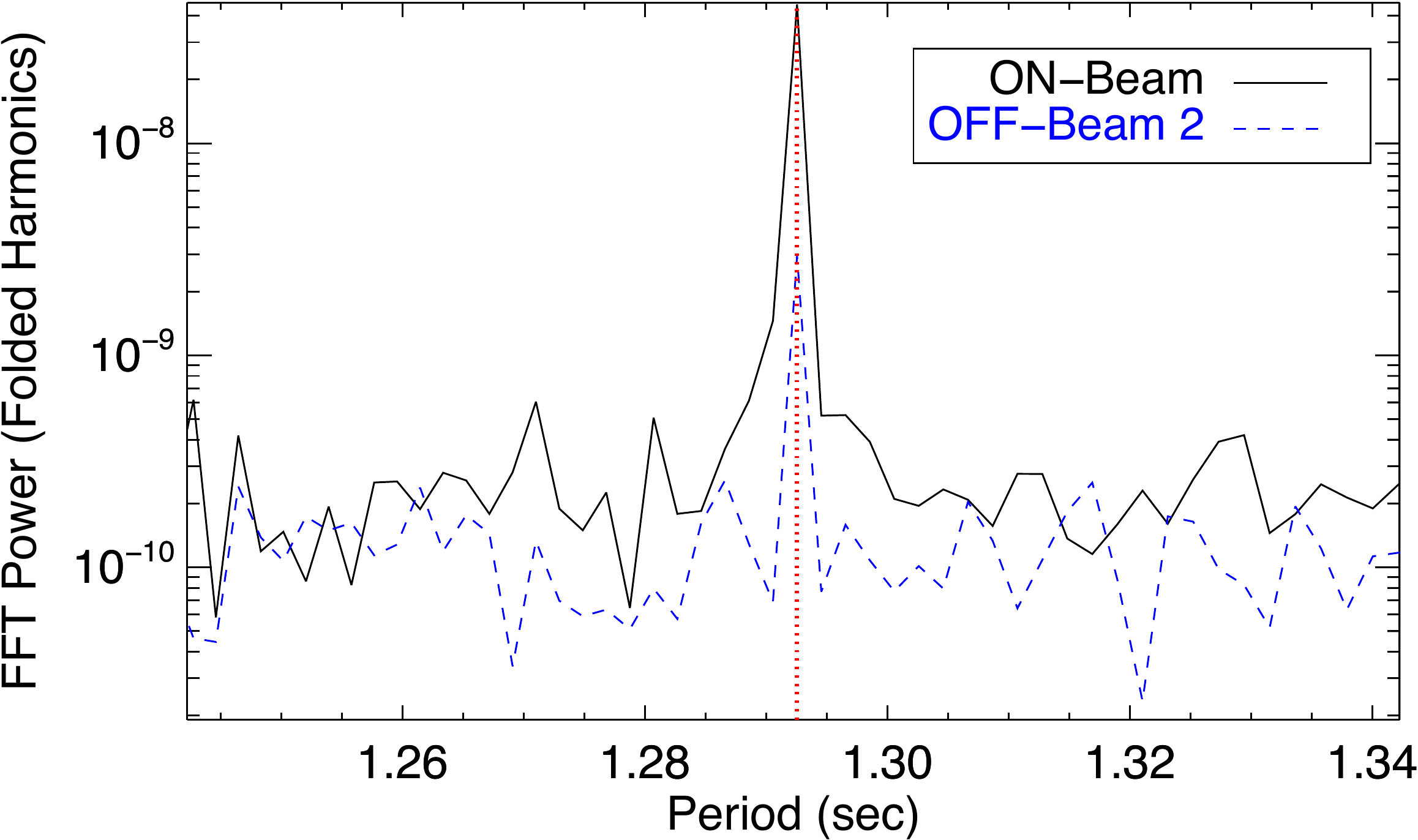} 
      \end{tabular}
    \caption{\editsjmg{FFT of the pulsar B0809+74 (observation L570723) for the ON-beam (black-line) and OFF-beam 2 (dashed blue line). The known period of the pulsar is marked as a dashed red line. 
    } 
    } 
    \label{fig:FFT_pulsar}
\end{figure}

\begin{figure}[!tb]
    \centering
    \includegraphics[width=0.48\textwidth]{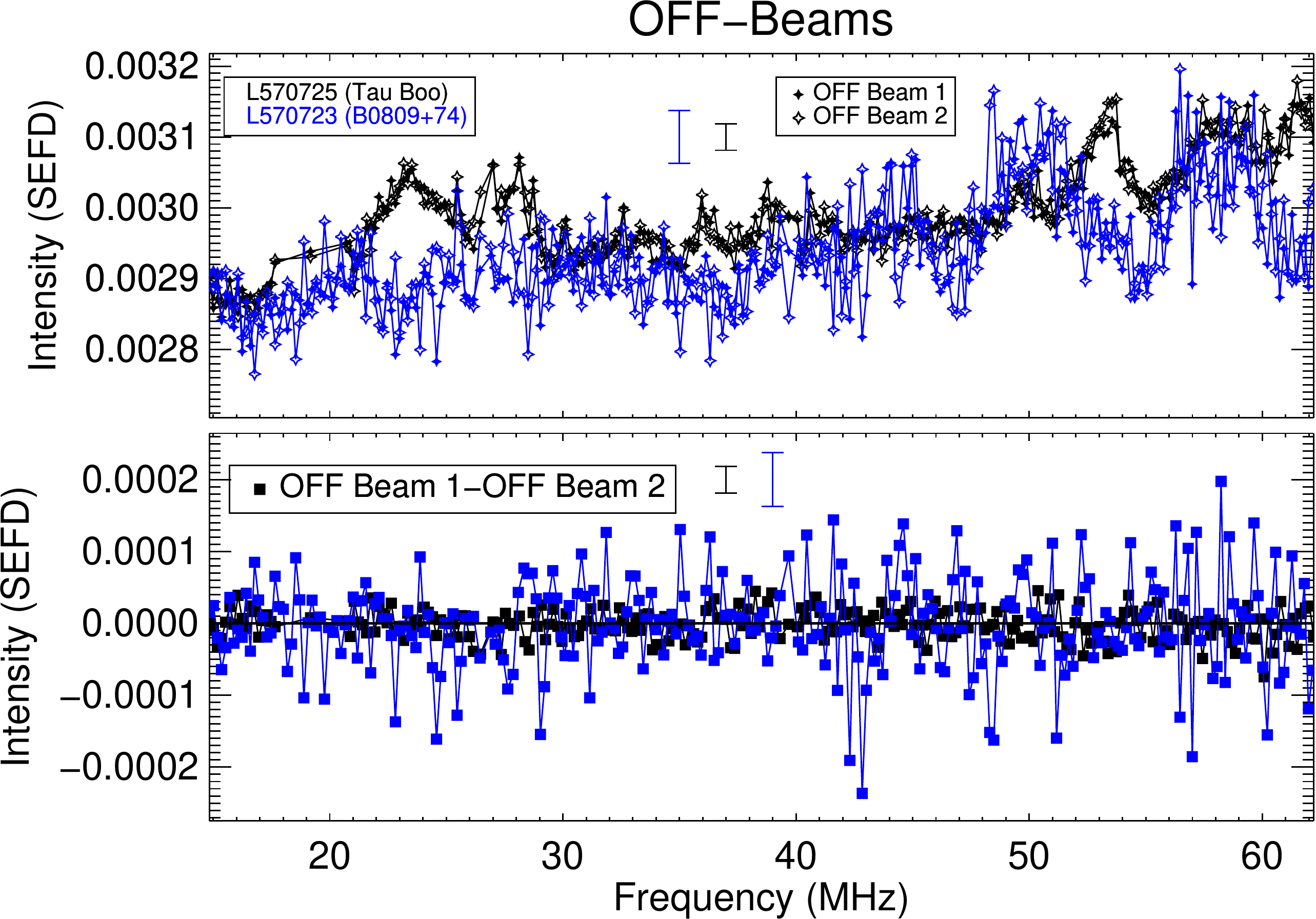}
      
    \caption{Integrated spectrum (Q1b) for the OFF beams on $\tau$~Boo (observation L570725) and the pulsar B0809+74 (observation L570723, taken 15 minutes before observation L570725). For observation L570725 only the first 15 minutes of data were used to calculate Q1b to allow for a more consistent comparison with the pular observation. The two observations have similar features in Q1b and above 30 MHz are consistent with each other assuming Gaussian error bars. \edits{We performed a K-S test on the two curves above 30 MHz and find that the probability that the two curves are drawn from the same parent distribution (the null hypothesis) is 97$\%$.} The fact that the two beams are pointed at completely different parts of the sky and still have similar overall flux levels suggests that the majority of the signal in the beams is instrumental. \edits{The 20-30 MHz feature in the OFF beam of $\tau$~Boo observation L570725 is the faint signal replicated from the ON-beam (Section \ref{sec:L570725}; Figure \ref{fig:Dynspec_Detection})}.
    } 
    \label{fig:OFFbeams_pulsar}
\end{figure}

\editsjmg{The observation of the pulsar B0809+74}
can be used to independently study the possible systematics (e.g imperfect phasing and low-level noise) affecting the $\tau$~Boo observation L570725 (Section \ref{sec:L570725}). The pulsar observation L570723 was taken 15 minutes before observation L570725. In order to detect B0809+74, 
\editsjmg{we use the same procedure as in} \citet{Turner2017pre8}. 
The FFT was performed on the data after running it through the \texttt{BOREALIS} pipeline and de-dispersing the observations at the known dispersion measure. To get the final power spectrum in the FFT, the 6 first harmonics were folded together. The FFT was computed from the range 30-55 MHz. The FFT of the ON-beam, OFF-beam 1, and OFF-beam 2 \editsjmg{are shown} in Figure \ref{fig:FFT_pulsar}.

\editsjmg{We find that the pulsar is detected with a signal–to–noise ratio (SN$_\text{FFT}$) of $\sim$628 in the ON-beam and with SN$_\text{FFT}$ of $\sim$23 and $\sim$57 in the OFF beams, respectively.} \editsjmg{This result confirms that a replica signal of the ON-beam can appear in the OFF beam, as
was also found in the $\tau$~Boo observation L570725 (Section \ref{sec:L570725}; Figure \ref{fig:Dynspec_Detection}).}

\subsection{\editsjmg{Comparison of integrated spectra (Q1b)}}

\editsjmg{We also compare the integrated spectra (observable quantity Q1b) of
one of the observations}
\edits{of the pulsar B0809+74
\editsjmg{to those of the $\tau$ Boo observation.}
\editsjmg{Figure \ref{fig:OFFbeams_pulsar} shows Q1b for both OFF beams for the two observations
(B0809+74 observation L570723 and $\tau$ Boo observation L570725).} We only used the first 15 minutes of data from the $\tau$ Boo L570725 observation to derive Q1b to allow for a more consistent comparison. 
\editsjmg{Above 30 MHz, the two observations} are consistent with each other within error bars. Additionally, we performed a K-S statistical test on the two curves above 30 MHz \edits{and find that the probability that the two curves are drawn from the same parent distribution (the null hypothesis) is 97$\%$.} 
\editsjmg{In the frequency range 20-30 MHz, the $\tau$ Boo observation shows large-scale features in both OFF beams; this feature is absent in the pulsar observation.}
Therefore, the likely source of the 21-30 MHz feature in the ON-beam of the $\tau$~Boo observation L570725  (Figure \ref{fig:L570725_detection}b) is \editsjmg{either} excess flux in the beam or an unknown time-dependent instrumental effect.}

\section{Station inspection plots}\label{App:staion_inspection_plots}

\begin{figure*}[!t]
\centering
    \includegraphics[width=\textwidth]{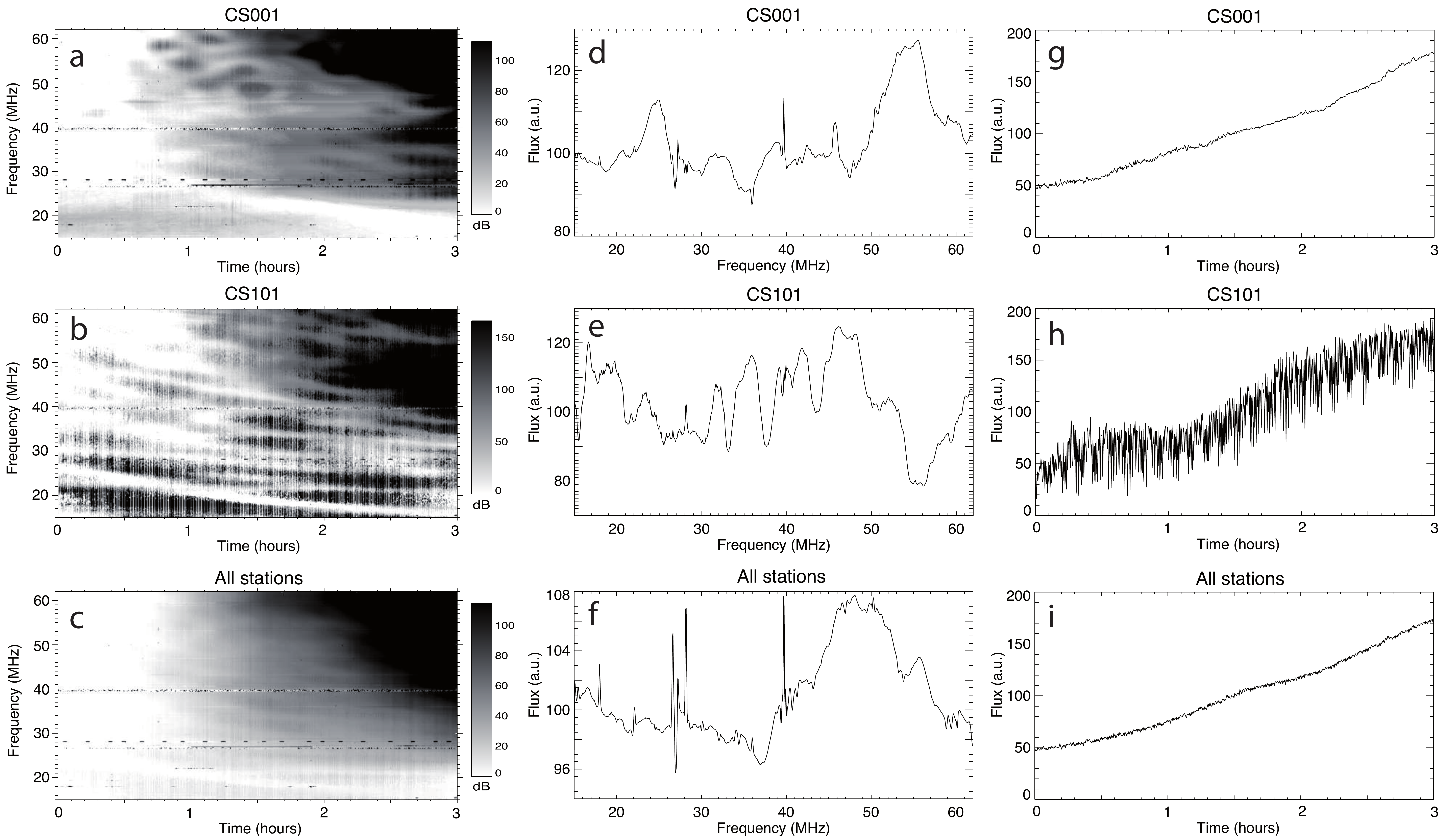}  
  \caption{\editsjmg{
  Analysis of the station inspection plots for observation L570725.
  Top row: Good station CS001 (\textit{panels a, d, g}). 
  Middle row: Bad station CS101 (\textit{panels b, e, h}).
  Bottom row: Incoherent (not phased) sum of all stations.
  Left column (panels a, b, c): Station inspection plots. 
  Each plot has been normalized by an average background. Large scale features are seen foe the bad station CS101.
  Middle column (panels d, e, f): Integrated spectra (Q1b) derived from the station inspection plots.
  Right column (panels g, h, i): Time-series (Q1a) derived from the station inspection plots.
    No large bumps are seen within 20-30 MHz for the bad station
    (CS101), therefore they do not seem to be cause of the L570725 ON-beam signal seen in Figure \ref{fig:L570725_detection}.}
   }
  \label{fig:stations_new}
\end{figure*}

During each observation, station summary plots are produced by ASTRON 
\editsjmg{to monitor telescope stability and data quality.}
We examined these station inspection plots to search for a possible non-planetary cause of the signal seen in the ON-beam of observation L570725. We discovered that in each of our LOFAR observations at least one station was misbehaving (summarised in Table \ref{tb:obs}). 

In \editsjmg{Figure \ref{fig:stations_new}}, we show 
\editsjmg{a few examples for}
station inspection plots (left-hand panels) and some derived quantities
\editsjmg{for the $\tau$~Boo observation L570725.}
\editsjmg{We compare a good station (CS001), a bad station (CS101), and the incoherent sum of all stations.}

To directly compare to the observed signal, we digitized the dynamic spectra 
and calculated the 
\editsjmg{integrated spectrum (Q1b, middle panels) and time-series (Q1a, right-hand panels)} for 
the stations mentioned above.
We do not see any large-scale bumps in the range 20-30 MHz in Q1b in any of the good or bad stations and also when we combine all 24 stations together (this is an incoherent sum since the phases were not taken into account). Our findings suggest that the spurious behavior seen in the stations plots are not likely the origin of the detected signal \editsjmg{discussed in Section \ref{sec:L570725}}.

\section{Dynamic spectrum differences of the $\tau$~Boo observations}\label{App:dynspecdiff_tauboo}
\begin{figure*}[!htb]
\centering
\begin{tabular}{cc}
    \vspace{-1em}
  \begin{subfigure}[c]{0.45\textwidth}
        \centering
        \caption{}
   \includegraphics[page=2,width=\textwidth]{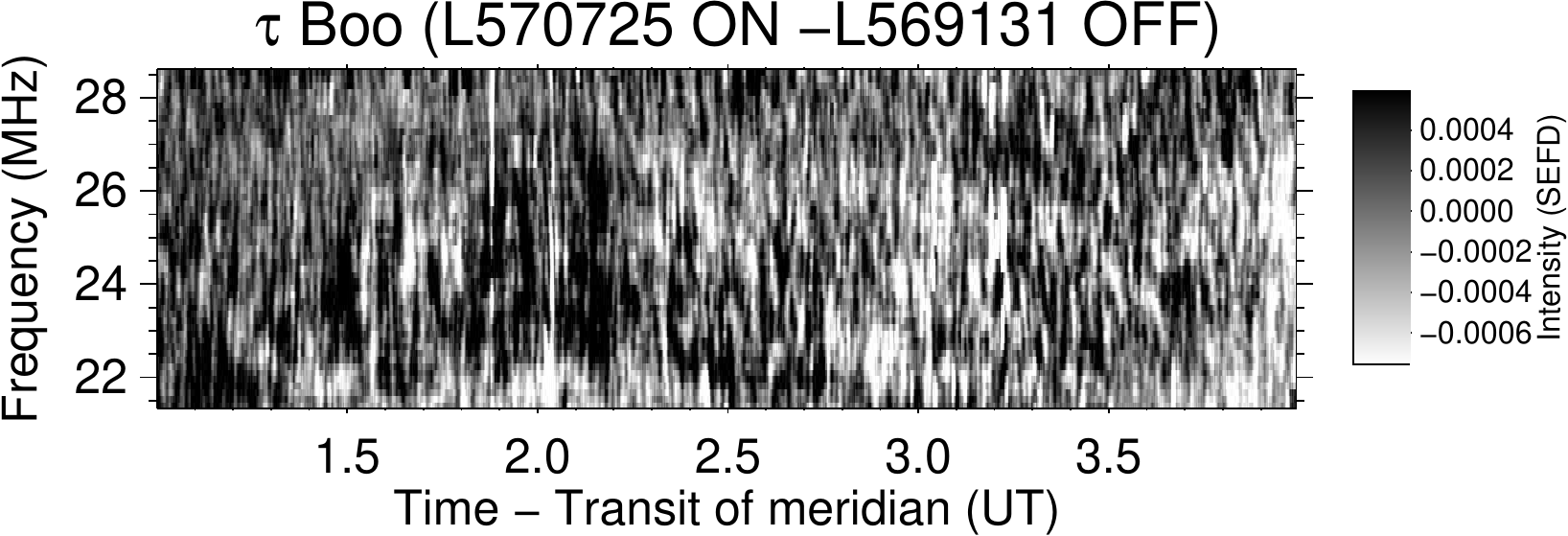} 
                  \label{}
    \end{subfigure}%
    \hspace{5mm}

        \begin{subfigure}[c]{0.45\textwidth}
        \centering
        \caption{}
         \includegraphics[page=3,width=\textwidth]{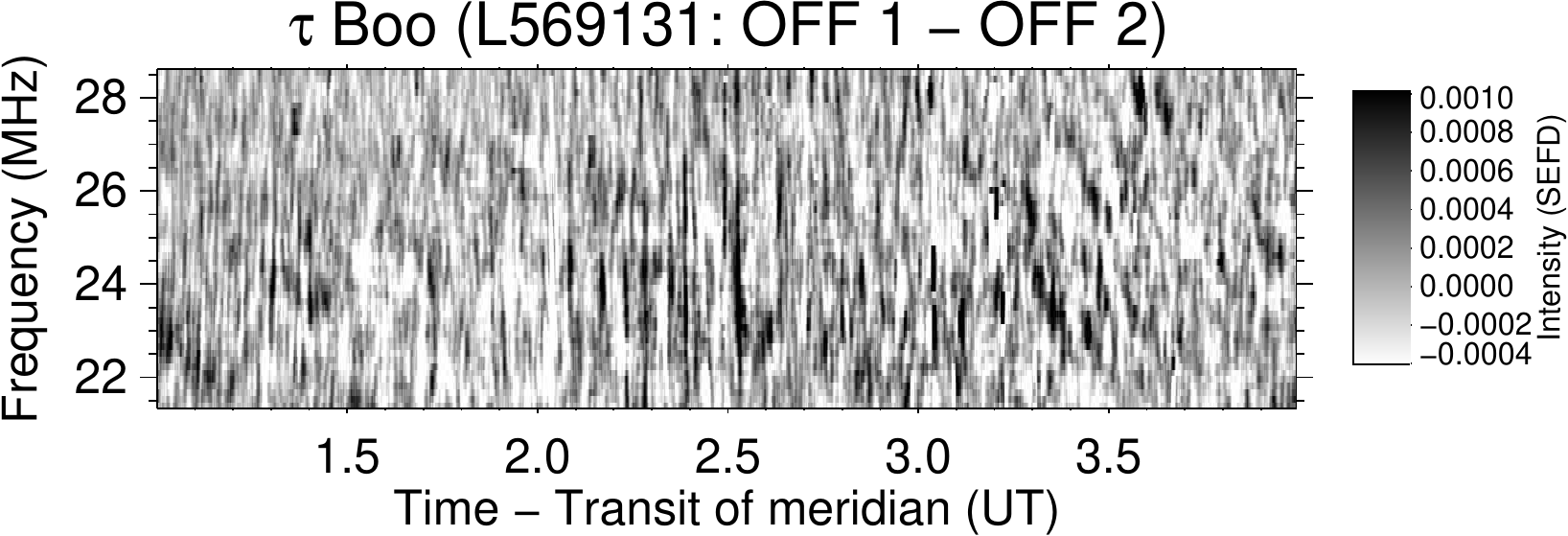}              
         \label{}
     \end{subfigure}%
    \\
    \vspace{-1em}
  \begin{subfigure}[c]{0.45\textwidth}
        \centering
        \caption{}
   	\includegraphics[page=2,width=\textwidth]{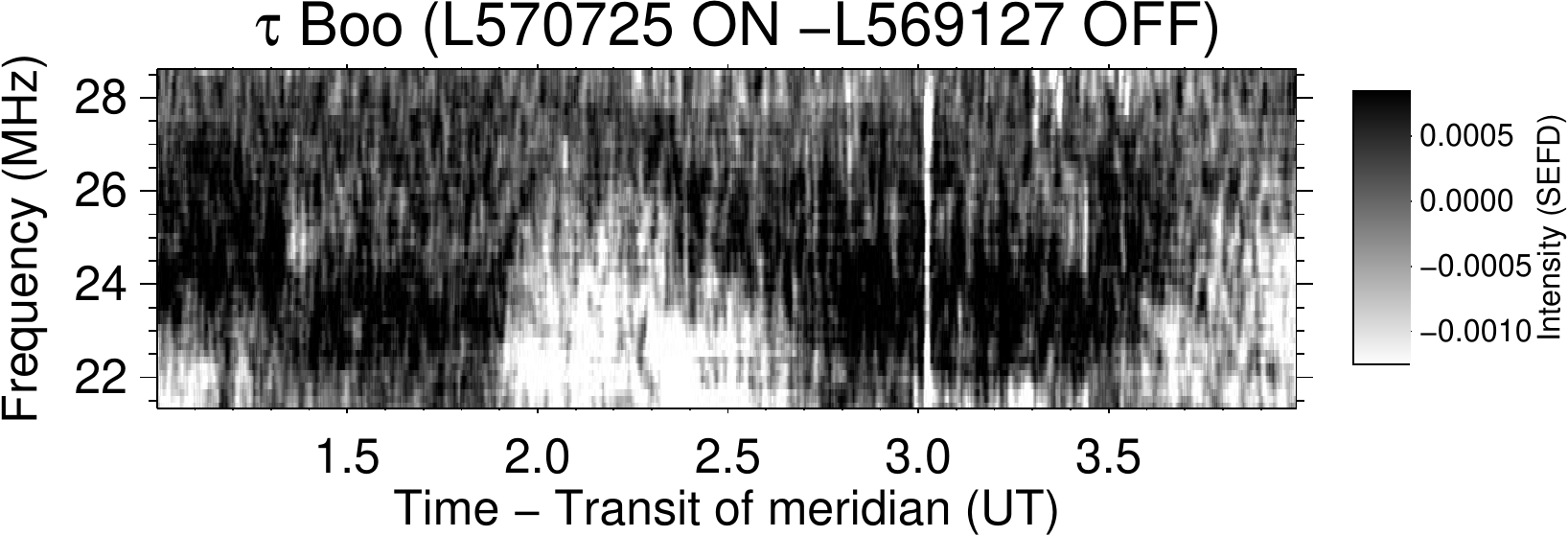} 
                  \label{}
    \end{subfigure}%
    \hspace{5mm}

        \begin{subfigure}[c]{0.45\textwidth}
        \centering
        \caption{}
  	  \includegraphics[page=3,width=\textwidth]{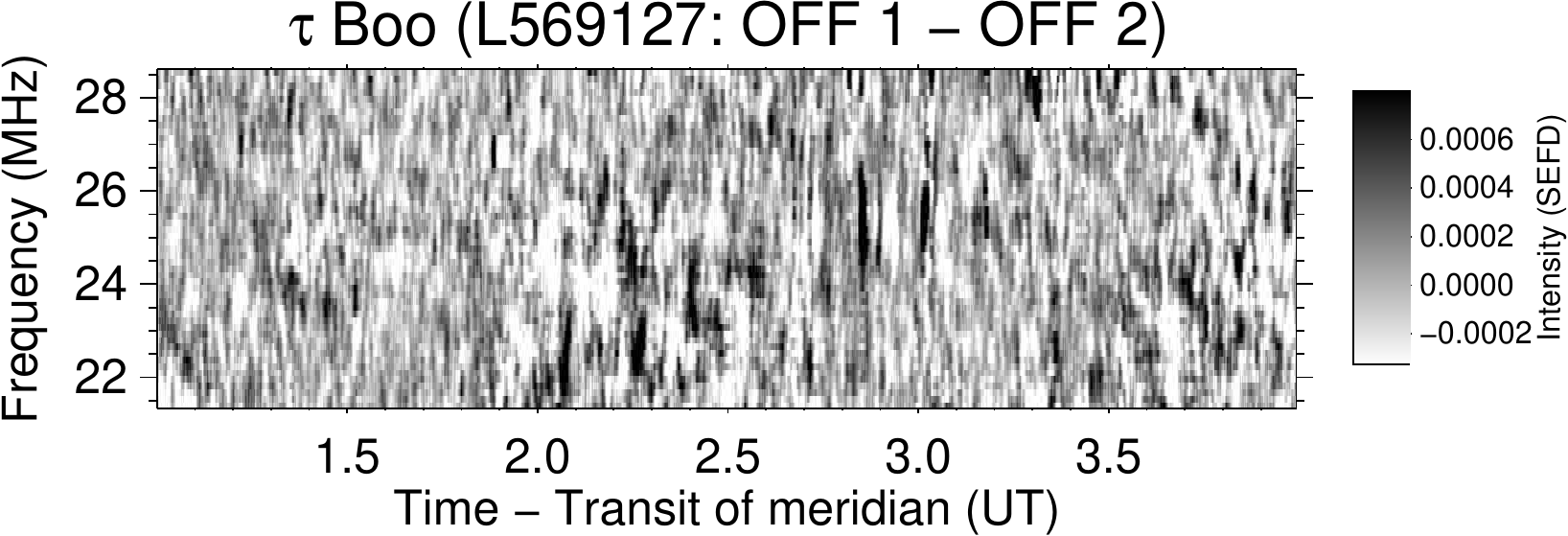}
     \end{subfigure}%
    \\
        \vspace{-1em}
   \begin{subfigure}[c]{0.45\textwidth}
        \centering
        \caption{}
   	 \includegraphics[page=2,width=\textwidth]{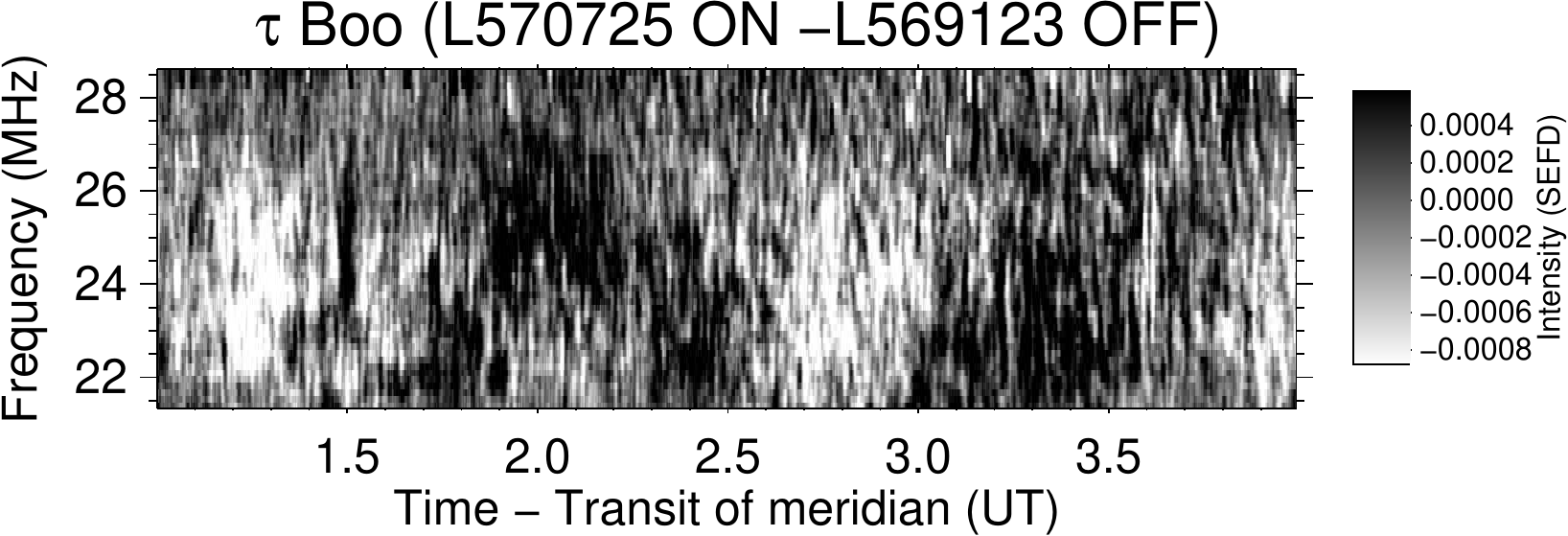}
                  \label{}
    \end{subfigure}%
    \hspace{5mm}

        \begin{subfigure}[c]{0.45\textwidth}
        \centering
        \caption{}
  	  \includegraphics[page=3,width=\textwidth]{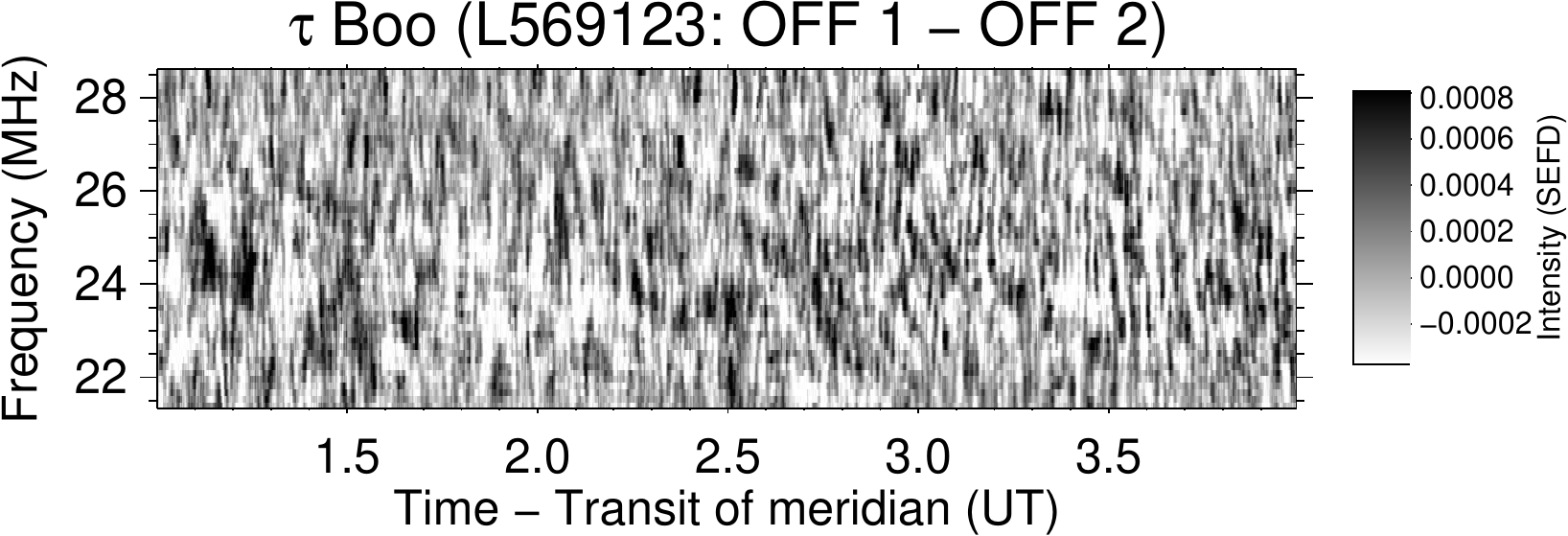}  
     \end{subfigure}%
    \\
        \vspace{-1em}
     \begin{subfigure}[c]{0.45\textwidth}
        \centering
        \caption{}
   \includegraphics[page=2,width=\textwidth]{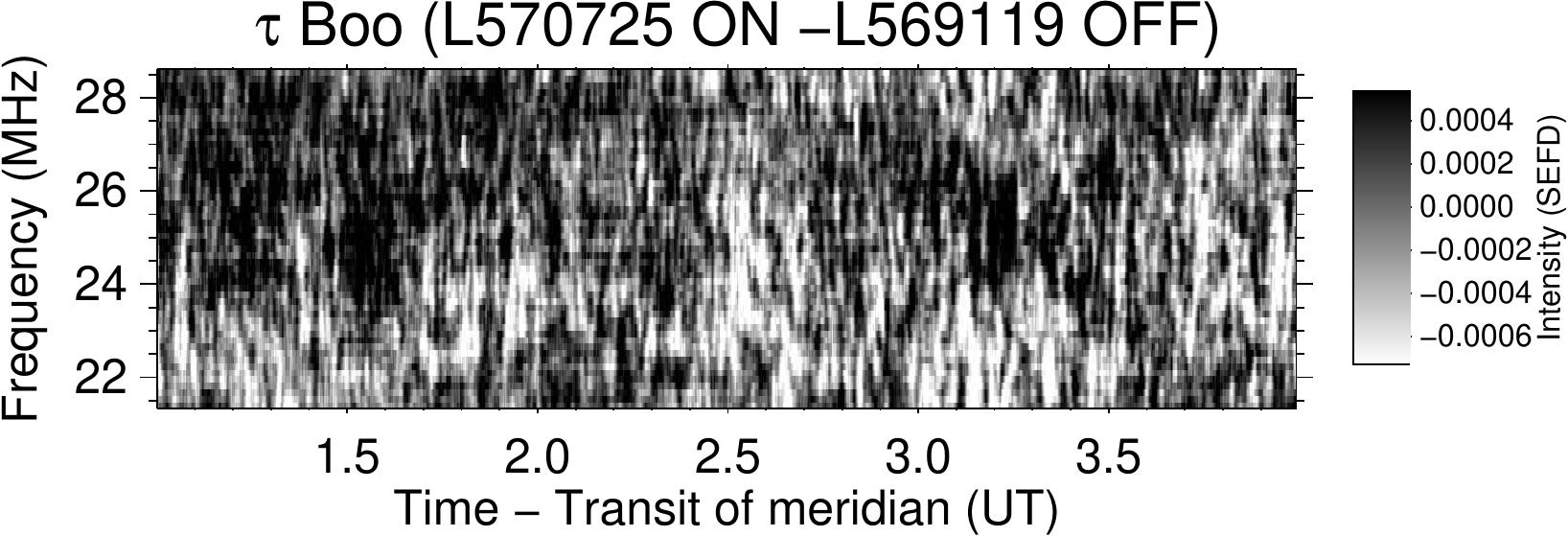}
                  \label{}
    \end{subfigure}%
    \hspace{5mm}

        \begin{subfigure}[c]{0.45\textwidth}
       	 \centering
       	 \caption{}
  	 \includegraphics[page=3,width=\textwidth]{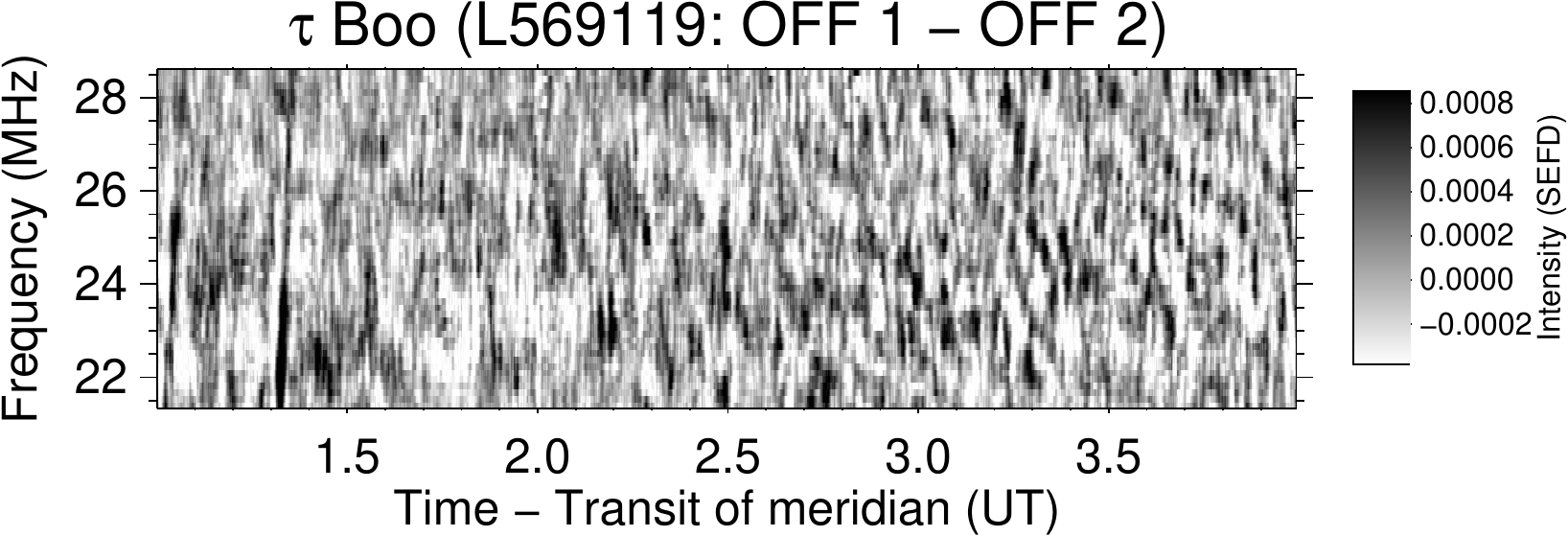}
    	 \end{subfigure}%
    \\
        \vspace{-1em}
         \begin{subfigure}[c]{0.45\textwidth}
        \centering
        \caption{}
    \includegraphics[page=2,width=\textwidth]{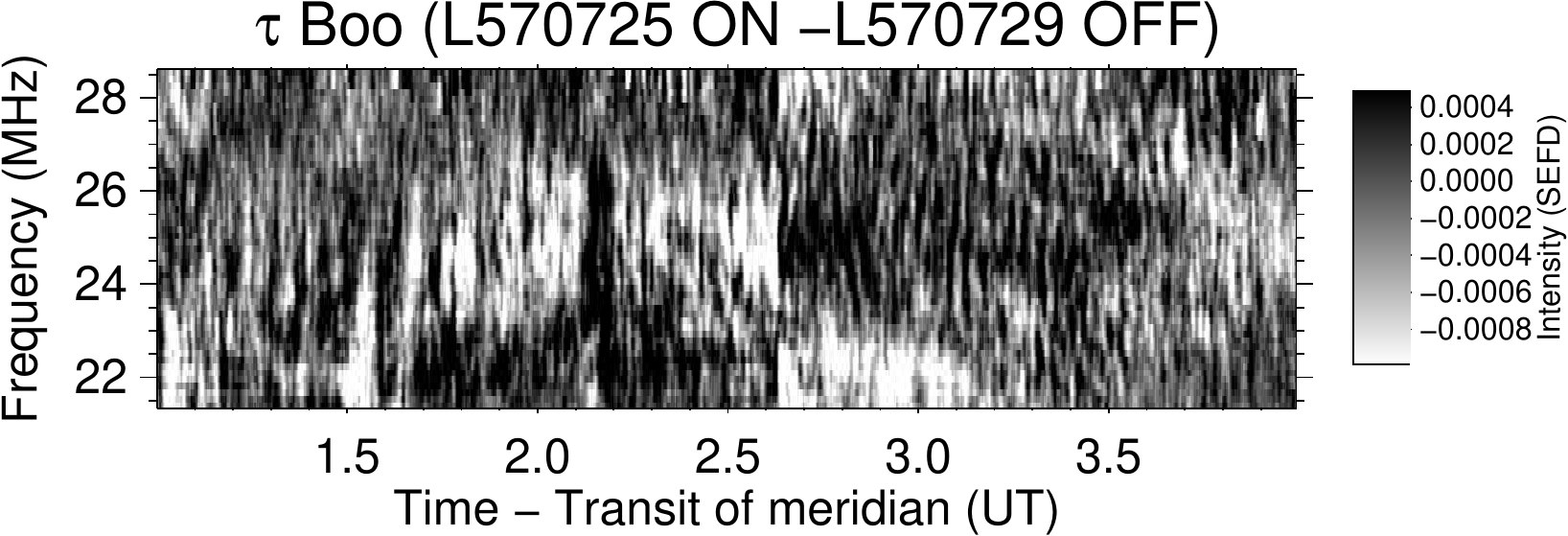}
                  \label{}
    \end{subfigure}%
    \hspace{5mm}

        \begin{subfigure}[c]{0.45\textwidth}
       	 \centering
       	 \caption{}
      \includegraphics[page=3,width=\textwidth]{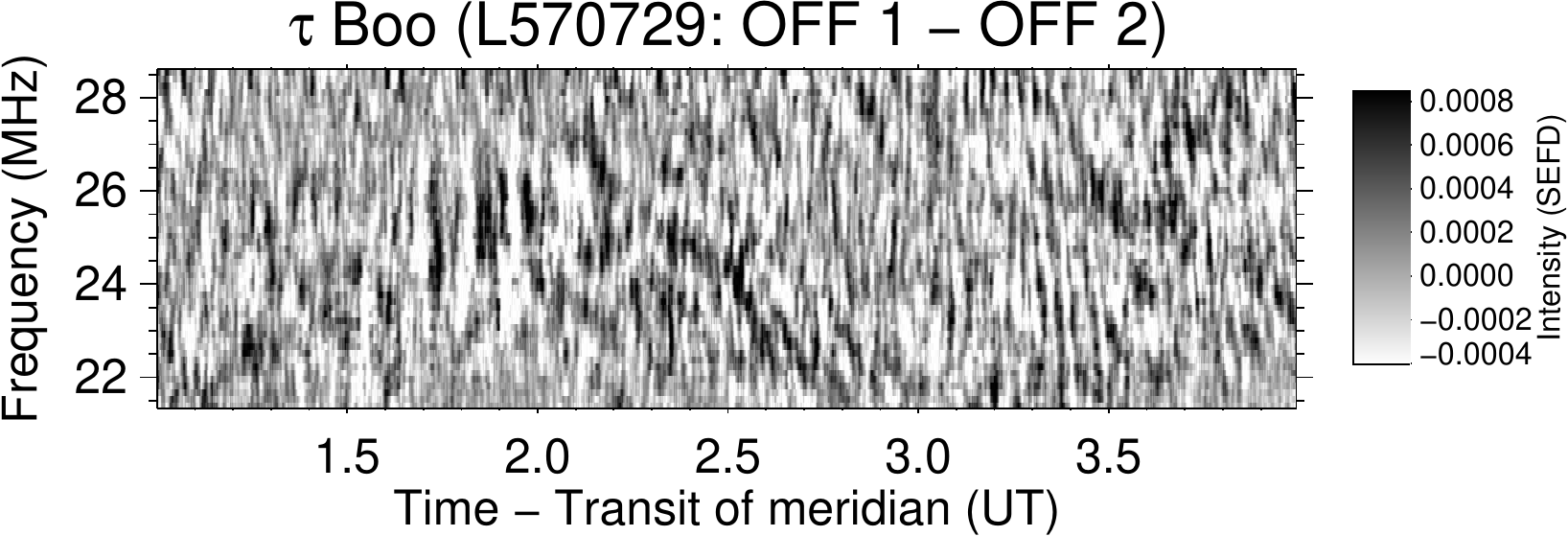}
    	 \end{subfigure}%
    \\
        \vspace{-1em}
     \begin{subfigure}[c]{0.45\textwidth}
        \centering
        \caption{}
  	 \includegraphics[page=2,width=\textwidth]{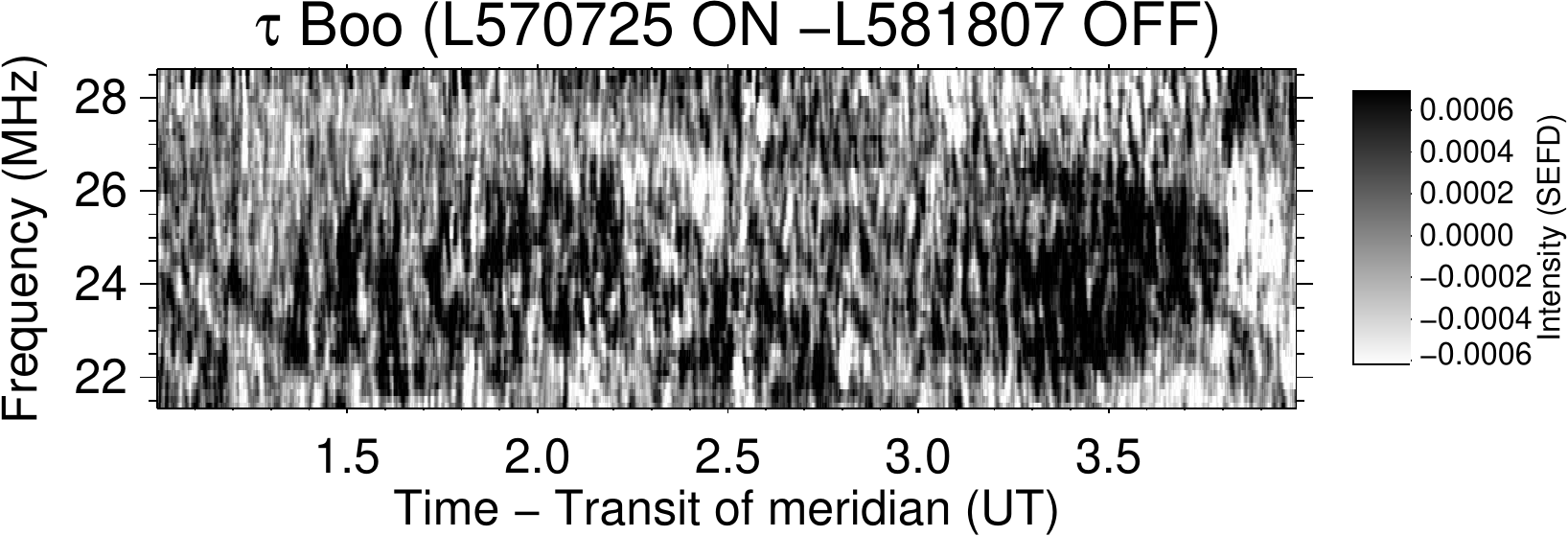}
                  \label{}
    \end{subfigure}%
    \hspace{5mm}

        \begin{subfigure}[c]{0.45\textwidth}
       	 \centering
       	 \caption{}
  	 \includegraphics[page=3,width=\textwidth]{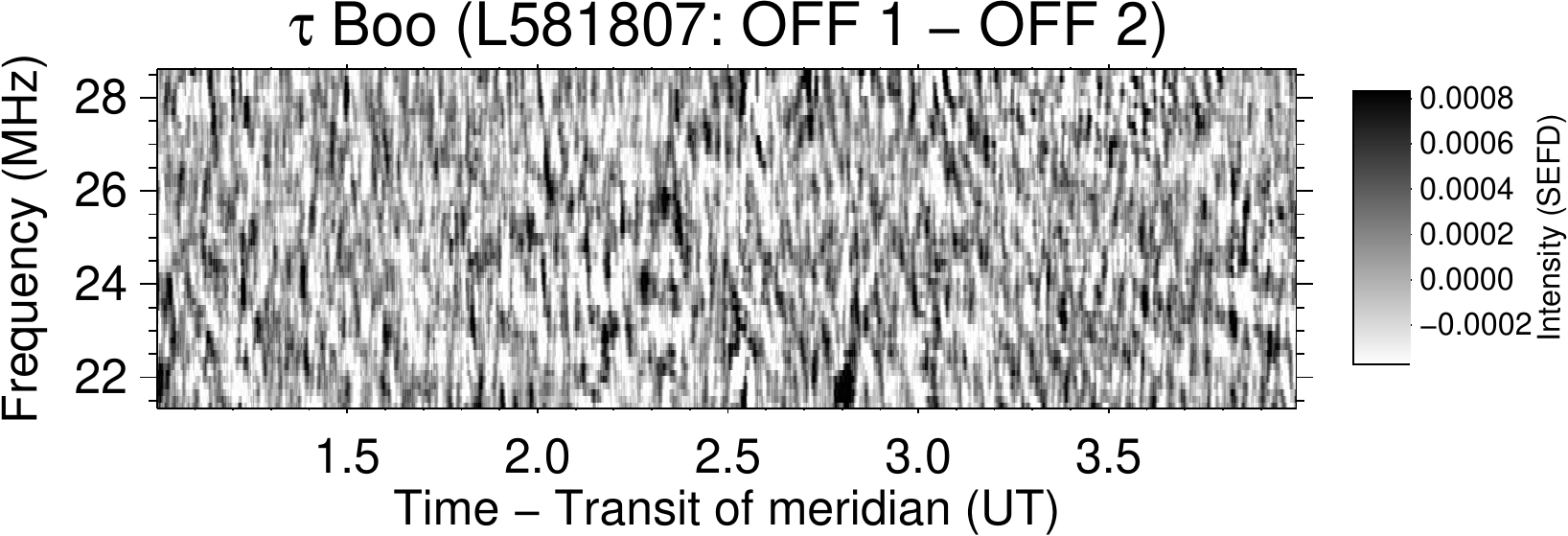}
    	 \end{subfigure}%
 \end{tabular}
  \caption{Dynamic spectrum differences of $\tau$~Boo in Stokes-V ($|V^{'}|$) in the ON-beam in L570725 with the OFF beams in L569131 (\textit{panel a}), L569127 (\textit{panel c}), L569123 (\textit{panel e}), L569119 (\textit{panel g}), L570729 (\textit{panel i}), and L581807 (\textit{panel k}). Dynamic spectrum differences of the two OFF beams for each date can be found in the right column (\textit{panels b, d, f, h, j, and l}). The data from the range 21-29 MHz is shown. The x-axis for all plots is the UT time of the observation subtracted by the transit time of the meridian. Moving to this reference frame ensures that even though the observations were taken at different dates, the characteristics of the beams (elevation, main beam pattern, and side lobes) at each time step are exactly the same. Thus, ionospheric variations, differential instrumental effects, and any remaining low-level RFI are the only difference between beams. The structured differences between the ON and OFF beams suggest that the signal in L570725 does not have the same time-frequency characteristics as the features seen in OFF beams of all the other dates. We do not see any obvious structure in the OFF beam difference plots. \editsjmg{This} suggests that for each observation the OFF beams are similar in their time-frequency emission structures. We find a consistent conclusion with what we find when comparing Q1b (integrated spectrum) for the two OFF beams for each individual observation (Figure \ref{fig:Q1b_all}).  
  }
  \label{fig:Dynspec_OtherDiff}
\end{figure*}

\edits{We compare the dynamic spectrum of the ON-beam of \editsjmg{observation} L570725 to the OFF beams of all the other $\tau$~Boo observations to determine the structure of the emission. Figure \ref{fig:Dynspec_OtherDiff} shows the subtraction of the ON-beam of L570725 by the \editsjmg{different} OFF beams. For all panels in this figure, the x-axis is the observation UT time subtracted by the transit time of the meridian of $\tau$~Boo. Moving to this reference frame is important because it ensures that observations taken at different dates will have the same beam characteristics (elevation, main beam pattern, and side lobes) at each time step in the new frame. The only differences between beams should be ionospheric variations, differences in instrumental systematics, and any remaining low-level RFI. In most cases, a structured feature persists. Therefore, the ON-beam signal in observation L570725 is inherently different than all OFF beams for every $\tau$~Boo observation. Hence, even though the integrated spectra of the different dates may look similar (Appendix \ref{App:Q1ball}, Figure \ref{fig:Q1b_all}), their actual emission (and thus their emission sources) is not the same. In the right column of Figure \ref{fig:Dynspec_OtherDiff}, we also show the difference of the OFF beam dynamic spectrum for each date. As with the integrated spectra (Figure \ref{fig:Q1b_all}), the dynamic spectrum for the two OFF beams in each individual date are very similar and only random noise remains. }

\newpage 
\clearpage
\onecolumn
\section{$\upsilon$~Andromedae marginal signal}\label{App:UpsAndr}

\begin{figure*}[!htb]
\centering
 \begin{tabular}{cc}
    \begin{subfigure}[c]{0.45\textwidth}
        \centering
        \caption{}
       \includegraphics[width=0.8\textwidth,page=1]{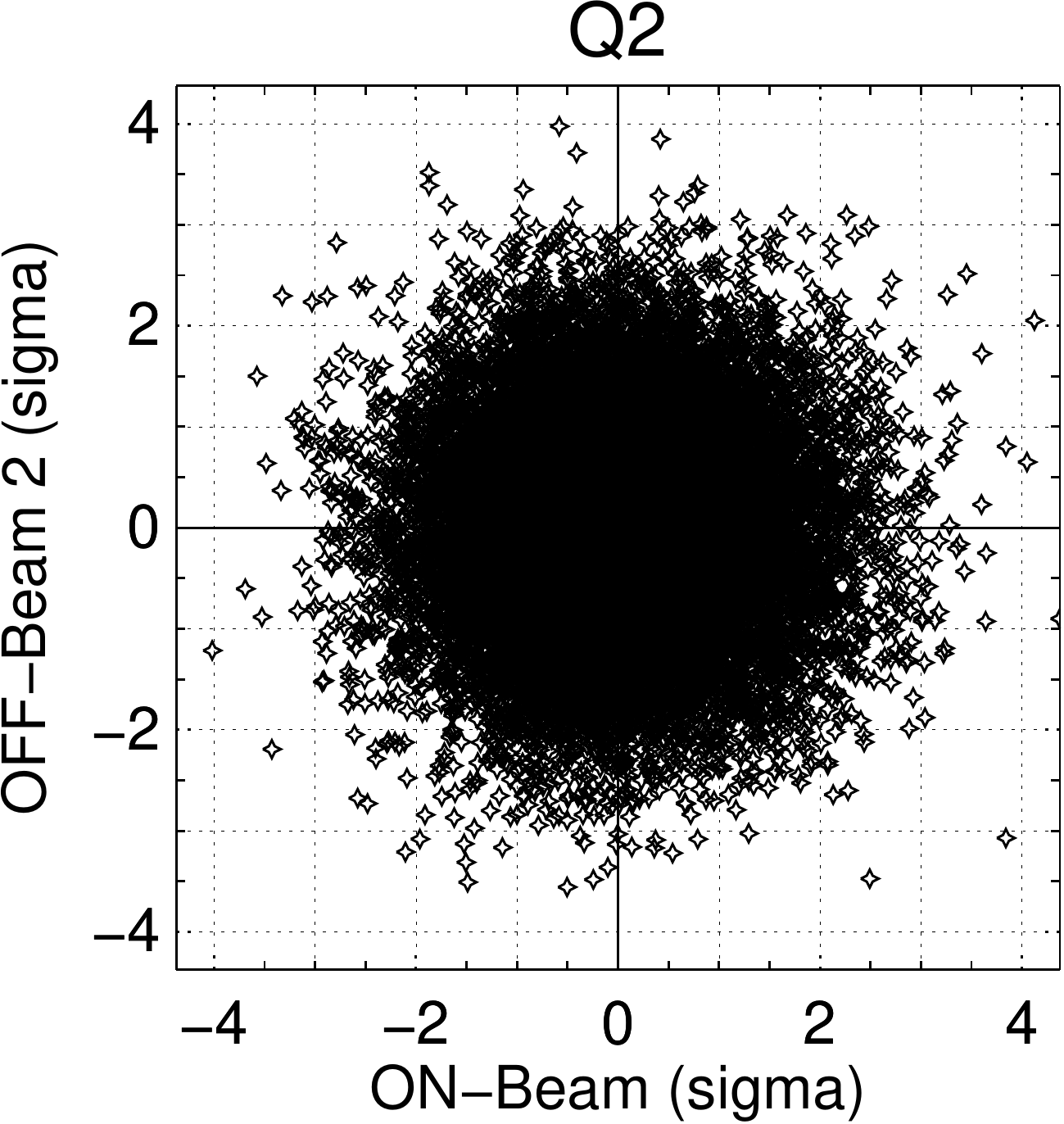}           
              \label{}
    \end{subfigure}%
    \hspace{5mm}

        \begin{subfigure}[c]{0.45\textwidth}
       	 \centering
       	 \caption{}
 		 \includegraphics[width=0.8\textwidth,page=2]{L545197_detection.pdf} 
      	 \end{subfigure}%
    \\
    \begin{subfigure}[c]{0.45\textwidth}
        \centering
        \caption{}
      \includegraphics[width=\textwidth,page=4]{L545197_detection.pdf} 
               \label{}
    \end{subfigure}%
    \hspace{5mm}

        \begin{subfigure}[c]{0.45\textwidth}
       	 \centering
       	 \caption{}
	  \includegraphics[width=\textwidth,page=6]{L545197_detection.pdf}
        	 \end{subfigure}%
    \\
    \begin{subfigure}[c]{0.45\textwidth}
        \centering
        \caption{}
 	   \includegraphics[width=\textwidth,page=8]{L545197_detection.pdf} 
               \label{}
    \end{subfigure}%
    \hspace{5mm}

        \begin{subfigure}[c]{0.45\textwidth}
       	 \centering
       	 \caption{}
  		  \includegraphics[width=\textwidth,page=10]{L545197_detection.pdf} 
        	 \end{subfigure}%
   \end{tabular}
  \caption{
  Q2 (\textit{panels a and b}) and beam differences for Q4a (\textit{panel c}), Q4b (\textit{panel d}), Q4e (\textit{panel e}), and Q4f (\textit{panel f}) for $\upsilon$~And in observation L545197 in the range 14-38 MHz in Stokes-V ($|V^{'}|$). The \editsjmg{tentative} signal is most clearly seen in Q4f, \editsjmg{which is} distinctly different \editsjmg{for the ON-beam (black curve) than for} the OFF beams \editsjmg{(red curve)}. The other comments are the same as Figure \ref{fig:L569131_detection}. \edits{The probability to reproduce the ON-beam curve by chance is} 1.3$\%$ or 2.2$\sigma$, \editsjmg{whereas} the false-positive probability \editsjmg{for} the OFF beams is 59$\%$. Additionally, we performed the Kolmogorov–Smirnov statistical test on the two curves for Q4f in \textit{panel f} and find that the probability to reject the null hypothesis (that the two curves are drawn from the same parent distribution) is 76$\%$. Therefore, the signal is \edits{possibly} a false-positive. 
  }
  \label{fig:L545197_detection}
\end{figure*}

\end{appendix}

\end{document}